\documentclass[fleqn,usenatbib]{mnras}

\usepackage{newtxtext,newtxmath}
\usepackage[T1]{fontenc}

\DeclareRobustCommand{\VAN}[3]{#2}
\let\VANthebibliography\thebibliography
\def\thebibliography{\DeclareRobustCommand{\VAN}[3]{##3}\VANthebibliography}

\usepackage{graphicx}	% Including figure files
\usepackage{amsmath}	% Advanced maths commands

\usepackage{lineno}
\usepackage{gensymb}
\usepackage[pdfpagelabels=false]{hyperref}	% Hyperlinks
\hypersetup{colorlinks=true,linkcolor=blue,citecolor=blue,filecolor=blue,urlcolor=blue,}
\usepackage[table,xcdraw]{xcolor}
\usepackage{soul}
\usepackage{comment}
\usepackage{lineno}
\usepackage[normalem]{ulem}

\usepackage{eso-pic}% http://ctan.org/pkg/eso-pic

\AddToShipoutPictureBG*{%
  \AtPageUpperLeft{%
    \hspace{0.75\paperwidth}%
    \raisebox{-5\baselineskip}{%
      \makebox[0pt][l]{\textnormal{DES-2022-0748}}
}}}%

\AddToShipoutPictureBG*{%
  \AtPageUpperLeft{%
    \hspace{0.75\paperwidth}%
    \raisebox{-6\baselineskip}{%
      \makebox[0pt][l]{\textnormal{FERMILAB-PUB-23-422-PPD}}
}}}%

\newcommand{\rdmp}{\textsc{redMaPPer}~}
\newcommand{\auto}{\textsc{Auto}~}

\title[ICL redshift 0.2 to 0.5]{Dark Energy Survey Year 6 Results: Intra-Cluster Light from Redshift 0.2 to 0.5}
\author[DES Collaboration]{
\parbox{\textwidth}{
Yuanyuan~Zhang$^{1,2}$ \thanks{E-mail: yuanyuan.zhang@noirlab.edu},
Jesse~B.~Golden-Marx$^{3}$,
Ricardo~L.~C.~Ogando$^{4}$,
Brian~Yanny$^{5}$,
Eli~S.~Rykoff$^{6,7}$,
Sahar~Allam$^{5}$,
M.~Aguena$^{8}$,
D.~Bacon$^{9}$,
S.~Bocquet$^{10}$,
D.~Brooks$^{11}$,
A.~Carnero~Rosell$^{12,8,13}$,
J.~Carretero$^{14}$,
T.-Y.~Cheng$^{15}$,
C.~Conselice$^{16,17}$,
M.~Costanzi$^{18,19,20}$,
L.~N.~da Costa$^{8}$,
M.~E.~S.~Pereira$^{21}$,
T.~M.~Davis$^{22}$,
S.~Desai$^{23}$,
H.~T.~Diehl$^{5}$,
P.~Doel$^{11}$,
I.~Ferrero$^{24}$,
B.~Flaugher$^{5}$,
J.~Frieman$^{5,25}$,
D.~Gruen$^{10}$,
R.~A.~Gruendl$^{26,27}$,
S.~R.~Hinton$^{22}$,
D.~L.~Hollowood$^{28}$,
K.~Honscheid$^{29,30}$,
D.~J.~James$^{31}$,
T.~Jeltema$^{28}$,
K.~Kuehn$^{32,33}$,
N.~Kuropatkin$^{5}$,
O.~Lahav$^{11}$,
S.~Lee$^{34}$,
M.~Lima$^{35,8}$,
J. Mena-Fern{\'a}ndez$^{36}$,
R.~Miquel$^{37,14}$,
A.~Palmese$^{38}$,
A.~Pieres$^{8,4}$,
A.~A.~Plazas~Malag\'on$^{6,7}$,
A.~K.~Romer$^{39}$,
E.~Sanchez$^{36}$,
M.~Smith$^{40}$,
E.~Suchyta$^{41}$,
G.~Tarle$^{42}$,
C.~To$^{29}$,
D.~L.~Tucker$^{5}$,
N.~Weaverdyck$^{42,43}$
\begin{center} (DES Collaboration) \end{center}
}
\vspace{0.4cm}
\\
}
\date{Accepted XXX. Received YYY; in original form ZZZ}

\pubyear{2023}

\begin{document}
\label{firstpage}
\pagerange{\pageref{firstpage}--\pageref{lastpage}}
\maketitle

% Abstract of the paper
\begin{abstract}

Using the full six years of imaging data from the Dark Energy Survey, we study the surface brightness profiles of galaxy cluster central galaxies and intra-cluster light. 
We apply a ``stacking'' method to over four thousand galaxy clusters identified by the redMaPPer cluster finding algorithm in the redshift range of 0.2 to 0.5. This yields high signal-to-noise radial profile measurements of the central galaxy and intra-cluster light out to 1 Mpc from the cluster center. Using redMaPPer richness as a cluster mass indicator, we find that the intra-cluster light brightness has a strong mass dependence throughout the 0.2 to 0.5 redshift range, and the dependence grows stronger at a larger radius. In terms of redshift evolution, we find some evidence that the central galaxy, as well as the diffuse light within the transition region between the cluster central galaxy and intra-cluster light within 80 kpc from the center, may be growing over time. At larger radii, more than 80 kpc away from the cluster center, we do not find evidence of additional redshift evolution beyond the cluster mass dependence, which is consistent with the findings from the IllustrisTNG hydrodynamic simulation.  We speculate that the major driver of intra-cluster light growth, especially at large radii, is associated with cluster mass growth. %In the central galaxy and in the transition region between the cluster central galaxy and intra-cluster light, there may be some growth driven by galaxy stripping and disruption. 
Finally, we find that the color of the cluster central galaxy and intra-cluster light displays a radial gradient that becomes bluer at a larger radius, which is consistent with a stellar stripping and disruption origin of intra-cluster light as suggested by simulation studies.

\end{abstract}

% Select between one and six entries from the list of approved keywords.
% Don't make up new ones.

%%%%%%%%%%%%%%%%%%%%%%%%%%%%%%%%%%%%%%%%%%%%%%%%%%

%%%%%%%%%%%%%%%%% BODY OF PAPER %%%%%%%%%%%%%%%%%%
\section{Introduction}
Galaxy clusters contain a diffuse stellar component of intra-cluster light (ICL). First discovered more than half a century ago \citep{1951PASP...63...61Z, 1952PASP...64..242Z}, ICL is abundant around the cluster central galaxies (CGs) or the brightest cluster galaxies (BCGs), and contains stars dispersed into the intra-cluster space. It has been  studied using optical or infrared imaging, and spectroscopic observations, which have been reviewed in  \cite{2017ASSL..434..333A, 2017PhDT.......215D,  2021Galax...9...60C, 2022NatAs...6..308M, 2022arXiv221209569A}. %This diffuse stellar component b to  with the cluster galaxies, thus often known as "Intra-Cluster Light". %In recent years, Intra-cluster light is also studied spectrocopcially by Integral Field Unit.% Extensive reviews on this topic can be found in \cite{2021Galax...9...60C, 2022NatAs...6..308M}
Because of the ICL's faint brightness and the overall difficulties of studying low surface brightness features \citep{2017ASSL..434..333A, 2019arXiv190909456M}, ICL has remained a poorly understood subject % in the context of studying galaxy cluster formation and evolution
until recently, when the number of studies jumped with refreshed interests due to new data \citep{2022ApJ...940L..51M}, simulations \citep{2022ApJ...934...43S}, and techniques \citep{2022MNRAS.514.3082M}. %{\color{red} some pieces of ICL formation and evolution especially in the context of how it connects to the formation and evolution of the galaxy cluster system are still missing.}

% icl in ads  https://ui.adsabs.harvard.edu/search/filter_database_fq_database=AND&filter_database_fq_database=(database%3A%20(physics%20OR%20astronomy))&filter_database_fq_database=database%3A%22astronomy%22&filter_property_fq_property=AND&filter_property_fq_property=property%3A%22refereed%22&fq=%7B!type%3Daqp%20v%3D%24fq_database%7D&fq=%7B!type%3Daqp%20v%3D%24fq_property%7D&fq_database=((database%3A%20(physics%20OR%20astronomy))%20AND%20database%3A%22astronomy%22)&fq_property=(property%3A%22refereed%22)&q=((title%3A(%22intracluster%20light%22))%20AND%20year%3A1949-2023)&sort=date%20desc%2C%20bibcode%20desc&p_=0

Simulation and semi-analytical studies have investigated ICL formation in many different channels \citep[e.g.,][]{2006MNRAS.369..958S, Rudick_2006, 2009JApA...30....1B, 2009ApJ...699.1518R, 2010MNRAS.406..936P, 2014MNRAS.437.3787C, 2012ApJ...757...48M} including galaxy disruption, stellar stripping, merging \citep{2004ApJ...607L..83M, 2007MNRAS.377....2M, 2018MNRAS.479..932C} and preprocessing \citep{2022arXiv221202510C}. The ICL's formation is often studied together with the evolution of cluster CGs or even the cluster's overall galaxy distribution due to difficulties in separating the two \citep[e.g., ][]{2007ApJ...668..826C, 2006ApJ...652L..89M, 2015MNRAS.451.2703C, 2018MNRAS.475..648P, 2020MNRAS.494.4314C}. 
Different channels of ICL formation carry implications for the ICL's observational properties and their redshift evolution in terms of age, color, and metallicity \citep{2017MNRAS.467.4501H, 2019ApJ...871...24C}, fraction of ICL in the cluster stellar light \citep{2007ApJ...666...20P, 2007MNRAS.377....2M, 2014MNRAS.437..816C, 2018ApJ...859...85T}, morphology \citep{Rudick_2006}, or scaling relation to cluster mass or mass distribution \citep{2020MNRAS.494.1859A, 2021ApJ...915..106C}. For example, \cite{2019ApJ...871...24C} analyzed ICL color and metallicity using semi-analytical models which contain ICL formed through tidal stripping of cluster satellite galaxies as well as through merging relaxation; they found a negative radial color and metallicity gradient. From hydrodynamic simulations, \citet{2018MNRAS.475..648P}  found that ICL stellar mass strongly correlates with the host halo mass, but this correlation appears to evolve little from redshift 1 to 0.

%\cite{2012A&A...537A..64G} has two bands for restricted redshift range but does not conclude anything regarding color nor its evolution :/ 

% bcg is  

In observational studies, the formation and evolution of ICL has been studied using its color \citep{1992ApJ...400...65M, 2006AJ....131..168K, 2007AJ....134..466K, 2014ApJ...794..137M, 2015MNRAS.448.1162D, 2017ApJ...851...75I, 2017ApJ...846..139M, 2018MNRAS.474.3009D, 2018MNRAS.474..917M, 2018ApJ...862...95K, 2021ApJ...910...45M, 2021A&A...651A..39R, 2021MNRAS.508.2634Y, 2022arXiv220905519G, 2023MNRAS.518.1195M}, stellar mass \citep{2011ApJ...735...76K, 2012MNRAS.425.2058B, 2020MNRAS.491.3751D, 2020A&A...639A..14S, 2022ApJ...930...25B}, stellar population spectroscopic studies, \citep[e.g.,][]{2010A&A...519A..95C, 2010MNRAS.407L..26C, 2011A&A...533A.138C, 2011A&A...528A..24V, 2012A&A...545A..37A, 2015Galax...3..212L, 2016MNRAS.461..230E,  2016A&A...589A.139B} %,2018A&A...620A.111L, 2018ApJ...864...36L, 2018A&A...619A..70H, 2020ApJ...894...32G, 2022MNRAS.511..201P} 
 and is often investigated together with BCG evolution \citep[e.g., ][]{2005ApJ...618..195G, 2016ApJ...816...98Z, 2022ApJ...928...28G}.
For example,  \cite{2020MNRAS.491.3751D} studied the stellar mass of BCG and ICL between redshift 0.05 to 1.75 and found its growth rate to be greater than that of the cluster by a factor of two. They also found that the core of the BCG formed early while the BCG outskirt and ICL were built at later times. On the other hand, detailed analysis of local BCG and ICL stellar populations by \cite{2020MNRAS.491.2617E} indicate that while the stellar population in the ICL is old, it is still younger ($\approx 9$ Gyr) than the BCG ($\approx 13$ Gyr), pointing towards a late and continuous formation of ICL through minor merging. %Hence, one should be able to follow this evolution? 4Gyr is enough to tell? how does it overlap with our redshift range? how much of history do we probe? 
% Spectroscopic Constraints on the Buildup of Intracluster Light in the Coma Cluster, Meng Gu
%Similarly, \citet{2020ApJ...894...32G} looked at 3 different regions of Coma cluster using IFU spectra and concluded that they are not only old (ages between 6.7 and 12.7 Gyrs) but metal poor, indicate the accretion of quiescent low-mass galaxies or the tidal stripping of massive galaxies -  but noting that Coma cluster is a double system.

In this work, we continue the observational study of ICL evolution by examining its properties in the redshift range of 0.2 to 0.5.  Our work is based on thousands of galaxy clusters and the full six years of observations from the Dark Energy Survey (DES) \citep{2021ApJS..255...20A}, a wide-field imaging survey \citep{DES2005} designed to probe cosmic structures in the late Universe \citep[e.g., ][]{2020PhRvD.102b3509A, 2022PhRvD.105d3512A, 2022PhRvD.105b3520A, 2022arXiv220705766D}.  We use a ``stacking'' method \citep{2004MNRAS.347..556Z, z05, 2011ApJ...731...89T, 2019ApJ...874..165Z, 2021MNRAS.501.1300S, 2022MNRAS.514.2692C, 2023MNRAS.518.3685A} with the DES galaxy cluster sample to reduce measurement noise. Our goal is to acquire high signal-to-noise measurements of the ICL surface brightness profile, color, and luminosity and quantify their evolution between redshift 0.2 and 0.5.  This paper presents one of the largest ICL redshift evolution studies, based on a cluster sample a few times larger than that in \cite{2022arXiv220905519G} which used a Cosmic Microwave Background (CMB) selected cluster sample from the Atacama Cosmology Telescope.
%\green{you could also compare this to in that to my work by saying you have x times more clusters than it}

%There are many difficulties in observational studies of ICL \citep[see reviews in][]{2017PhDT.......215D, 2022NatAs...6..308M}. 
One challenge to ICL studies is the difficulty in disentangling ICL from cluster CGs \citep[see discussions in][]{2010MNRAS.405.1544D, 2022ApJ...928...99C}. Although stars in the intra-cluster space may have different stellar composition or dispersion dynamics  \citep[e.g., ][]{2015A&A...579A.135L, 2018A&A...620A.111L, 2018ApJ...864...36L, 2018A&A...619A..70H, 2020ApJ...894...32G, 2022MNRAS.511..201P} than cluster CGs or BCGs -- from imaging data alone, it is often difficult to separate the ICL from the low-surface brightness outskirts of those galaxies. %\green{is there a reason your using cluster central galaxies?}{\color{red}I was thinking of using the BCG acronym to refer to cluster central galaxies are getting confusing since many people still use it to refer to brightest cluster galaxies. Hence CG since it's a bit less ambiguous.} \green{That makes sense to me} 
Different separation methods have been suggested \citep{2011ApJ...732...48R}, including analytical decomposition, using a machine learning algorithm \citep{2022MNRAS.514.3082M}, surface brightness limits \citep{2014A&A...565A.126P}, or using physical distance apertures to separate those components. In this paper, we follow the practice of \citet{2018MNRAS.475..648P} who analyzed CG and ICL as the ``diffuse light'' of galaxy clusters. We use the phrase diffuse light interchangeably with CG+ICL in this paper. When needed, we use a physical aperture to separate CG and ICL, with ICL defined as the diffuse light beyond 30 kpc from the CG center (an outer radius limit is defined according to the context), while CG is defined as the diffuse light component within 30 kpc. 

The remainder of this paper is organized as the following. In section~\ref{sec:data} we review our data sets and the methods. Section~\ref{sec:surface} presents our measurements of the diffuse light surface brightness, while Section~\ref{sec:lum_stellar} quantifies the cluster mass and redshift dependence of the diffuse light luminosities. %, and section~\ref{sec:color} shows our measurements of the BCG and ICL color. 
Section~\ref{sec:systematics} discusses observational effects that may impact the interpretation of our results, and Section~\ref{sec:comparisons} discusses our results in the context of simulations and other observational studies. Section~\ref{sec:summary} summarizes our findings. Throughout this paper, we assume a flat $\Lambda$CDM cosmology with $\Omega_m = 0.3$, and $h = 0.70$.

%\green{since I haven't read the rest yet, I'm not sure if it's mentioned, but should you provide the cosmology that's assumed?}

\section{Data and Methods}
\label{sec:data}
\subsection{The \rdmp Galaxy Cluster Catalog}

% If you use beamer only pass "xcolor=table" option, i.e. \documentclass[xcolor=table]{beamer}
\begin{table*}
\caption{The Galaxy Cluster Sample in this Analysis}
\label{tbl:clusters}
\begin{tabular}{|ll|lllllll|}
\hline
    Redshift ($z$) Bin    &         Richness ($\lambda$) Bin         & Number Counts & Median $z$     & Mean $z$ & Median $\lambda$ & Mean $\lambda$ & \begin{tabular}[c]{@{}l@{}}$R\_\lambda$ (Mpc/h)\\ Based on Mean $\lambda$\end{tabular} & \begin{tabular}[c]{@{}l@{}} Masking limit\\ ($z-$band) {\tt mag\_auto}\end{tabular}\\
\hline 
\rowcolor[HTML]{EFEFEF} 
0.2-0.3 &                  & 1169          & 0.256                         & 0.255         & 28.55         & 33.61       &       &   20.67                                                                          \\
        & 20-30 & 656           & 0.254                         & 0.253         & 23.72         & 24.17       & 1.03    &                                                                          \\
        & 30-45 & 326           & 0.259                         & 0.256         & 35.51         & 35.97       & 1.23   &                                                                           \\
        & 45-60 & 121           & 0.257                         & 0.255         & 50.51         & 51.16       & 1.44  &                                                                            \\
        & 60+     & 66            & 0.274                         & 0.263         & 73.55         & 83.58       & 1.80   &                                                                           \\
        \hline 
\rowcolor[HTML]{EFEFEF} 
0.3-0.4 &                  & 1556          & 0.359                         & 0.355         & 27.33         & 32.92       &         &       21.38                                                                   \\
        & 20-30 & 942           & 0.359                         & 0.354         & 23.70         & 24.11         & 1.03    &                                                                          \\
        & 30-45 & 399           & 0.358                         & 0.355         & 35.44         & 36.01       & 1.23    &                                                                          \\
        & 45-60 & 115           & 0.360                         & 0.356         & 52.56         & 52.06       & 1.45    &                                                                          \\
        & 60+     & 100           & 0.361                         & 0.355         & 75.25         & 81.57       & 1.78   &                                                                           \\
        \hline 
\rowcolor[HTML]{EFEFEF} 
0.4-0.5 &                  & 1357          & 0.449                         & 0.449         & 27.13         & 32.10       &       &       21.87                                                                     \\
        & 20-30 & 836           & 0.449                         & 0.449         & 23.44         & 23.97        & 1.03   &                                                                           \\
        & 30-45 & 349           & 0.451                         & 0.449         & 35.14         & 35.90       & 1.23   &                                                                           \\
        & 45-60 & 96            & 0.454                         & 0.452         & 51.08         & 51.78       & 1.45   &                                                                           \\
        &  60+     & 76            & 0.440                         & 0.445         & 69.83         & 79.17       & 1.76   &                                              \\
        \hline                            
\end{tabular}
\end{table*}

% https://cdcvs.fnal.gov/redmine/projects/des-clusters/wiki/RedMaPPer_on_Y3A2
% redmapper_y3a2_6.4.22v2 
% y3_gold_2.2.1_wide_sofcol_run2_redmapper_v6.4.22+2_lgt20_vl02_catalog.fit 
% Y3A2 Gold 2.2.1 sof colors, redMaPPer v6.4.22+2, Volume Limited, lambda>20
% https://arxiv.org/pdf/2110.02418.pdf 
% y3 volume limited

The red sequence Matched-filter Probabilistic Percolation cluster finder algorithm (\rdmp) \citep{2014ApJ...785..104R} has been used by the DES Collaboration to derive galaxy cluster catalogs from the Science Verification data,  \citep{2016ApJS..224....1R}, the Year 1 observations \citep{2018arXiv180500039M}, and the Year 1 to Year 3 observations \citep{2021arXiv211002418O}.  
\rdmp is a red-sequence based algorithm that provides excellent cluster richness ($\lambda$) and photometric redshift estimates \citep{2014ApJ...785..104R}. It also provides a random point catalog that tracks the sky footprint and depth covered by the cluster-finding algorithm. 

This paper is based on the \rdmp cluster catalog,  version 6.4.22+2, derived from the DES Year 3 Gold data sets \citep{2021ApJS..254...24S}. A relevant difference between this catalog and the DES Year 1 version \citep{2018arXiv180500039M} is the much larger sky coverage. As a result, this catalog contains more than 21,000 galaxy clusters with richness above 20, which approximately corresponds to a halo mass threshold of  $10^{14.1}$M$_\odot$ \citep{2018arXiv180500039M, 2019MNRAS.490.3341F}.  These galaxy clusters are detected from the DES single-object fit (SOF) catalog \citep{2018ApJS..235...33D, 2021ApJS..254...24S} which contains objects detected and deblended by \textsc{Source Extractor} \citep{1996A&AS..117..393B}, while the photometry was derived from single-object fitting using the {\tt ngmix} algorithm on multi-epoch image stamps of each object, and the deblended nearby objects are masked on each single-epoch image. For the DES Year 3 data processing campaign, the SOF photometry measurements are preferred in many applications because of the tighter photometry constraints compared to the \textsc{Source Extractor} measurements derived using the coadded images.

Of particular importance to this analysis, \rdmp provides CG (Central Galaxy) candidates for each cluster. Unlike algorithms that aim to select the BCG (Brightest Cluster Galaxies), \rdmp aims to select a relatively luminous cluster galaxy that is nearest to the cluster's gravitational center, and the goal of this selection is to find the central galaxy of the massive dark matter halo as defined in simulation modeling studies \citep[e.g.][]{2007MNRAS.375....2D, 2008ApJ...676..248Y}. \rdmp provides five CG candidates for each cluster. We use the most likely CG candidate; multi-wavelength studies have shown that this candidate is the correct one with an $\sim 80\%$ frequency \citep{2015MNRAS.454.2305S, 2019MNRAS.487.2578Z, 2020ApJS..247...25B}.

%We selected XXX galaxy clusters in the redshift range of $0.2 < z < 0.6$ and  
%describe cluster catalog. 
%random catalog description 

For this diffuse light analysis, we apply a few additional selection criteria to the \rdmp clusters as well as to the \rdmp random points. (1) Around the cluster center (or a random point), in a circular region with a radius of 0.15 $\mathrm{deg}$, we require at least one DES exposure image in each of the $g$, $r$, $i$, and $z$ filters. %This requirement yields $\sim$4869 $\mathrm{deg}^2$ of DES data. 
(2) Around the cluster center (or a random point), in a circular region with  a radius of 0.15 $\mathrm{deg}$, we require the 10~$\sigma$ depth magnitude of the DES \auto measurements to be deeper than a redshift dependent ``masking'' magnitude (see next section for details). This selection criterion ensures that our diffuse light measurements are comparable between different redshift slices. The cut has minimal effect on the clusters/randoms below redshift 0.4 but excludes a small fraction of the clusters between redshift 0.4 and 0.5 and most of the clusters above redshift 0.5. (3) Around the cluster center (or the random point), in a circular region with a radius of 0.2 $\mathrm{deg}$, we exclude areas containing famous or bright stars (the Yale bright stars or 2MASS stars of $J$ < 8) \footnote{This cut requires HEALPix values in the DES foreground map file to be less than 2.}, nearby galaxies including the Large Magellanic Cloud, and globular clusters to reduce scattered light in the images.% and it excludes {\color{red} $\sim$ X \% } of the clusters. 

After applying these selection criteria, we are left with a sample of over 4000 clusters in the redshift range of 0.2 to 0.5. The number of clusters in each richness/redshift bin is listed in Table ~\ref{tbl:clusters}. We also include clusters in the redshift range of 0.5 to 0.6 in some of the analyses in this paper. However, because of the DES depth limit, we are concerned that our galaxy masking procedure (see Section~\ref{sec:method}) may be incomplete in this redshift range\footnote{The redshift 0.5 to 0.6 clusters will be masked to 22.32 mag in z-band in {\tt mag\_auto}. Only $\sim 1\%$ of the redMaPPer clusters reach 22.3 mag in the DES 10$\sigma$ z-band depth map continuously in the whole 0.15 deg$^2$ region around them and a depth-based cut would eliminate most of the clusters. For comparison, The redshift 0.4 to 0.5 clusters will be masked to 21.9 mag in z-band in {\tt mag\_auto}, while 67\% of the redMaPPer clusters reach 21.9 mag in the DES 10$\sigma$ z-band depth map continuously in the whole 0.15 deg$^2$ region around them. Note that the DES coadd catalogs are generally over 95\% complete above 23.7 mag in z-band, and the decision of a depth cut is made out of an abundance of caution.}  and above. Therefore, the measurements of the 0.5 to 0.6 clusters are presented only for illustrative purposes and are not included in our quantification of diffuse light evolution. 
%\green{I'm not sure if it's commonplace, but I referred to it as BCG+ICL, could you call it CG+ICL?}

%DES DM or MOF as source? SOF? comment it against.
%galaxy clusters pruning following DW18. radius
%depth cut ahead 
%what is the number of clusters? mass range? 
%entire sample? complete? 
%discuss redshift measurement

%\begin{figure*}
	% To include a figure from a file named example.*
	% Allowable file formats are eps or ps if compiling using latex
	% or pdf, png, jpg if compiling using pdflatex
	%\includegraphics[width=2.0\columnwidth]{figures/rdmp_z_lambda.png}
    %\caption{}
    %\label{fig:lambda_z}
%\end{figure*}

We note that a redMaPPer galaxy cluster catalog based on the full six years of DES observations is also internally available to the DES collaboration. However, we opt to use the Year 3 version described here because its richness definition has better consistency with the Year 1 version in \cite{2018arXiv180500039M}, which provides the richness-mass relation used in our estimations. The redMaPPer catalog based on Year 1 to Year 6 observations goes to higher redshift ($z\sim0.9$) than the Year 3 version in this paper (which is based on Year 1 to Year 3 observations), but both versions have excellent redshift coverage in the 0.2 to 0.5 redshift range studied in this paper.

\subsection{DES Object Catalogs and Images}

% y6 des coadd images without background subtraction 
% https://des.ncsa.illinois.edu/releases/dr2 
% this is y6 coadd from DR2
% (Sevilla-Noarbe et al. 2021) 
In this paper, we use the DES images and catalogs produced by the Dark Energy Survey Data Management (DESDM) project \citep{2011arXiv1109.6741S, 2018PASP..130g4501M}. A detailed description of the DESDM pipeline can be found in \cite{2018ApJS..239...18A}.  To summarize, the DESDM pipeline takes raw images from the Dark Energy Camera (DECam) \citep{2015AJ....150..150F}, performs instrumental signature removals and corrections \citep{2014PASP..126..750P, 2015JInst..10C5032G},  flat-field corrections \citep{2018PASP..130e4501B}, full-focal-plane background subtractions \citep{2017PASP..129k4502B}, as well as photometric \citep{2018AJ....155...41B} and astrometric \citep{2017PASP..129g4503B} calibrations to produce science-ready single exposure images. Those images are coadded \citep{2002ASPC..281..228B} into multi-epoch coadd images, which are used to produce object catalogs and photometry measurements by the \textsc{Source Extractor} software \citep{1996A&AS..117..393B}. The science-ready single exposure images are also used as input for photometry measurements such as the \textsc{ngmix} photometry measurements mentioned in the previous section.

For this work, we benefit from the full 6 years of DES operations \citep{2018SPIE10704E..0DD}, and the DES Data Release 2 (DR2) processing campaign \citep{2021ApJS..255...20A} which includes not only more data, but also improved  processing since the previous data release. Changes and improvements relevant to our analysis include: 
%tiles being combined with more exposures and at least half coverage in all bands, improving data uniformity; the coadded images set that we use relies only on single-epoch full-focal-plane background subtraction and does not include the Swarp coaddition background subtraction applied to the main released dataset, since it can produce spurious measurements close to extended objects like clusters; combining $riz$ images into detection image as an average created more robust faint objects detection, specially close to diffuse light; finally, source detection threshold went from 10$\sigma$ to $5\sigma$.
coadded images based on single-epoch full-focal-plane background subtraction, which do not include local background subtraction as applied to previous versions of DES coadd images (we use the ``\_nobkg'' version of the coadd images, which do not have the local short-scale  sky background subtracted as mentioned in Section~\ref{sec:method}); combining DES $r$, $i$ and $z$ band images into detection images as an average to create more robust faint objects detection; finally, changing the source detection threshold from 10$\sigma$ to $5\sigma$ to produce more complete object catalogs. The DR2 coadd images have a combined sky coverage of 4913 $\mathrm{deg}^2$ in DES $g$, $r$, $i$, $z$ and $Y$ bands, and the 95\% completeness of the coadd catalogs reaches 23.7 magnitude in the DES $z$-band, with a 10 $\sigma$ magnitude limit of 23.1.

We use both DES images and catalogs in this paper. Other than the \rdmp cluster catalog, the coadd object catalogs used in this paper are constructed from DES coadd images using the \textsc{Source Extractor} software to detect and deblend  objects. Moreover, we make use of the object's \auto measurements from \textsc{Source Extractor} to determine each object's masking area, which is based on Kron apertures and magnitudes \citep{1980ApJS...43..305K}. %We also apply a  10$\sigma$ depth cut to the object catalogs in order to improve homogeneity across redshifts. 
The images used in this paper include both the science-ready single exposure images and the multi-epoch coadd images. In the next section, we describe how we use these images and catalogs in our workflow.

%previously, RM Y1 on DES Y3 data. Now RM Y3 on DES Y6 data. 

\subsection{The Averaging/Stacking method}
\label{sec:method}

In this paper, we again use a ``stacking'' method described in \citet{2019ApJ...874..165Z}, which has also been adopted in \cite{2020RNAAS...4..174L, 2021MNRAS.501.1300S, 2022arXiv220905519G}. We present ICL and CG properties averaged over a large cluster sample. The ``stacking'' method proceeds as the following: %\green{should my paper be cited here as well? Or not since I don't use the stacked measurements and randoms.} {\color{red} yeah, I am not sure yet, but maybe we should just be a bit more general here, how about the phrasing now?} \green{I think the above phrasing works well}

\begin{enumerate}
    \item Coadd images of each galaxy cluster, are downloaded from the DESDM database\footnote{\url{https://des.ncsa.illinois.edu/desaccess/}}. For each image, we mask all objects above a $z$-band magnitude limit determined by the cluster's redshift. We exclude the \rdmp CG from masking to preserve the CG light. The masking magnitude is chosen to be 0.2$L*$ with $L*$ being the characteristic luminosity of a cluster red galaxy luminosity function measurement \citep{2019MNRAS.488....1Z}. Assuming a faint end slope of -1, this masking limit would remove 82\% of the light from cluster galaxies. 
    \item After masking, the radial surface brightness (SB) profile is derived from the masked images, as the mean pixel value in radial annuli on the images. 
    \item Similarly, radial SB profiles are derived for a sample of random points that cover the same sky area as the cluster catalog. In later steps, those random profiles are subtracted from the cluster profiles to eliminate residual backgrounds in the cluster profiles.
    \item Using a Jackknife resampling method \citep[see examples of applications in][]{2009MNRAS.396...19N, 2017MNRAS.469.4899M}, we divide the cluster samples, and the random points into 40 subsamples according to their sky coordinates\footnote{\url{https://github.com/esheldon/kmeans_radec}}. For each sky coordinate subsample, we derive the CG+ICL SB profile by averaging the profiles of clusters and randoms, and then subtracting the random profiles from that of the clusters. We then apply the Jackknife resampling method \citep{efron1982jackknife} to those sky coordinate subsamples to derive the means and uncertainties of the final CG+ICL measurements. 
    \item Additional quantities, such as the CG+ICL color and luminosities are further computed from the CG+ICL SB profiles. In this paper, we also analyze the radial SB profiles of the cluster total light including CG, ICL and cluster satellite galaxies. Those measurements are derived using the same procedures listed here, but without the objects masking described in step (i).
\end{enumerate}

We highlight one difference between this paper and  \citet{2019ApJ...874..165Z, 2020RNAAS...4..174L, 2021MNRAS.501.1300S, 2022arXiv220905519G}. In Step (i), for each cluster, the previous analyses processed and coadded single exposure images from DESDM. For this work, we make use of the already coadded images from the DESDM database. Those DESDM images are based on the single exposure images, but coadded without applying local background subtraction steps in the \textsc{swarp} and \textsc{Source Extractor} software. In section~\ref{sec:background}, we compare the profiles derived from those coadded images and those based on the single exposure images \citet{2019ApJ...874..165Z, 2020RNAAS...4..174L, 2021MNRAS.501.1300S, 2022arXiv220905519G}. The results are highly consistent.

The computational resources needed for this method are not trivial.  The masking of a $0.15 \times 0.15$ $\mathrm{deg}^2$ region centered on one galaxy cluster, depending on the masking magnitude, can take a few minutes to hours with a single CPU processor. For this work, the masking and profile measurements of each cluster/random point is performed on the Open Science Grid\footnote{\url{https://opensciencegrid.org}}, a High Throughput Computing Consortium. The processing of tens of thousands of clusters or random locations is distributed to thousands of parallel processes on the Open Science Grid in an ``opportunistic'' mode. %, with each process taking from a few minutes to days depending on the difficulty level of the task and the efficiency of the allocated CPU processor. 
Given the need to test and validate the analyses with different set-ups, we estimate that up to hundreds of thousands of CPU hours have been used in this work. 
% cool! 
% Interesting!  I didn't realize you didn't run it on Fermigrid

\section{ICL Surface Brightness}
\label{sec:surface}

\subsection{Richness and Redshift Dependence}

\begin{figure*}
	% To include a figure from a file named example.*
	% Allowable file formats are eps or ps if compiling using latex
	% or pdf, png, jpg if compiling using pdflatex
	\includegraphics[width=2.3\columnwidth]{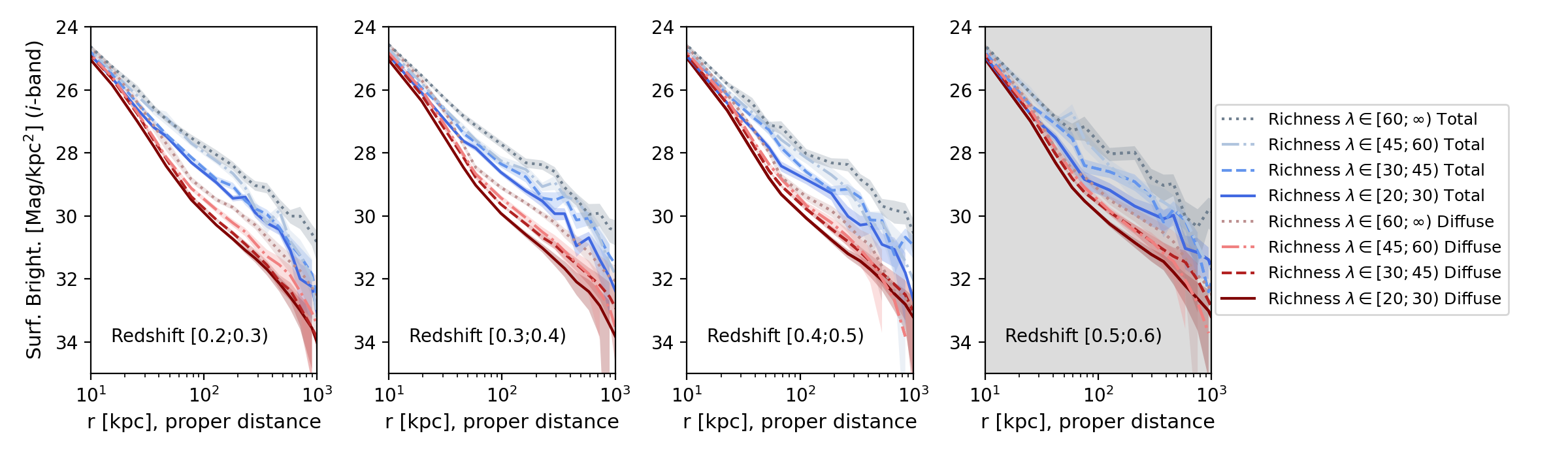}
	\includegraphics[width=2.2\columnwidth]{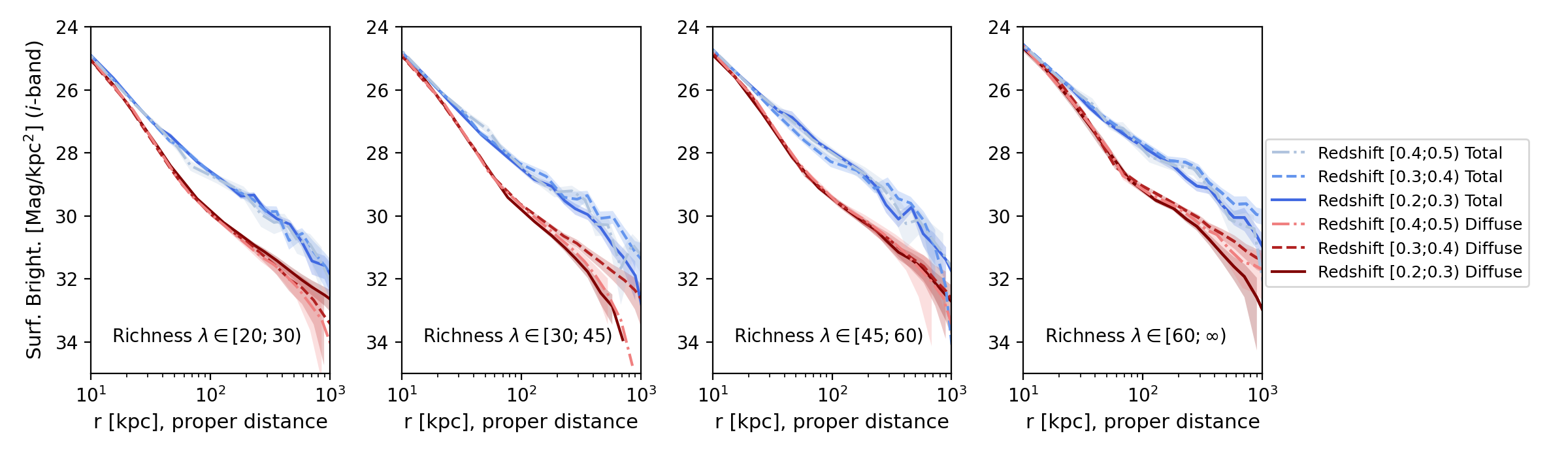}
	\includegraphics[width=2.3\columnwidth]{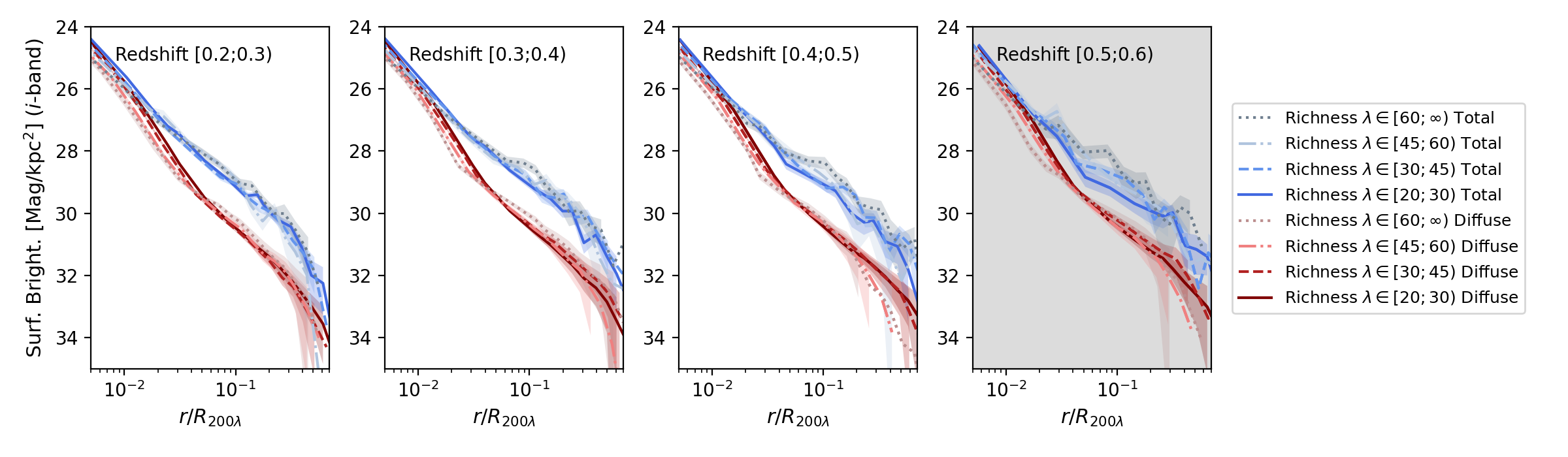}
    \caption{Surface brightness of the clusters in richness/redshift ranges. Upper Row: Clusters in the same redshift ranges in each panel, with different lines representing different richness subsamples. We show both the surface brightness of the diffuse light (CG+ICL, red hues) and also the surface brightness of total cluster light (blue hues). Both profiles display strong richness dependence across the four redshift panels. Middle Row: Clusters in the same richness ranges in each panel, with different lines representing different redshift subsamples. Again, we show both the surface brightness of the diffuse light (CG+ICL, red hues) and also the surface brightness of total cluster light (blue hues). We do not observe consistent redshift trends across the richness panels, indicating weak, if any, signs for redshift evolution. We quantify the significance of the redshift/richness trends in the next Section. Lower Row: Surface brightness profiles after the cluster's radius has been scaled by a percolation radius (corresponding to the cluster subsample's average $R_{200m}$). The radial profiles of the diffuse light as well as the clusters' total stellar contents are ``self-similar'' after radial scaling.}
    \label{fig:sb_mass}
\end{figure*}

Our first analysis is the surface brightness (SB) radial profiles of the diffuse light (CG+ICL), as well as the SB of the total cluster stellar content including the rest of the cluster galaxies. The goal of this analysis is to visually examine the shapes of those profiles and their general redshift/richness trends. 
The galaxy clusters are split into redshift and richness sub-samples, and those surface brightness profiles are presented in Figure~\ref{fig:sb_mass}.

%We first examine the clusters in fixed richness and redshift bins.
We use three redshift bins from 0.2 to 0.5 to analyze the clusters. In each redshift bin, the clusters are further divided into four richness bins, with the richness binning defined in previous DES cluster-lensing studies \citep{2018arXiv180500039M} and listed in Table~\ref{tbl:clusters}. As mentioned in Section~\ref{sec:data},  we apply the ``stacking'' procedure to each redshift/richness binned cluster subsample. The residual background for each subsample is derived from random points, but the masking magnitude limit for the random points is adjusted according to the cluster subsample's redshift range. For each redshift bin, we apply distance corrections %and K-corrections using a stellar population model \citep{2003MNRAS.344.1000B, 2012PASP..124..606M} constructed from a single starburst \citep{2003PASP..115..763C} of metallicity $Z=0.008$ at $z=3.0$, 
so that the measurements are shifted to be in the observer frame of redshift 0.25. Those measurements are presented in Figure~\ref{fig:sb_mass}. In each of those richness and redshift bins, we measure the diffuse light profiles up to 1 Mpc from the center. Our measurements agree with previous studies which show the radial extension of ICL up to several hundreds of kpc, or even one Mpc from the cluster center \citep{2007AJ....134..466K, z05, 2021ApJS..252...27K, 2022MNRAS.515.5335L,  2022MNRAS.514.2692C}.

The upper panels of Figure~\ref{fig:sb_mass} show the SB profiles of the galaxy clusters first split by redshift and then by richness. In each subpanel, the redshift range of the clusters is fixed to be the same and each line represents a different richness range. Those radial SB profiles show a clear richness dependence: richer galaxy clusters generally are brighter in SB, while less rich clusters are fainter. The trends are observed for both the diffuse light, and for the total light including cluster satellite galaxies. Moreover, the distinctions between the different richness subsamples are present throughout the three redshift bins, indicating robust richness dependence across the 0.2 to 0.5 redshift range. Our result supports previous findings that detect strong ICL correlations with cluster mass \citep[e.g.,][]{2005ApJ...618..195G, 2019MNRAS.482.2838M, 2020MNRAS.492.3685H, 2021MNRAS.501.1300S, 2021ApJS..252...27K, 2022MNRAS.515.4722H, 2022arXiv220905519G, 2022arXiv221206164R, 2022MNRAS.514.2692C}. Further, the SB richness dependence in Figure~\ref{fig:sb_mass} grows more prominent with enlarging radius. 

In the middle panel of Figure~\ref{fig:sb_mass}, we present the SB profiles, first split by richness and then by redshift. In each subpanel, the richness range of the clusters is fixed to be the same and each line represents a different redshift range. Interestingly, those redshift-divided profiles appear similar within their uncertainty ranges; while fixing the cluster's richness range, we do not observe a consistent trend of the SB profiles being either brighter or fainter at a lower redshift. The lack of a consistent trend does not necessarily indicate that there is no redshift evolution for a richness-fixed sample, but potentially an evolution that is too small to be noticeable in those SB figures. In Section~\ref{sec:lum_stellar}, we further quantify the redshift-related trends using their luminosity measurements.
%\green{I'm not sure I entirely follow the redshift argument.  Is it just that it might be too small to detect because of the error bars?}{\color{red} How about now?} \green{yup! much clearer}

\subsection{ICL ``Self-Similarity''}

Previously we have noted the remarkable similarity of the ICL SB profiles after scaling by the cluster's radius in \citet{2019ApJ...874..165Z, 2021MNRAS.501.1300S}. Similarly, we investigate this effect with a much larger sample in this analysis. 

Given the relatively small richness range of each bin, we use a similar procedure as described in \citet{2021MNRAS.501.1300S} to scale the radial profiles. For each richness bin, we rescale the clusters' SB profiles by one radius determined by the average richness of each richness bin. Because there are no weak-lensing mass measurements for the galaxy cluster samples studied in this paper, we scale their radial SB profiles by the redMapper percolation radius which is a function of richness, $R_\mathrm{200\lambda} = 1.95 \times (\lambda / 100) ^ {0.45} \mathrm{Mpc}/h$. This radius relation is based on the $R_\mathrm{200m}$ to richness relation derived using the DES Year 1 richness-mass relation \citep{2018arXiv180500039M}, and meant to be an approximation of $R_\mathrm{200m}$ derived from cluster richness. We note that the richnesses are not necessarily consistent between different versions of the redMaPPer catalogs based on different DES data releases (a small difference has been found in preliminary comparisons). The percolation radii used here is a close but not necessarily accurate estimation of the clusters average $R_{200m}$.

The last row of Figure~\ref{fig:sb_mass} shows the SB radial profiles after scaling by the percolation radius. This row is meant to be compared to the top row of the same figure (without radial rescaling). These scaled profiles, both of the diffuse light and the cluster light, indeed appear to be much more similar across the richness bins, especially outside a transition radial range around 0.04 $R_\mathrm{200\lambda}$. In the central regions, the profiles do  not appear to be similar after rescaling, suggesting that the CG SB profiles can not be well-described by scaled cluster radii. This phenomenon can be explained by an inside-out growth scenario \citep{2010ApJ...725.2312O, 2010ApJ...709.1018V}, such that CG stellar cores formed early at $z >2$, while the accretion of CG outskirts and the ICL profiles happen later and are more influenced by the galaxy cluster's mass accretion process.

\subsection{Volume-Limited Cluster Sample}
\label{sec:density_fixed}

\begin{figure}
	% To include a figure from a file named example.*
	% Allowable file formats are eps or ps if compiling using latex
	% or pdf, png, jpg if compiling using pdflatex
	\includegraphics[width=1.0\columnwidth]{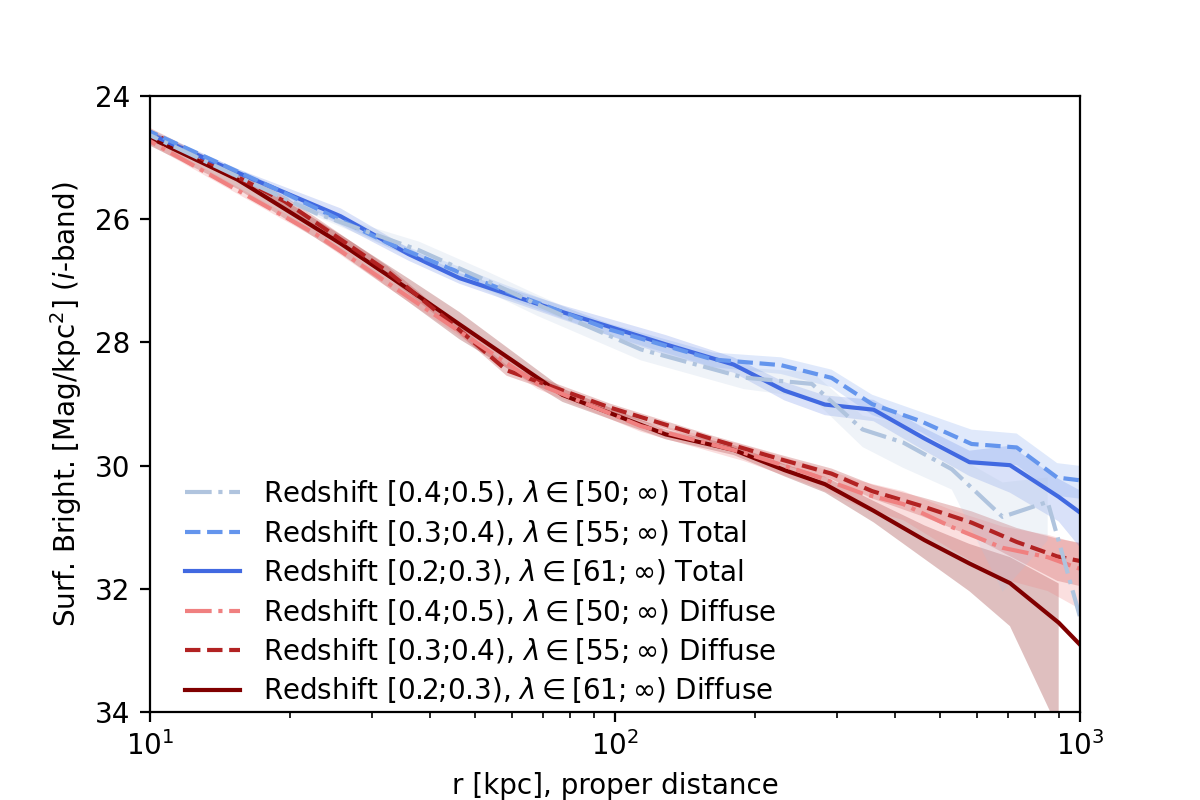}
    \caption{Surface brightness of a volume-limited cluster sample. The radial profiles of these redshift subsamples appear to be consistent within the measurement uncertainties, indicating a redshift evolution that is below a detection level. Again, we show both the surface brightness of the diffuse light (CG+ICL, red hues) and the total cluster light (blue hues). We will further quantify the redshift evolution of the cluster luminosities in the next section.}
    \label{fig:sb_volume}
\end{figure}
 
In the previous subsection, we have shown that when fixing the cluster's richness, clusters in different redshift ranges have similar SB profiles. However, this does not answer the question of how ICL and CGs evolve with time in a single galaxy cluster whose richness will also evolve with time -- more likely, their richnesses/masses will increase over time because of ongoing merging events. 

In this sub-section, we account for cluster richness evolution by constructing a volume-limited cluster sample in different redshift bins. Specifically, in the highest redshift bin of 0.4 to 0.5, we compute the cosmic volume contained in this redshift bin, and select the clusters above richness of 50. For the lower redshift bins, 0.3 to 0.4 and 0.2 to 0.3, we again compute their respective cosmic volumes, and then adjust the richness thresholds for the cluster selections, so that each redshift sub-samples have the same cluster density given their different cosmic volumes. For the redshift slice of 0.3 to 0.4, the richness threshold becomes 55 and for the redshift slice of 0.2 to 0.3, the richness threshold becomes 61. Only clusters above those richness thresholds are selected for comparison in this sub-section. These selections ensure that the cluster samples have the same spatial densities in each redshift bin. A similar volume-limited selection method is also use in \cite{2022arXiv220905519G}, with the distinction that \cite{2022arXiv220905519G} select clusters based on SZ-computed masses, while this analysis is using optical richness. 

The SB-profiles of those volume-limited samples are presented in Figure~\ref{fig:sb_volume}.  We again do not observe significant differences between the redshift sub-samples, as the previous subsection has already noted no visible differences between redshift subsamples within fixed richness ranges. In this exercise, we limit the analysis to a high richness threshold which tends to have lower richness-to-mass scatter \citep{2019MNRAS.490.3341F, 2020MNRAS.495..686A} and is less subjective to potential systematic effects such as line-of-sight projections \citep{2019MNRAS.482..490C, 2020PhRvD.102b3509A, 2022MNRAS.514.4696W, 2022MNRAS.515.4471W}. For the same reason, we also do not sub-divide the clusters according to richness. The consequently smaller sample size lowers the significance of a possible redshift trend. We further quantify the redshift-related trend in Section~\ref{sec:lum_stellar}.

%\section{Additional ICL properties}

\subsection{Color Radial Profiles}
\label{sec:color}

\begin{figure}
	% To include a figure from a file named example.*
	% Allowable file formats are eps or ps if compiling using latex
	% or pdf, png, jpg if compiling using pdflatex
	\includegraphics[width=1.0\columnwidth]{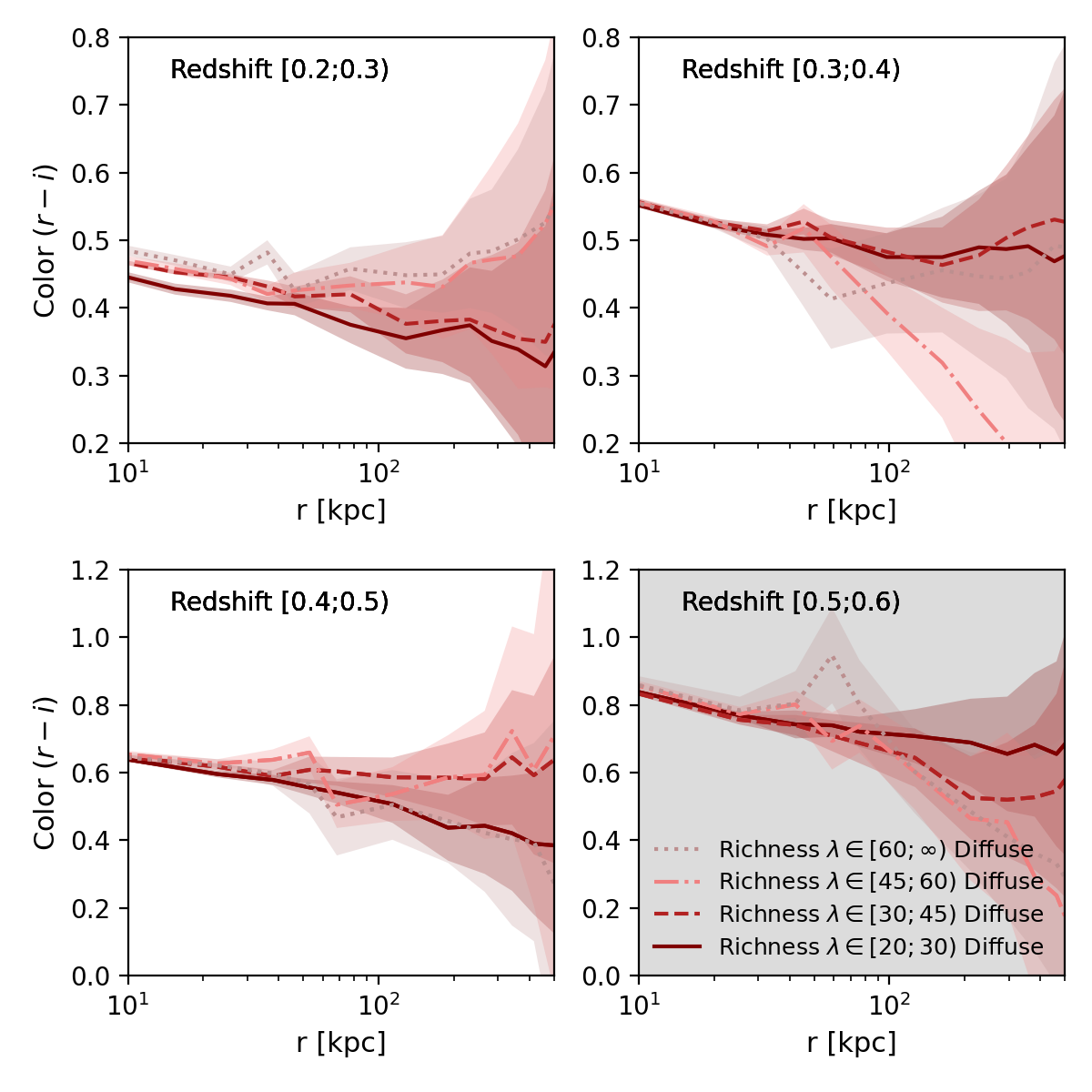}
    \caption{The DES $r-i$ color profile of the diffuse light in cluster subsamples of different redshift and richness ranges. The diffuse light color is consistent with the average color of the cluster's total stellar content in the center. In addition, a radial gradient can be seen in all of the redshift and richness ranges, such that the diffuse light becomes bluer at larger radii. This is consistent with previous studies that ICL consists of more metal-poor and younger stars \citep{2020MNRAS.491.2617E} and suggests an ICL origin from galaxy disruption and tidal stripping \citep{2019ApJ...871...24C}.}
    \label{fig:sb_color}
\end{figure}

\begin{figure}
	% To include a figure from a file named example.*
	% Allowable file formats are eps or ps if compiling using latex
	% or pdf, png, jpg if compiling using pdflatex
	\includegraphics[width=1.0\columnwidth]{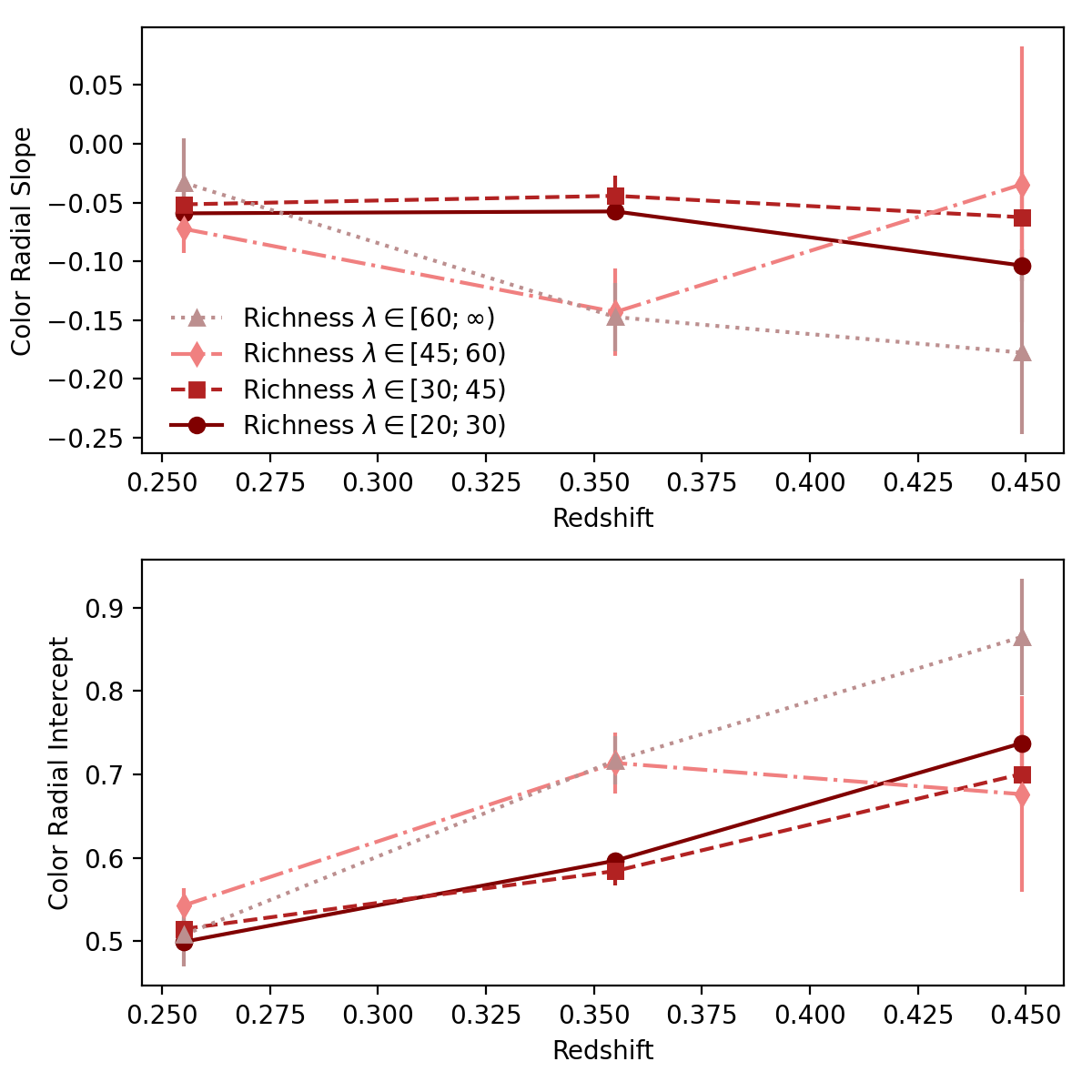}
    \caption{We fit the color profiles in Figure 3 to a linear model with radius $\mathrm{Color}(R) = a\times log(R)+b$ to further quantify those profiles. The radial slope parameter $a$ (upper panel) is negative for all of the richness and redshift bins, which is a robust detection of the radial gradient of the diffuse light's color profile. The $b$ parameter (lower panel), which is the diffuse light's color at $R=1$ kpc, appears to be slightly larger for richer clusters, possibly reflecting a more massive and redder satellite population in richer clusters.}
    \label{fig:sb_color_slope}
\end{figure}

We derive the $r-i$ color of the ICL radial profiles using measurements of the DES $r$ and $i$ band SB profiles. These color measurements are shown in Fig.~\ref{fig:sb_color}. Given that such measurements require highly-significant ICL SB profiles in two bands, we only show color measurements out to a radius slightly beyond $400$ kpc.

The color profiles are presented in redshift subpanels with clusters further divided in richness bins. 
As previously noted in the literature, the color of the diffuse light displays a radial gradient, becoming bluer at a larger radius. 
Interestingly, we also notice a consistent, although not significant, richness trend in those colors, which appear to be redder in richer clusters. In addition to the diffuse light color profiles, we have acquired the color measurements of the cluster's total stellar content, but the measurements are much more uncertain because of Poissonian noise. However, we note that the average color of the cluster's total stellar content is generally consistent with the color of the diffuse light.

To further quantify the colors, we fit those measurements as a function of radius:
\begin{equation}
\mathrm{Color}(R) = a\times \log(R)+b
\end{equation}
Here, $a$  is the radial slope of the colors, and $b$ is the intercept of the profile at $R=1.0$ kpc. The fitted parameters are shown in Figure~\ref{fig:sb_color_slope}. In each redshift/richness bin, $a$ appears to be negative, indicating a robust detection of a radial gradient. This is consistent with previous measurements of ICL radial color gradient \citep[e,g.,][]{z05, 2022MNRAS.514.2692C, 2018MNRAS.474.3009D, 2021MNRAS.508.2634Y} and that the ICL consists of more metal-poor and younger stars \citep[e.g., ][]{2020MNRAS.491.2617E} compared to the CG. As mentioned and analyzed in \cite{2019ApJ...871...24C}, the color radial gradient suggests that the ICL's origin is from galaxy disruption and stripping: if clusters acquire ICL mainly through merging, ICL would have a relatively uniform color because of stellar population mixing. On the other hand, the disruption and stripping of cluster satellite galaxies would produce a radial gradient because of the radial dependence of those processes. 

In addition to the radial gradient, we also detect a possible richness dependence in color, as the intercept ($b$)  of the fitting results appears to be redder (higher positive value) in richer clusters. This is possibly related to richer and thus more massive clusters containing a higher fraction of red sequence galaxies than the less massive clusters \citep{2009ApJ...699.1333H, 2018A&A...613A..67S, 2020MNRAS.498.4303R, 2023MNRAS.524.4455G}.
%\green{this is really interesting.  I think the explanation makes a lot of sense too}

\section{ICL Luminosity}
\label{sec:lum_stellar}

\subsection{Richness and Redshift Dependence}
\begin{figure*}
	% To include a figure from a file named example.*
	% Allowable file formats are eps or ps if compiling using latex
	% or pdf, png, jpg if compiling using pdflatex
	\includegraphics[width=2.3\columnwidth]{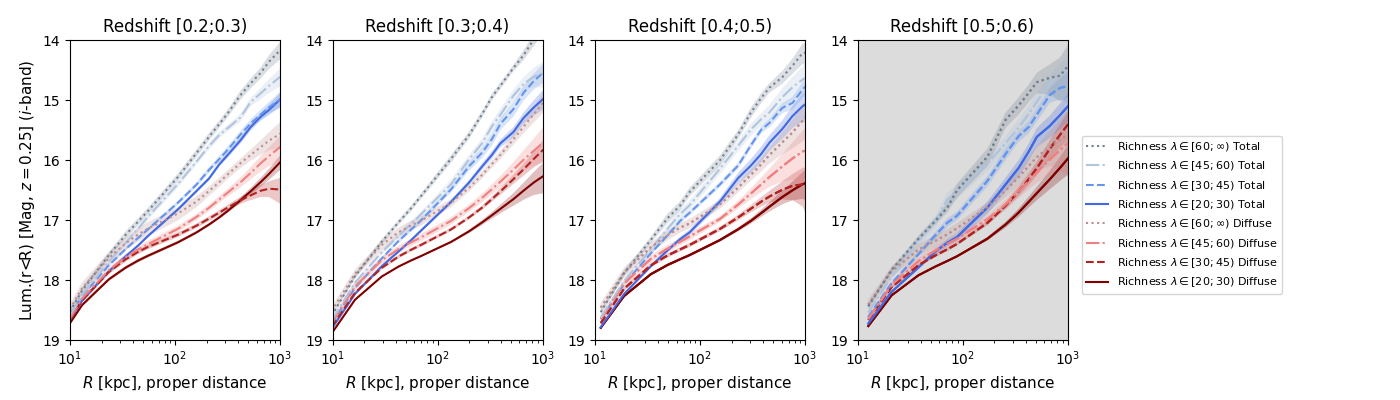}
	\includegraphics[width=2.3\columnwidth]{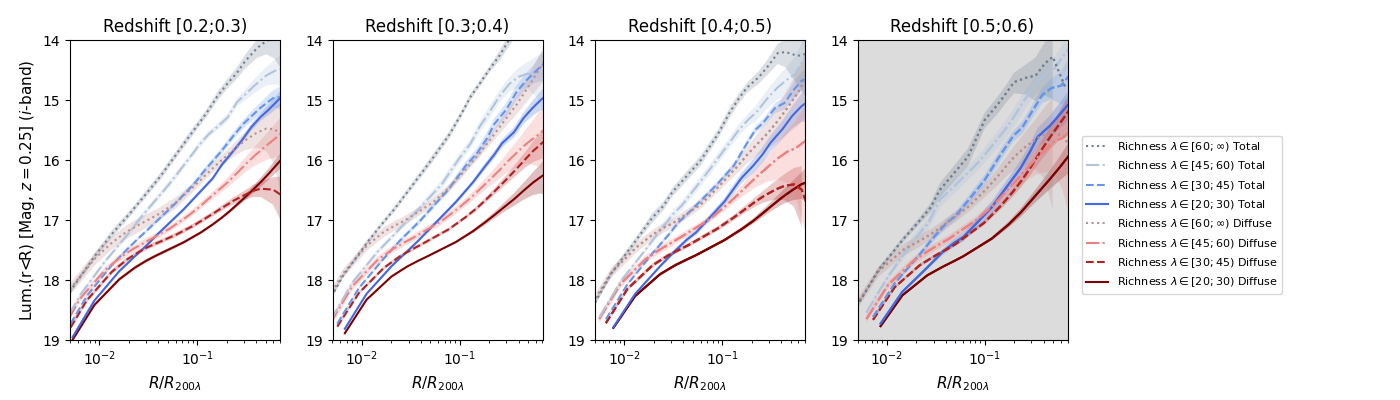}

    \caption{Integrated luminosity (K-corrected to be the apparent magnitude in the observer frame of $z=0.25$) as a function of radius (upper panels), or radius scaled by $R_{200\lambda}$ (lower panels). This figure illustrates the luminosity measurements used in our analyses and shows that diffuse luminosity (red lines) and total cluster stellar luminosity (blue lines) increase as the radial range increases. See Section 4 for discussions on the trends manifested in this figure. Each panel shows a different redshift range, while the last redshift panel (redshift 0.5 to 0.6) is coded in a gray color, indicating that this redshift slice may be less reliable because of potential incomplete masking. We note that the last redshift range is not used in  quantitative analyses. }
    \label{fig:Lum_mass}
\end{figure*}

To further quantify the ICL's richness and redshift dependences, we examine the luminosities of the diffuse light and the cluster's total stellar content. Those luminosities are derived by integrating the SB profile in radial ranges as the following:
\begin{equation}
    L(r<R, z=0.25) = \int_0^{R} S(r, z=0.25) \pi r \mathrm{d} r
\end{equation}
Here, $S(r, z=0.25)$ is the SB measurements presented in the previous section, which have been distance-corrected as if it was observed at redshift 0.25. Thus, $L(r<R, z=0.25)$ is the luminosity measurements enclosed within radius $R$, but shown as the apparent magnitude in the observer frame of redshift 0.25.  We choose this pivot redshift because it is close to the median redshift value in the lowest redshift subsample.  Figure~\ref{fig:Lum_mass} shows the radial profiles of the integrated luminosity as a function of the outbounding $R$.
%\green{Is there a reason you did this to z=0.25?  I'm assuming it's for a comparison to your previous paper, but if not, just wondering}{\color{red} added.}

There are two significant trends in those luminosity profiles. First, richer and thus more massive clusters contain more diffuse and total light. The richness dependence is observed across the four redshift ranges. Second, at small radii, within $\sim$ 50 kpc, diffuse light contributes to the bulk of the cluster total stellar light. Outside $\sim50$ kpc, the cluster's total stellar light increases significantly because of the contribution from cluster satellite galaxies. As a result, diffuse light appears to grow less significantly than the total stellar content with radius. We also show the luminosity radial profiles as a function of radius scaled by $R_{200\lambda}$. The richness/mass dependence of the luminosity becomes even more pronounced in those scaled radius plots. % \green{Since your not showing the ratio in this figure, this may want to be phrased slightly differently so it's more clear where this is coming from} {\color{red} how about now?} \green{this is great now}

\begin{figure*}
	% To include a figure from a file named example.*
	% Allowable file formats are eps or ps if compiling using latex
	% or pdf, png, jpg if compiling using pdflatex
	\includegraphics[width=2.0\columnwidth]{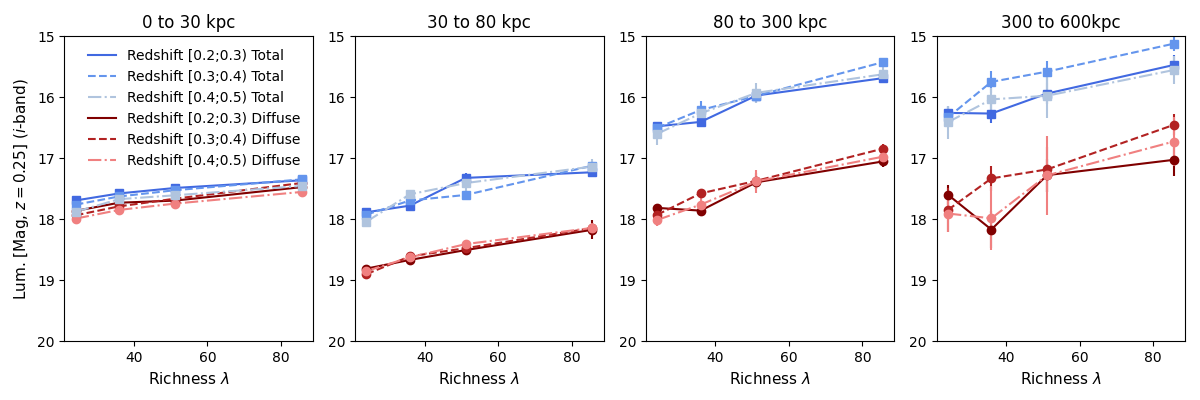}

    \caption{Luminosities (distance-corrected to be the apparent magnitude in the observer frame of $z=0.25$)  enclosed within 4 radial bin (0 to 30 kpc, 30 to 80 kpc, 80 to 300 kpc and 300 to 600 kpc). We examine how the luminosities change with redshift and richness. The luminosities of the  total cluster stellar content (blue lines) and the cluster diffuse light (CG+ICL, red lines), increase significantly as the cluster's richness increases. However, those luminosities do not appear to change significantly with redshift, except in the lowest richness range and within 30 kpc. See Section 4 for quantitative analyses.}
    \label{fig:Lum_aper_all}
\end{figure*}

We further investigate how these luminosities change with redshift and radial apertures. For clusters in a fixed richness range, we  derive their luminosities enclosed within 30 kpc and in 30 to 80, 80 to 300 and 300 to 600 kpc annuli. The innermost 30 kpc radial bin is chosen to match the CG size, while the second radial bin, out to 80 kpc (we have experimented with 50 kpc, 75 kpc and 100 kpc, and found 80 kpc to be most representative in terms of the redshift trends), is chosen to probe the CG to ICL transition range. Finally, the 80 to 300 and 300 to 600 kpc annuli are chosen to probe the extended components of ICL. 
Those luminosity measurements in apertures/annuli are shown in Figure~\ref{fig:Lum_aper_all}. 

Interestingly, with the lowest richness sample -- the clusters in the richness range of 20 to 30 -- we notice that both their diffuse light and total light appear to get brighter towards lower redshift, indicating redshift evolution. In some of the higher richness bins, the luminosities of both the cluster's total and stellar light appear to be getting brighter or unchanged towards lower redshift within 30 kpc. However, outside of 30 kpc, there is no sign of consistent redshift evolution.

\begin{comment}
Before IR
\begin{table}[]
\caption{constraints on parameters in $m(r<R_0) = a\times \mathrm{log}_{10}\frac{\lambda_0}{20} + b \times \mathrm{log}_{10}\frac{1+z_0}{1.25} + c.$, which is used to quantify the relation between stellar luminosity and clustr richness and redshift. }\label{tbl:linear_relation}
\begin{tabular}{|llll|}
\hline
                         & $a$               & $b$            & $c$         \\
\hline
\rowcolor[HTML]{EFEFEF} 
$r \leq$ 50 kpc Total    & $-0.907\pm0.061$  & $1.37\pm0.45$  & $17.382\pm0.021$ \\
\rowcolor[HTML]{EFEFEF} 
$r \leq$ 50 kpc Diffuse  & $-0.922\pm 0.070$ & $1.48\pm0.46$  & $17.709\pm0.021$ \\
$r \leq$ 150 kpc Total    & $-1.31\pm0.08$    & $0.10\pm0.61$  & $16.585\pm0.028$ \\
$r \leq$ 150 kpc Diffuse  & $-1.10\pm0.07$    & $0.75\pm0.46$  & $17.293\pm0.022$ \\
\rowcolor[HTML]{EFEFEF} 
$r \leq$ 300 kpc Total   & $-1.54\pm0.09$    & $-0.72\pm0.75$ & $16.079\pm0.039$ \\
\rowcolor[HTML]{EFEFEF} 
$r \leq$ 300 kpc Diffuse & $-1.31\pm0.09$    & $0.66\pm0.66$  & $16.995\pm0.032$ \\
$r \leq$ 600 kpc Total   & $-1.63\pm0.13$    & $-0.66\pm1.17$ & $15.480\pm0.057$ \\
$r \leq$ 600 kpc Diffuse & $-1.54\pm0.17$    & $-0.37\pm1.53$ & $16.64\pm0.070$  \\
\hline
\end{tabular}
\end{table}
\end{comment}

\begin{table}
\caption{constraints on parameters in $L(R_0<r<R_1) = a\times \mathrm{log}_{10}\frac{\lambda_0}{20} + b \times \mathrm{log}_{10}\frac{1+z_0}{1.25} + c$ , which is used to quantify the relation between stellar luminosity and cluster richness and redshift. }\label{tbl:linear_relation}
\begin{tabular}{|llll|}
\hline
                         & $a$               & $b$            & $c$         \\
\hline
\rowcolor[HTML]{EFEFEF} 
$r \leq$ 30 kpc Total    & $-0.73\pm0.05$  & $2.23\pm0.34$  & $17.593\pm0.014$ \\
\rowcolor[HTML]{EFEFEF} 
$r \leq$ 30 kpc Diffuse  & $-0.79\pm 0.07$ & $1.76\pm0.46$  & $17.764\pm0.019$ \\
30 to 80 kpc Total    & $-1.37\pm0.09$    & $-0.32\pm0.80$  & $17.75\pm0.027$ \\
30 to 80 kpc Diffuse  & $-1.28\pm0.09$    & $-0.01\pm0.55$  & $18.68\pm0.023$ \\
\rowcolor[HTML]{EFEFEF} 
80 to 300 kpc Total   & $-1.76\pm0.12$    & $-0.66\pm1.02$ & $16.315\pm0.036$ \\
\rowcolor[HTML]{EFEFEF} 
80 to 300 kpc Diffuse & $-1.76\pm0.12$    & $-0.10\pm1.11$  & $17.699\pm0.039$ \\
300 to 600 kpc Total   & $-1.78\pm0.21$    & $-1.53\pm1.97$ & $16.117\pm0.070$ \\
300 to 600 kpc Diffuse & $-2.00\pm0.31$    & $-1.39\pm3.04$ & $17.578\pm0.105$  \\
\hline
\end{tabular}
\end{table}

To further quantify the richness dependence and redshift evolution in those different apertures, we fit the measurements to the following relation:
\begin{equation}
    L(R_0<r<R_1) = a\times \mathrm{log}_{10}\frac{\lambda_0}{20} + b \times \mathrm{log}_{10}\frac{1+z_0}{1.25} + c. 
\end{equation}
In this relation, the total amount of light ($L(R_0<r< R_1)$, in the unit of magnitudes) contained within an aperture or annulus, is fitted with a linear relation to the logarithmic values of the cluster subsample's average richness and redshift. The richness and redshift dependences are described by parameters $a$ and $b$ respectively, and a non-zero value would indicate detection of dependence. In the relation, the pivot richness is chosen as 33, which is the mean richness value of the sample, while the pivot redshift is chosen as 0.25, which is close to the median redshift value in the lowest redshift subsample. %\green{based on Table 1, I would have assumed it was closer to 0.35}{\ynzhang added that it is the median redshift of the lowest redshift subsample -- thanks(JGM)} 
Thus, the intercept of the relation, $c$, can be interpreted as the apparent magnitude of a richness 33 and redshift 0.25 cluster. The fitting of those parameters $a$, $b$, and $c$, is performed with a Markov Chain Monte Carlo (MCMC) method, and the likelihood is constructed from the $\chi^2$ values between the measurements and the relations (using the uncertainties of the measurements as the weighting). Table~\ref{tbl:linear_relation} shows the derived posterior values of the $a$, $b$, and $c$ parameters.

The fitted results confirm our fore-mentioned observations. First, the values of $a$ deviate from 0 at very significant levels, between $\sim$ 7 to 16 $\sigma$ levels in all of the analyzed apertures for both the diffuse and the total stellar content. This result confirms the significant richness, and thus cluster mass dependence for the diffuse and total stellar content. Further, the value of $a$ is increasingly negative at large radii, suggesting stronger richness, and thus mass, dependence for the diffuse light as well as the cluster total stellar content. 

Also consistent with our fore-mentioned observations, for the diffuse light, the value of parameter $b$, which quantifies redshift evolution, is consistent with 0 outside 30kpc. This indicates that when fixing cluster richness, the amount of diffuse light is not obviously increasing (or decreasing) towards lower redshift. Cluster mass may be the main driver for diffuse light evolution in a large radial bin. 
However, within 30 kpc, the amount of diffuse light appears to be increasing towards lower redshift, suggesting that the amount of stellar light associated with the CG is building up over time. According to the fitting results, the luminosity of the CG becomes brighter by 0.113 magnitude (calculated from $b\times \mathrm{log}_{10}(1.45/1.25)$) from redshift 0.45 to 0.25 within 30 kpc. This brightening corresponds to a flux increase of 11\%. The $b$ value of the diffuse light between 30 to 80 kpc is consistent with 0, but we have also adjusted the 80 kpc outbound range between 50 and 100 kpc, and found that the $b$ value is larger in apertures closer to 30 kpc, indicating possible additional redshift evolution closer to the CG. 
%However, within 80 kpc, the amount of diffuse light appears to be increasing towards lower redshift, suggesting that the amount of stellar light associated with the CG and its outskirt is indeed building up over time. This build-up happens more rapidly in the 30 to 80 kpc range than the central 30 kpc range, as indicated by the higher value of $b$. According to the fitting results, the luminosity of the diffuse light between 30 to 80 kpc becomes brighter by 0.23 magnitude (calculated from $b\times \mathrm{log}_{10}(1.45/1.25)$) from redshift 0.45 to 0.25, or brighter by 0.09 magnitude within 30 kpc. This brightening corresponds to a flux increase of 24\% and 9\% respectively. We have also adjusted the 80 kpc range between 50 and 100 kpc, and found that the redshift evolution is the strongest within 100 kpc. 

For the cluster's total light, $b$ is positive within 30 kpc, also indicating some growth with time (brightening towards lower redshift). Because CG is the dominating component of cluster total light within 30 kpc, the growth in this radial range mostly reflects CG growth. %Within 30 to 80 kpc range, the cluster total stellar light grows by 0.26 magnitude, and only a small fraction can be accounted for by the diffuse light growth; there must also be significant growth associated with the cluster satellite galaxies.
In annuli outside 30 kpc, $b$ is generally consistent with 0, indicating no evidence of significant redshift evolution. 

\subsection{Volume-Limited Cluster Sample}

\begin{figure}
	% To include a figure from a file named example.*
	% Allowable file formats are eps or ps if compiling using latex
	% or pdf, png, jpg if compiling using pdflatex
	\includegraphics[width=1.05\columnwidth]{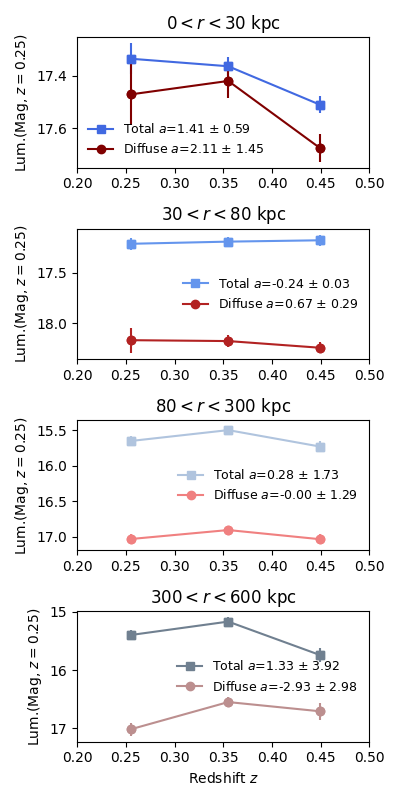}
    \caption{Luminosity (distance-corrected to be the apparent magnitude in the observer frame of $z=0.25$) as a function of redshift in a volume-limited cluster sample. Again we analyze the brightness enclosed within 4 radial bins (0 to 30 kpc, 30 to 80 kpc, 80 to 300 kpc, and 300 to 600 kpc, top to bottom panels) and examine how they change with redshift. The luminosities of the total cluster stellar content (blue lines) and the cluster diffuse light (CG+ICL, red lines) both show some signs of becoming brighter over time within 30 kpc, and between 30 to 80 kpc. Those trends indicate growth in the CG and in the CG to ICL transition region. In section 4, we include quantitative analyses of those trends.}
    \label{fig:lum_aper_vol}
\end{figure}

We further investigate the luminosity evolution of the volume-limited cluster sample discussed in Section~\ref{sec:density_fixed}, which helps answer the question of ICL growth when tracking the same cluster's evolution over time. Similarly, we calculate the luminosities enclosed within radial bins for both the cluster diffuse light and total light and show how they change with redshift in Figure~\ref{fig:lum_aper_vol}. A sign of redshift evolution can be seen within 80 kpc.  We do not find evidence of redshift evolution in the rest of the radial bins.  

To quantify the redshift evolution of this cluster sample, we also fit their measurements to the following relation:
\begin{equation}
  L(z) = L_0 + a \times \mathrm{log}_{10}(1+z)     
\end{equation}
In this relation, $a$ quantifies the redshift evolution of this cluster sample. A positive value would indicate brightening luminosity over time (towards lower redshift), while a negative value would indicate the opposite. The fitting procedure is performed with the {\texttt{curve\_fit}} function of \texttt{Scipy}, and the derived values and uncertainties of $a$ are noted in Figure~\ref{fig:lum_aper_vol}.

With both the diffuse light and the cluster total light, we detect positive values of $a$ above the significance level of $1\sigma$ within 30 kpc in the CG as well as in the CG to ICL transition region, between 30 to 80 kpc range. In both of the radial ranges, the redshift evolution, indicated by the positive value of $a$, appears to be larger in the diffuse light than in the cluster total light. This result, together with the results from the richness/redshift subsamples, indicates that the increase in stellar content with time associated with diffuse light is partially driven through the deposition of new material onto the CG and at the CG's outskirt. For the larger radial bins, we again find $a$ to be consistent with 0, and thus, do not find evidence of redshift evolution. It is possible though, that the redshift evolution in those radial bins falls below our measurement limit as the uncertainties of $a$ are large.
%When switching to using the scaled radius to calculate BCG and ICL brightness as well as the cluster stellar contents, the redshift evolution parameter becomes more positive within 0.03 $R_\lambda$. 
%We also note that, with less than $1\sigma$ differences between the values of parameters, the redshift evolution parameters may be more positive for the diffuse stellar content than the total stellar content, within 80 kpc, but becomes more negative outside 80kpc. %Given the insignificant level of those positive detections, it may be prudent to not over-interpret the results. 
%However, it is possible that the diffuse stellar content evolves fastest where it's close to the CG outskirt, potentially because of the more rapid stripping/disruption happening at the cluster center.

\subsection{Diffuse fraction}

\begin{figure}
	% To include a figure from a file named example.*
	% Allowable file formats are eps or ps if compiling using latex
	% or pdf, png, jpg if compiling using pdflatex
	\includegraphics[width=1.1\columnwidth]{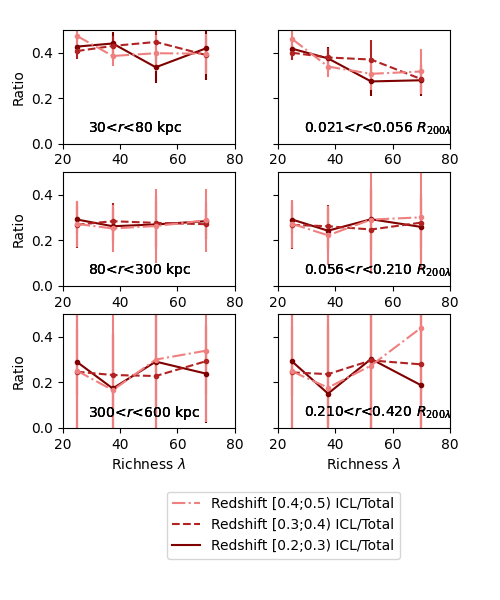}
    \caption{Diffuse fractions in the total cluster stellar content, calculated within physical radii (right column) and scaled radii by $R_{200\lambda}$ (left column).  We do not observe consistent redshift or richness-dependent trends (except in the 0.021 to 0.056 $R_{200\lambda}$ bin) in the measurements. However, the diffuse fractions appear to be dropping at larger radii. See Section 4.3 for more detailed discussions. }
    \label{fig:sb_frac}
\end{figure}

 The fraction of CG and ICL in the cluster's total stellar content is an important quantity \citep[e.g.,][]{2013ApJ...778...14G, 2013ApJ...770...57B, 2022arXiv221206164R, 2023Natur.613...37J}. The build-up of the ICL, CG, and cluster total light is not necessarily aligned over time. For example, the ICL and cluster total light may have gone through more or less significant growth in the recent era compared to the CG. Thus, one may observe a change over time in the CG/ICL to cluster total light ratio. 

Based on the luminosity measurements, we quantify the fractions of CG and ICL in the total cluster stellar content, with the following equation,

\begin{equation}
    \mathrm{Ratio}_\mathrm{Diffuse}(R_0<r<R_1)  = \frac{\mathrm{Lum}_\mathrm{diffuse} (R_0<r<R_1) }{\mathrm{Lum}_\mathrm{total} (R_0<r<R_1) } 
\end{equation}
In this equation, $\mathrm{Lum}_\mathrm{total} (R_0<r<R_1)$ enclosed within radius $R_0$ and $R_1$ is the luminosity of the cluster's total stellar content derived from its surface brightness, and $\mathrm{Lum}_\mathrm{diffuse} (R_0<r<R_1)$ is the luminosity of the diffuse light derived from its surface brightness. We refer to the ratio of these measurements as the diffuse fraction in this analysis.

Figure \ref{fig:sb_frac} shows those fractions derived for clusters in different redshift/richness ranges. %In radial bins going beyond 150 kpc, the CG fraction appears to decrease with richness. We have noted that CG surface brightness (within 30 kpc of the diffuse light measurements) has a milder richness dependence compared to the ICL/total stellar content.Thus, the ICL luminosity and total cluster stellar luminosity increase more quickly with richness than the BCG luminosity, resulting in a decreasing BCG fraction with richness. This trend is more prominent when integrating the ICL and total stellar luminosities in scaled radial bins, as richer clusters have a larger scale radius. %\green{I'm having trouble seeing this trend for the non-scaled radius}{\ynzhang the last two bins of r< 300kpc and r< 600 kpc?} \green{I think I was looking at the wrong bins/blue lines instead of the red lines.  This is clear now and probably always was.}
Other than the 0.021 to 0.056 $R_{200\lambda}$ radial range, the diffuse fraction appears to stay unchanged with richness and redshift, indicating similar richness and redshift dependence between diffuse light and cluster total stellar content. However, the diffuse  fraction does appear to decrease at a large radius, which is close to 40\% in the 30 to 80 kpc range, but decreases to $\sim20\%$ in the 300 to 600 kpc range. Between the 0.021 to 0.056$R_{200\lambda}$ radial range, the diffuse fraction does appear to decrease with richness, but this is likely because of the scaling of $R_{200\lambda}$ with richness; a richer cluster would have a higher $R_{200\lambda}$ value, which excludes more of the BCG outskirt with a 0.021$R_{200\lambda}$ cut. 

%At a small radius ($r<300\mathrm{kpc}$ or $r<0.30R_\mathrm{\lambda}$, the ICL fraction slightly decreases with richness, but this is likely because the total stellar content of clusters within a small radius is dominated by contributions from the CG and ICL. As CG luminosity has a milder richness dependence than ICL, the ICL fraction increases with richness because of its increasing contributions to the CG and ICL combination. 

In addition to those trends, Figure~\ref{fig:sb_frac} highlights the importance of selecting a radial  and cluster mass/richness range when studying diffuse fractions. The fractions appear to drop with an increasing radius, but also is dependent on whether or not the measurements are made in physical radii or radial units scaled by cluster radius. Given the discrepancies in literature reports on diffuse fractions, fair comparisons will need to be made between cluster samples of comparable masses in similar radial scales.

\section{Systematic Effects and Tests}
\label{sec:systematics}

\subsection{PSF effect}

\begin{figure}

	\includegraphics[width=1.0\columnwidth]{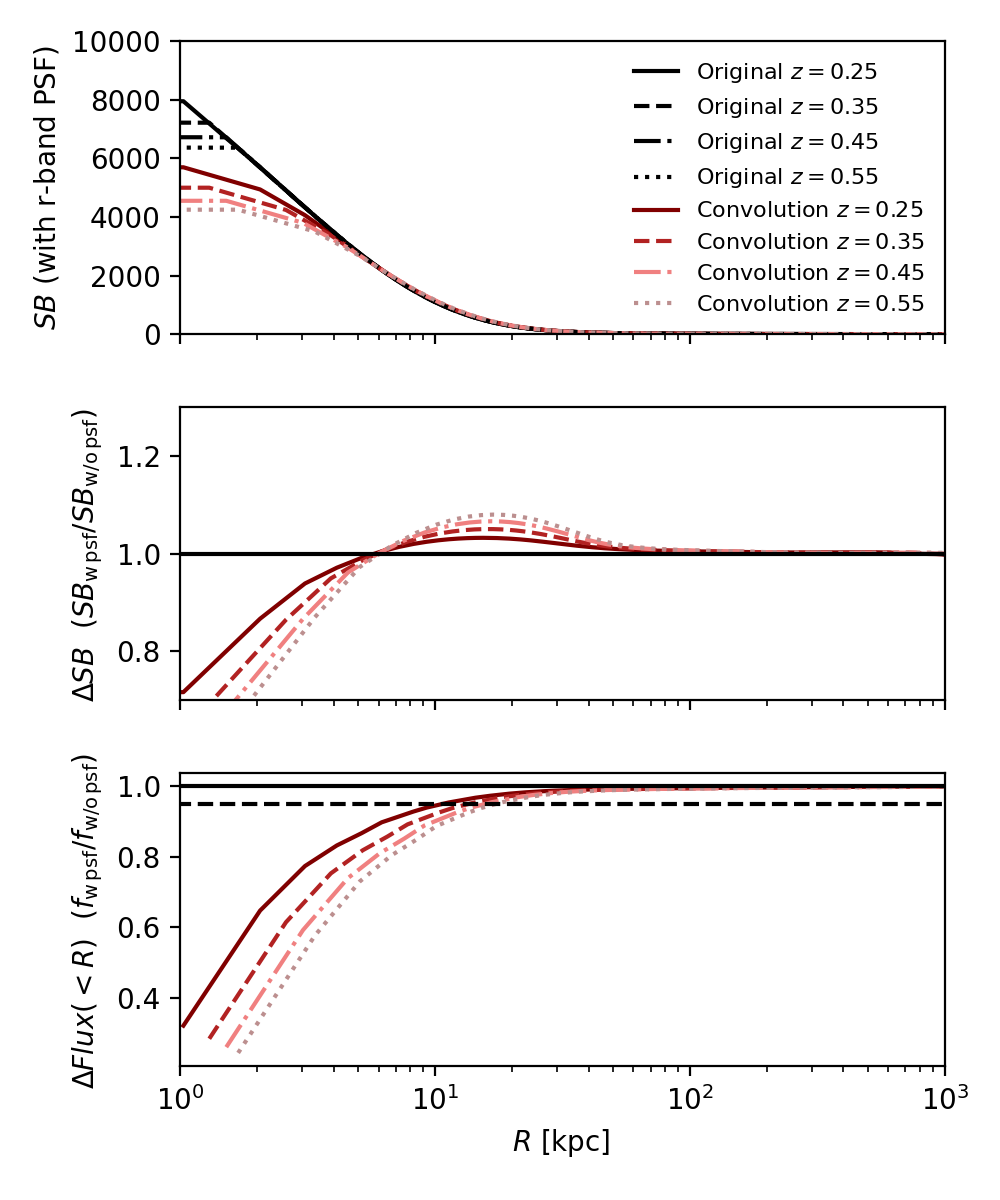}
    \caption{Testing the effect of PSF on SB and luminosity measurements. Upper panel: diffuse light SB models are convolved with a PSF model at different redshifts. The PSF flattens out the SB distribution in the center. Middle Panel: Relative change in SB after the diffuse light profile models are convolved with a PSF model. The changes are most significant within 10 kpc. Lower panel: Relative changes in luminosity (derived by integrating the SB profiles radially as described in Section 4) after a PSF model convolution. The integrated  luminosity is most affected within 20 kpc.}
    \label{fig:sb_psfr}
\end{figure}

Because the point spread function is known to have extended wings \citep{1969A&A.....3..455M, 1971PASP...83..199K, 1996PASP..108..699R, 2007ApJ...666..663B}, it would contribute to the extended low surface brightness features of galaxies or galaxy clusters. The radial scales we probe in this paper are significantly larger than the PSF FWHM of the DES images, therefore we expect minimal PSF contributions to the ICL detection (see discussion in \cite{2019ApJ...874..165Z}). On the other hand, those contributions may change with redshift given the change of angular distance scale with redshift.  Thus, we perform image simulations to probe the possible effect of PSF on the results presented in this paper. 

To do so, we convolve a PSF model with an analytical diffuse light profile model and examine the differences before and after PSF convolution. We generate mock 2D images of diffuse light using an analytical model, setting the angular scale of the analytical diffuse light  profile models at four redshifts, 0.25, 0.35, 0.45 and z = 0.55 (but without adjusting their surface brightness level as we are only looking at before-and-after PSF convolution differences). These 2D images are then convolved with a 2D PSF image model. Both the analytical models and the PSF models are based on the DES-Year1 measurements in \cite{2019ApJ...874..165Z} in r-band, as the PSF models have similar large radial behaviors outside 2 arcseconds. We then derive the SB measurements in radial bins before and after PSF convolution.  The results are shown in Figure~\ref{fig:sb_psfr}.  

The top panel shows the flux changes of the profiles before and after the PSF convolution for the three profiles at different redshifts. PSF convolution flattens the central regions of those profiles limited by the pixel scale of the images (0.263 arcsecond pixel scale). The middle panel of Figure~\ref{fig:sb_psfr} shows the relative changes in those profiles before and after convolution. Outside of 10 kpc, PSF convolution has a minor effect on SB measurements which change by less than 10\%, but the change depends on redshift. Outside of 100 kpc, PSF effects appear to be negligible for all of the four redshifts, which is less than 1\% at 100 kpc for $z=0.55$. 

As with the integrated (within radius) brightness measurements, similarly, the PSF effect appears to be negligible if integrating to 20 kpc, affecting less than $5\%$ of the flux measurement, or around 2\% if integrating to 30 kpc. Within 10 to 20 kpc, the PSF may affect the CG flux measurements by up to 12\%, depending on the redshift. Within 10 kpc, the integrated luminosity needs to be carefully interpreted due to the PSF effect.

We conclude that PSF effect alone can not account for the redshift evolution in the diffuse light luminosity measurement within 30 kpc, which shows a  change of $\sim 0.2~\mathrm{mag}$, or $\sim 20\%$ in flux from redshift 0.45 to 0.25 (Section~\ref{sec:lum_stellar}). With a carefully designed CG aperture (30 kpc in this analysis), our luminosity redshift evolution results should be minimally affected by the PSF effect.%There must be another cause of the redshift evolution we have observed in the CG and diffuse light measurements.

\subsection{Masking Magnitude Limit}

\begin{figure*}
	% To include a figure from a file named example.*
	% Allowable file formats are eps or ps if compiling using latex
	% or pdf, png, jpg if compiling using pdflatex
	\includegraphics[width=2.15\columnwidth]{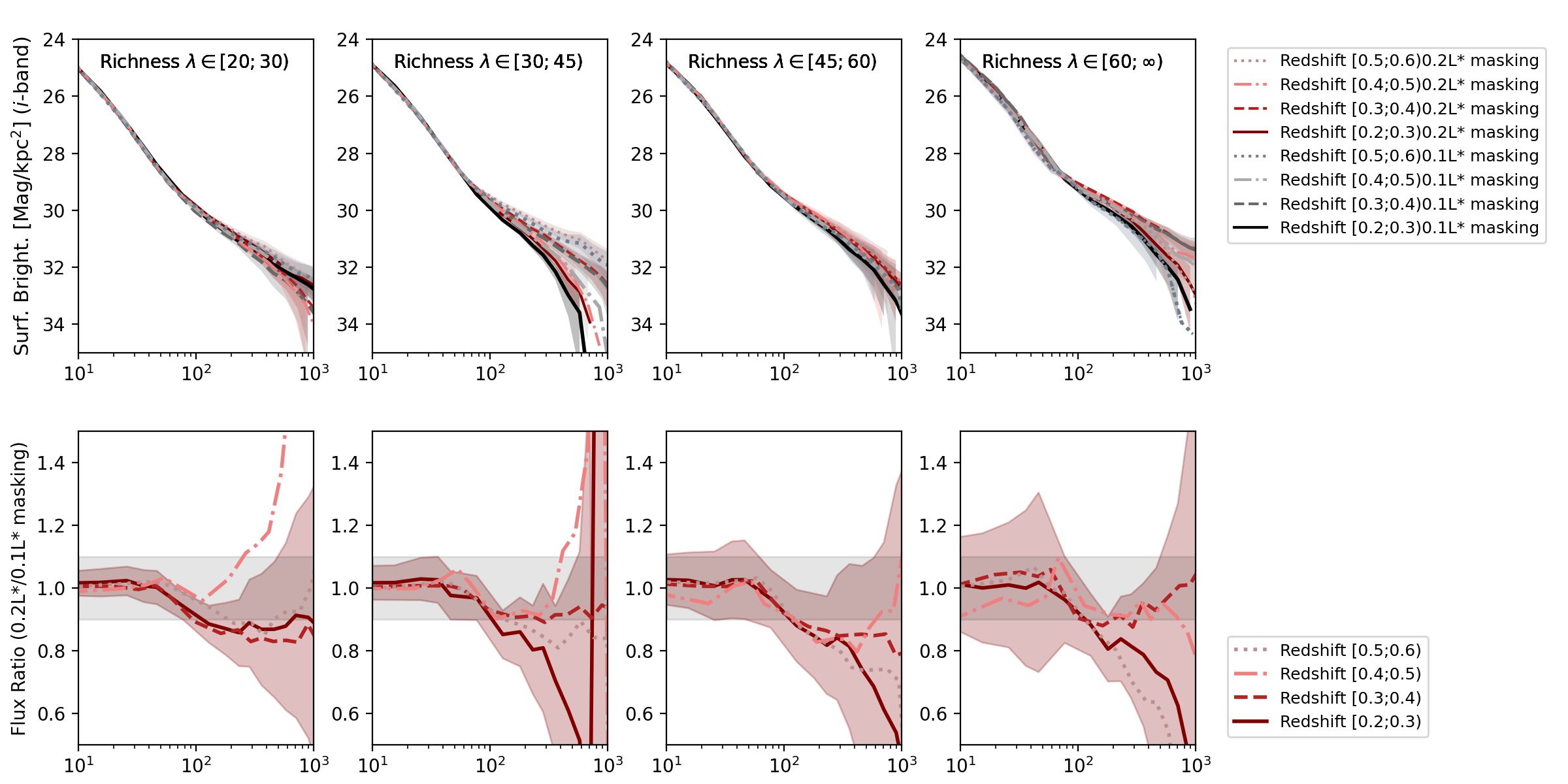}
    \caption{Diffuse light profiles derived when using a deeper masking magnitude limit (upper panels) and the relative differences to the fiducial measurements presented in previous sections (lower panels). The surface brightness measurements outside 100 kpc can be lowered by 10\% when using a deeper masking magnitude, which removes more contamination from the faint cluster satellite galaxies.}
    \label{fig:sb_Llim}
\end{figure*}

The masking magnitude limit we use for this work varies with redshift. This may affect the results of this paper when cluster galaxies below the masking limit contribute a noticeable amount of light to the diffuse light measurements. We acknowledge this issue as a limitation in our analysis, as we do not explicitly account for the contributions from the fainter cluster member galaxies below the masking limit.

We test how much our results may have been affected by these magnitude limits. In this test, we redo the measurements of the diffuse light using a masking magnitude that is fainter by 0.7526 mag (or masking to 0.1$L*$ of the cluster luminosity function), and compare the results to those from the fiducial analyses presented earlier. We have not applied this deeper magnitude limit in our fiducial analysis because of the increasingly incomplete galaxy detection associated with this magnitude limit, which would render the results in the redshift 0.4 to 0.5 bin less reliable. Nevertheless, we show the SB measurements with this deeper magnitude and the comparison to the fiducial analysis in Section~\ref{sec:lum_stellar}.

Indeed, using a deeper masking limit notably reduces the surface brightness measurements of the diffuse light throughout the redshift and richness bins. Outside of 100 kpc, the reduction in flux consistently reaches a $\sim 10\%$ level, although there are significant fluctuations as indicated by the uncertainties.  Given that satellite galaxies 2.5 to 5 times brighter than the ICL in this radial range (Section 4) are excluded, a reduction of 10\% in flux means that the deeper magnitude is only further removing 2\% to 4\% of the faint cluster satellite galaxy contribution. A deeper masking magnitude is unlikely to significantly further reduce ICL brightness unless there is a noticeable upturn in the cluster galaxy luminosity function at the faint end \citep[e.g.,][]{2016MNRAS.459.3998L}.

Other than the masking limit as well as the PSF effect, there are other additional effects that influence our results. %First, the K-correction we assume in this paper depends on SED assumptions, but the differences (depending on the model) are generally very small between redshift 0.25 to 0.45. 
Another issue related to masking is that the masking aperture does not enclose all of the light from cluster satellite galaxies. A galaxy's light can reach tens or even hundreds of kpcs. In \cite{2019ApJ...874..165Z}, we found that the aperture of masking only affects diffuse light measurements at a percentage level. In addition, in this analysis, we have adjusted the masking radius to be 3.5 Kron radii rather than 2.5 Kron radii which will further reduce the effect. 
Moreover, the cluster galaxy luminosity function may evolve with redshift. However, recent literature studies find that the redshift evolution of the cluster galaxy luminosity function is very mild at most \citep[e.g.,][]{2009ApJ...699.1333H,2018A&A...613A..67S,2019MNRAS.488....1Z,2021A&A...645A...9P}.

\subsection{Sky Background}
\label{sec:background}

\begin{figure}
	% To include a figure from a file named example.*
	% Allowable file formats are eps or ps if compiling using latex
	% or pdf, png, jpg if compiling using pdflatex
	\includegraphics[width=1.0\columnwidth]{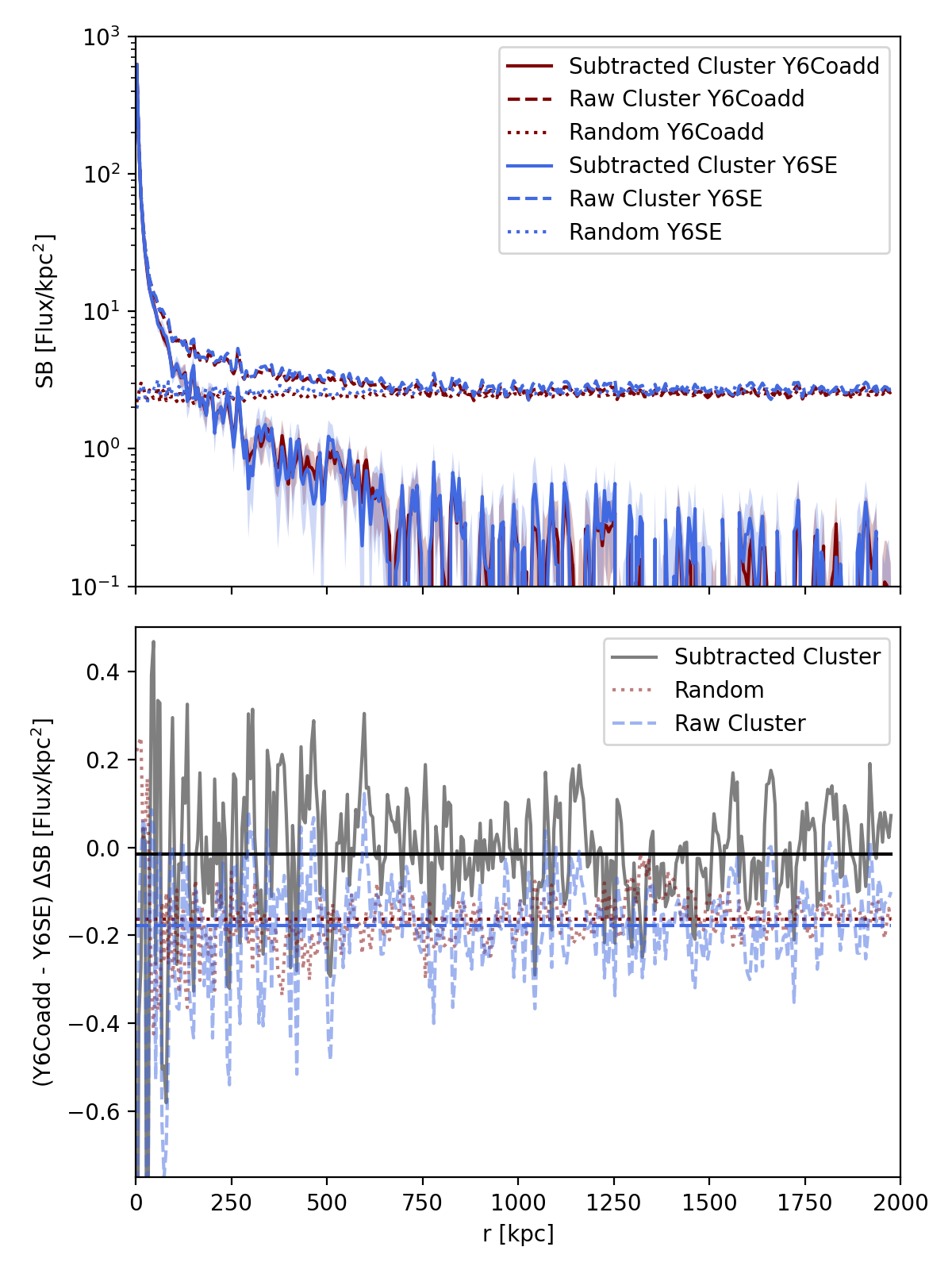}
    \caption{Upper Panel: Diffuse profiles derived from DES special coadded images (red lines, Y6ACoadd, fiducial results in this paper) vs. those derived from single epoch images as in Zhang et al. 2019 (Y6SE, blue lines) . Lower Panel: Differences in these profiles. These two approaches yield consistent surface brightness measurements to an accuracy level of over 30 mag/kpc$^2$ in terms of raw diffuse light and random profile measurements. After random profile subtractions, the differences vanish at a surface brightness level of 40.5 mag/kpc$^2$.}
    \label{fig:background_sub}
\end{figure}

\begin{figure}
	% To include a figure from a file named example.*
	% Allowable file formats are eps or ps if compiling using latex
	% or pdf, png, jpg if compiling using pdflatex
	\includegraphics[width=1.0\columnwidth]{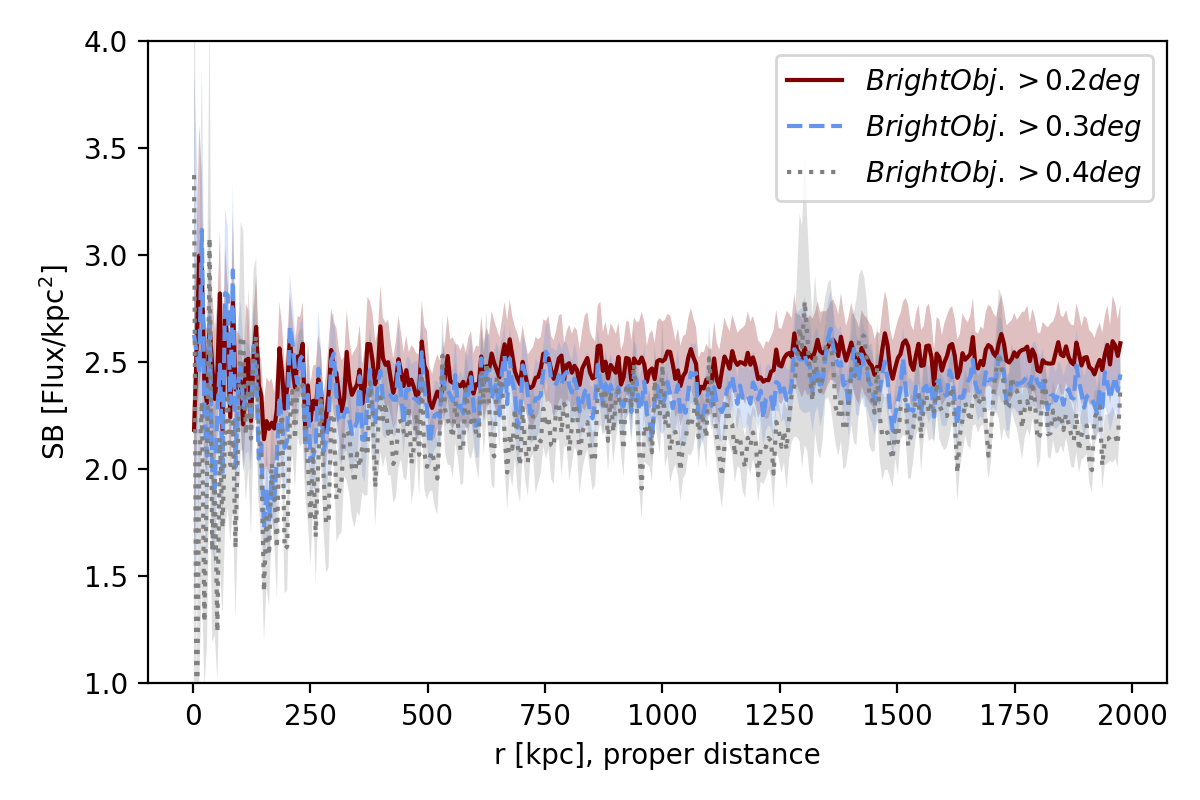}
    \caption{Surface brightness of the randoms when using different distance cuts to bright objects in the sky. Our analysis requires the bright objects identified in a DES masking file to be 0.2 deg away from the cluster center. Using further cuts would lower the surface brightness measurements of randoms because of less contamination from the bright objects.}
    \label{fig:sb_bkg}
\end{figure}

Accurate diffuse light measurements require accurate evaluation and removal of the sky background in optical images. Similar to \cite{2019ApJ...874..165Z}, in this paper, the images we use have removed sky background that is estimated over the whole field-of-view (FOV) of DECam, approximately 3 $\mathrm{deg}^2$, using a PCA method \citep{2017PASP..129k4502B}. Given that one galaxy cluster, even at redshift 0.2, only makes up a very small area in the DECam FOV, the sky background estimation is not sensitive to the presence of galaxy clusters, thus avoiding a background over-estimation issue that often plagues ICL measurements. 
\cite{2019ApJ...874..165Z} tested the DECam FOV PCA background evaluations for ICL measurements, and it was shown that the PCA sky estimations at the cluster centers and at a large cluster radius (1.36 arcmin from the cluster center) are highly consistent.

After removing the full FOV sky background level, the images still possess a residual background. Since we average the measurements for several hundreds and sometimes several thousands of clusters, we estimate a residual background for those averaged measurements, through processing ``sky randoms''  that track the area coverage of the cluster sample. A surface brightness profile of the sky randoms is acquired using the same procedure of the cluster measurements. Those ``random'' profiles are subtracted from the ``raw'' cluster measurements to acquire the final cluster-related measurements. The top panel of Figure~\ref{fig:background_sub} illustrates the procedure. 

In Figure~\ref{fig:background_sub}, because of the residual background, the ``raw'' cluster measurements still have a SB level of $\sim2$ /kpc$^2$ in flux measured at large radii ($\sim 2$ Mpc), but this residual is also present in the ``random'' measurements. After subtracting the randoms, the final cluster measurements fluctuate around 0 at very large radii ($\sim2$ Mpc). Note that in Figure~\ref{fig:background_sub}, we are showing the measurements in  DES ``flux'' ($10^{-12}$ of a maggy), where the ``flux'' used here is a linear measure of an object's brightness, as opposed to the logarithmic ``magnitude'' unit of brightness with the following relation $\mathrm{mag}=30-2.5\times \mathrm{log}_{10}(\mathrm{flux})$.

We note the importance of using random catalogs that faithfully trace the sky coverage of the redMaPPer cluster catalog. The raw profile measurements of randoms in Figure~\ref{fig:sb_bkg} are sensitive to the selection of the random catalogs (and thus the redMaPPer cluster catalogs). These two catalogs are selected to avoid sky regions that contain bright foreground galaxies and stars -- at least 0.2 deg away from their centers. If we adjust the distance cuts to 0.3 deg or 0.4 deg, the random's profile value would become lower, indicating different ``residual'' background levels.  

Finally,  a crucial difference between this paper and \cite{2019ApJ...874..165Z} is that we use the coadded images from the Dark Energy Survey directly, which is based on coadding single epoch images after the PCA sky background subtraction.  The DES  coadd images (the ``no-bkg'' coadd images in the DES data release, which did not subtract local background) are based on the procedure in \cite{2019ApJ...874..165Z} to better preserve low-surface brightness features. 

To test that the DES coadds are indeed suitable for detecting intra-cluster light, we separately process the redshift 0.2 to 0.35 clusters by coadding single epoch images using the same procedures in \cite{2019ApJ...874..165Z} and compare the measurements to the DES special coadd-based measurements. Their differences are shown in Figure~\ref{fig:background_sub}.
%\green{since you spent a lot of time on this, glad it's being included}

The raw SB measurements of those clusters and the randoms from both sets of images are offset at a flux level of 0.2, corresponding to a surface brightness level of 31.7 $\mathrm{mag/kpc}^2$. Those raw measurement differences between the two coadding procedures are likely caused by pixel weighting differences. After the random subtraction, the measurements are similar at a surface flux level of 0.015, which means that the two methods are similar to a surface brightness level of 40.5 $\mathrm{mag/kpc}^2$, and thus highly consistent.

\section{Discussion on Redshift Evolution}
\label{sec:comparisons}

\subsection{Comparison to Simulation}

\begin{figure*}
	% To include a figure from a file named example.*
	% Allowable file formats are eps or ps if compiling using latex
	% or pdf, png, jpg if compiling using pdflatex
	\includegraphics[width=2.1\columnwidth]{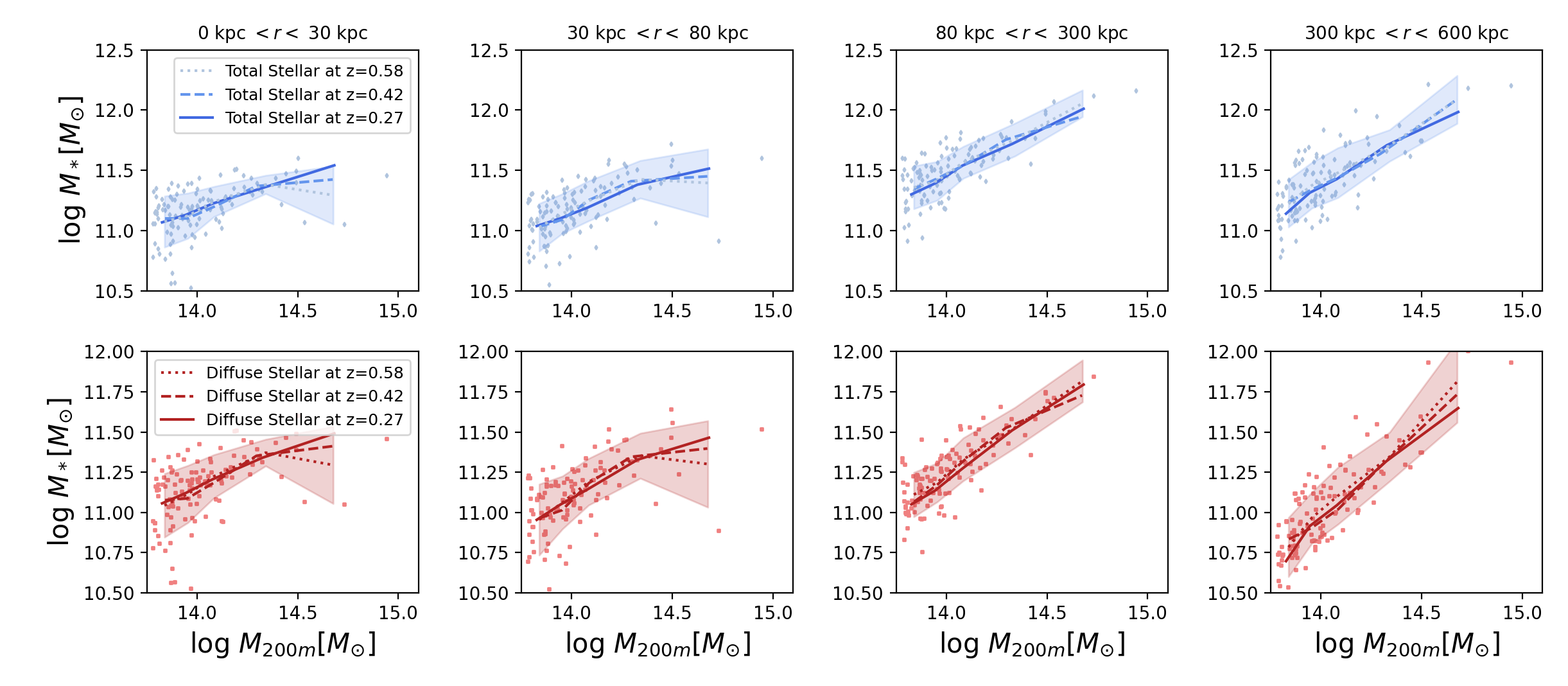}
    \caption{The stellar mass of diffuse light and of the cluster's total stellar content in the IllustrisTNG 300-1 simulation, as a function of the host halo's mass. The different lines indicate the running means in different redshift snapshots. These stellar mass-halo mass relations do not seem to vary with redshift. However, those relations depend on the apertures used to calculate the stellar masses, and are steeper in large radial ranges.}
    \label{fig:Lum_TNG}
\end{figure*}

%%%%https://arxiv.org/abs/2302.10943 this new paper needs to be cited and discussed

To gain theoretical insights into the evolution of ICL, we turn to the IllustrisTNG simulation suite \citep{2019ComAC...6....2N, 2018MNRAS.475..648P, 2018MNRAS.475..676S, 2018MNRAS.475..624N, 2018MNRAS.477.1206N, 2018MNRAS.480.5113M} to examine how the diffuse stellar components of galaxy clusters change with cluster mass and redshift. This has already been a subject of investigation in \cite{2018MNRAS.475..648P}.

Our analysis here is based on the TNG300-1 simulation, which has the largest volume (300 Mpc$^3$) in the IllustrisTNG simulation suite, and also the highest simulating resolution among the 300 Mpc$^3$ volume series. The TNG300-1 simulation contains 263 dark matter halos above the mass threshold of $6\times10^{13}\mathrm{M_\odot}/h$ at redshift 0.27. It has an advantage over the smaller-volume simulations  (for example the TNG 100 Mpc$^3$ and 50 Mpc$^3$ series) which contain much smaller samples of cluster-sized dark matter halos despite their higher simulation resolutions.

We select the redshift snapshots at 0.27, 0.42 and 0.58 for this analysis, to represent the redshift range studied in this paper. In each redshift snapshot, we select dark matter halos with $M_\mathrm{200m}$ above $6\times10^{13}\mathrm{M_\odot}/h$ as ``galaxy clusters''. After cutting dark matter halos that are too close to the simulation box boundaries (within 20 $\mathrm{cMpc/h}$\footnote{$\mathrm{c}$ in $\mathrm{cMpc/h}$ stands for comoving distance.}), we are left with 205, 155 and 115 dark matter halos respectively in the three redshift snapshots. Those dark matter halos will be referred to as galaxy clusters in the rest of this analysis. 
For each simulated cluster, centered on its weighted mass center, we select the diffuse stellar particles contained in 3D distance apertures and compute their total stellar masses. Those stellar masses are shown together with the host halo mass in each redshift snapshot in Figure~\ref{fig:Lum_TNG}. In addition to the diffuse stellar component, we also included the dark matter halo's total stellar content (subhalo+diffuse) within those radial apertures for comparison.

In this simulation, both the diffuse and total stellar components of galaxy clusters steadily increase as the galaxy cluster mass increases. This mass dependence grows steeper in larger radial apertures. On the other hand, examining the mean of those stellar masses ($M_*$) as a function of halo mass ($M_\mathrm{200m}$), there does not appear to be tangible differences in different redshift snapshots, indicating no redshift evolution. 

To further quantify the mass dependence and redshift evolution in the simulation, we fit the halos's stellar masses $M_\mathrm{*}$, halo masses $M_\mathrm{200m}$ and redshifts $z_0$ to the following stellar-mass and halo-mass relation.  
\begin{equation}
    \mathrm{log}_{10} M_\mathrm{*} = a\times (\mathrm{log}_{10} M_\mathrm{200m} - 14.0) + b \times \mathrm{log}_{10}\frac{1+z_0}{1.25} + c. 
\end{equation}
This relation is similar to the one adopted in Section~\ref{sec:lum_stellar}, substituting richness-dependence for mass-dependence. 
%\green{This is a great approach.  Definitely something I could have looked at.}

The fitted constraints on the relation are listed in Table~\ref{tbl:linearrelationsims}. In the relation, parameter $a$ quantifies the mass-dependence of the stellar masses. Its value is positive in all of the radial bins, and becomes even more positive at larger radii, in agreement with the steeper mass dependence we have seen in Figure~\ref{fig:Lum_TNG}.  On the other hand, parameter $b$ quantifies the redshift evolution. Its value is consistent with 0 in all of the bins, indicating a non-detection of redshift evolution. Overall, our results confirm the findings in \cite{2018MNRAS.475..648P} \citep[as well as in ][]{2022arXiv220905519G} that did not find redshift evolution in the stellar-mass to halo-mass relation of galaxy clusters, in either the diffuse component or the subhalo component. However, both stellar mass components scale strongly with halo mass. We note that in small radial ranges, the properties of halo central galaxies or massive galaxies in the simulation do not always match observations \citep[e.g., ][]{2018MNRAS.475..648P, 2020MNRAS.491.3751D, 2019MNRAS.490.2124L, 2023MNRAS.520.5651C}. Nevertheless, those simulation results qualitatively agree with our measurements in the large radial ranges outside of 80 kpc. 

\begin{table}
\caption{Constraints on parameters in the simulation stellar mass to halo mass relation 
$\mathrm{log}_{10} M_\mathrm{*} = a\times (\mathrm{log}_{10} M_\mathrm{200m} - 14.0) + b \times \mathrm{log}_{10}\frac{1+z_0}{1.25} + c$
}\label{tbl:linearrelationsims}
\begin{tabular}{|llll|}
\hline
                         & $a$               & $b$            & $c$         \\
\hline
\rowcolor[HTML]{EFEFEF} 
$r \leq$ 30 kpc Total    & $0.51\pm0.20$  & $ -0.08\pm1.22$  & $11.17\pm0.07$ \\
\rowcolor[HTML]{EFEFEF} 
$ r \leq$ 30 kpc Diffuse  & $0.51\pm 0.21$ & $-0.11\pm1.21$  & $11.16\pm0.07$ \\
$30  \le  r \leq$ 80 kpc Total    & $0.63\pm0.21$    & $0.17\pm1.22$  & $11.14\pm0.07$ \\
$30 \le  r \leq$ 80 kpc Diffuse  & $0.67\pm0.21$    & $0.03\pm1.21$  & $11.08\pm0.07$ \\
\rowcolor[HTML]{EFEFEF} 
$80 \le r \leq$ 300 kpc Total   & $0.81\pm0.20$    & $0.36\pm1.22$ & $11.45\pm0.07$ \\
\rowcolor[HTML]{EFEFEF} 
$80 \le r \leq$ 300 kpc Diffuse & $0.89\pm0.21$    & $0.44\pm1.19$  & $11.20\pm0.07$ \\
$300 \le r \leq$ 600 kpc Total   & $0.99 \pm0.21$    & $0.53\pm1.21$ & $11.34\pm0.07$ \\
$300 \le  r \leq$ 600 kpc Diffuse & $1.11\pm0.21$    & $0.65\pm1.22$ & $10.93\pm0.07$  \\
\hline
\end{tabular}
\end{table}

\subsection{Comparison to Literature}

%Many BCG modeling studies have predicted significant ICL growth to explain a relatively slow BCG growth in observations. For example, Golden-Marx, Zhang and others. ...

%ICL studies over both a low and high redshift range to study its redshift evolution is relatively scarce. redshift 1.0 studies. 

Perhaps the most surprising result from this paper is the relative lack of ICL evolution at a radius larger than 80 kpc.  Many analyses characterizing CG and ICL growth, including the work of the co-authors of this paper, have predicted significant growth of the ICL \citep[e.g., ][]{2013ApJ...770...57B, 2016ApJ...816...98Z, 2018MNRAS.479..932C, 2022ApJ...928...28G} as a mechanism to explain the relatively slow CG growth observed below redshift 1.5 \citep[e.g., ][]{2010ApJ...718...23S, 2012MNRAS.427..550L, 2013ApJ...771...61L}.  However, we do find signs of diffuse light redshift evolution in the CG as well as in the CG to ICL transition within 80 kpc. 

Prior to this analysis, there have been few works that analyze large samples of ICL profiles over a broad range of redshift to directly quantify their redshift evolution. 
One of the most comparable literature studies to our work is presented in \cite{2020MNRAS.491.3751D}, which analyzed 42 clusters in the redshift range of 0.05 to 1.75. \cite{2020MNRAS.491.3751D} measured the BCG and ICL growth out to about 100 kpc from the cluster center, and found that the stellar masses of BCG and ICL increase more rapidly than the cluster's total mass from redshift 1.5 to the present. They conclude that BCG+ICL growth is not solely driven by cluster mass growth. % If this holds true for our observations even when fixing the cluster masses, we would have observed BCG and ICL luminosities increasing towards lower redshift (or a negative $b$ parameter in equation) within 100 kpc. 
In this analysis, we indeed observe that the CG and ICL luminosity increases mildly within 30 kpc.%, but find no evidence for ICL redshift evolution outside of 80 kpc. 

There are some differences between our work and that of \cite{2020MNRAS.491.3751D}. \cite{2020MNRAS.491.3751D} find that the ICL grows by a factor of $1.08\pm0.21$ from redshift 1.55 to 0.4 when examining the 10 kpc to 100 kpc range. While in this paper, we find evidence of ICL growth by 11\% within the 30 kpcfrom redshift 0.45 to 0.25. %, but when looking at redshift below 0.4, they find very little evidence of ICL redshift evolution. 
In our analysis, the results are derived for clusters in a time span of roughly 1.7 Gyr (redshift 0.45 to 0.25). The ICL growth observed in \cite{2020MNRAS.491.3751D} occurs over an extended period of 4.97 Gyr from redshift 1.55 to 0.4. Interpolating from their measurements, the ICL measured between 10kpc to 100 kpc can grow by 37\% in 1.7 Gyr, significantly higher than the 11\% measured in our work in the 30 kpc range where we see the most growth. On the other hand, \cite{2020MNRAS.491.3751D} have noted a slow-down in the BCG and ICL growth after redshift 0.4, that there's no change in the diffuse light stellar mass (between 10 and 100 kpc) to halo mass relation from redshift 0.4 to 0.1.

Furthermore, in our work, we do not find signs of ICL growth outside 80 kpc. On the other hand, \cite{2022arXiv220905519G} studied ICL growth from redshift 0.8 to 0.2, but also do not find much evidence for ICL growth. \cite{2022arXiv220905519G} defines ICL with a large radial aperture of between 50 to 300 kpc and those results are based on the same imaging data set and processing method as in this paper. In both \cite{2022arXiv220905519G} and this work, we are limited by the PSF resolution (as discussed in Section 5) to probe a smaller radial range such as 10 to 30 kpc. Combining the findings from \cite{2022arXiv220905519G, 2020MNRAS.491.3751D} and this work, we speculate that the CG, as well as the region close to the CG within 100 kpc, rather than the ICL at a very large cluster radius, holds the key for explaining CG and ICL growth. However, the growth may not be very noticeable below redshift 0.45. 
 
Another comparable analysis is from \cite{2021MNRAS.502.2419F}, which studied ICL growth over the redshift range of 0.1 to 0.5, using 18 X-Ray selected clusters with Hyper Suprime Cam Subaru Strategic Program observations.  Using a radial aperture of $R_{500}$ and a surface brightness limit of $25 \mathrm{mag/arcsec}^2$, \cite{2021MNRAS.502.2419F} find that the ICL fraction increases by a factor of $2-4$ over the 0.1 to 0.5 redshift range with no obvious mass dependence. %\sout{Given the use of a significantly larger radial aperture, $R_{500}$ ranges from 300,\kpc to 900,\kpc, than any of the previous studies it's possible that this result suggests that along with growth around 100,\kpc, there may be growth at even larger radii. }
However given that the ICL definition in \cite{2021MNRAS.502.2419F} is based on a surface brightness limit, a radius aperture of $R_{500}$ that scales with cluster mass, as well as a ``divot'' correction due to background subtraction in the images, it is possible that the ICL definitions in their and our analyses are not directly comparable.

\section{Summary and Prospects}
\label{sec:summary}

In this paper, we present measurements of the CG and ICL radial profiles using the full 6 years of DES data. The major findings from those measurements can be summarized as the following:

(1) the diffuse light (CG+ICL) extends to 1 Mpc in the redshift range of 0.2 to 0.5 investigated in this analysis. Prior to this analysis, multiple studies have already detected ICL in the several hundreds of kpc to Mpc radial range, which includes both ``stacking'' based analysis like this paper \citep[e.g., ][]{z05, 2022MNRAS.514.2692C}, and deep imaging studies of individual galaxy clusters \citep[e.g., ][]{2007AJ....134..466K,  2021ApJS..252...27K, 2022ApJ...928...28G}. Our finding again showcases the wide radial reach of ICL. There may be much to study with the radial properties of ICL.

(2) We find that the diffuse light surface brightness and luminosity strongly depend on richness -- a galaxy cluster mass proxy. This dependence is stronger at large radii outside of 50 kpc from the cluster center. The richness and thus cluster mass dependence appears to be the major factor behind the differences between diffuse light observations in different subsamples, as their radial profiles scale well with the cluster's radius ($R_\mathrm{200\lambda}$) and their fractions in the cluster's total stellar luminosity appears to be richness-independent. The results agree with previous studies that find a strong mass correlation between ICL luminosity or stellar mass, or a possible correlation between the cluster mass distribution and ICL surface brightness \citep[e.g.,][]{ 2019MNRAS.482.2838M, 2020MNRAS.492.3685H, 2021MNRAS.501.1300S, 2021ApJS..252...27K, 2022arXiv221206164R}. 

Perhaps most interesting of all to cluster cosmology studies, this finding, again, suggests the potential of ICL as a cluster mass proxy \citep{2022arXiv220905519G}, or to help improving cluster finding algorithms \citep{2022MNRAS.515.4722H}.  Cosmology studies based on galaxy cluster abundance measurements have long emphasized the importance of developing accurate and precise cluster mass proxies (i.e., galaxy cluster observables that scale well with masses), because a mass proxy with low scatter to the cluster's true mass can significantly reduce the requirement for follow-up observations, and thus reduce the derived uncertainties on cosmological parameters such as $\Omega_m$ and $\sigma_8$ \citep{2010ApJ...708..645R}. Further, the precision of those cosmology studies also depends on having an accurate mass proxy that is not affected by the cluster's large-scale structure environment \citep{2022MNRAS.515.4471W}. It will be particularly interesting to incorporate diffuse light quantities in developing cluster mass proxies or cluster finding algorithms \citep{2022MNRAS.515.4722H}. 

(3) Perhaps with a bit of a surprise, we find that the diffuse light at large cluster radii (outside of 80 kpc from the cluster center) is not evolving much with redshift in the 0.2 to 0.5 range. Closer to the cluster center, within 80 kpc, we have found some evidence that the diffuse light's luminosity increases with time (towards lower redshift). We speculate that ICL build-up may be more pronounced closer to the CG, while at large radii, on the scale of hundreds of kpc, ICL build-up is more in tune with the cluster mass build-up, which also explains the stronger cluster mass dependence at large radii. 
%\green{Great explanation}

In the context of CG and ICL co-evolution studies, many (including the authors of this paper) have speculated a more rapid growth of ICL than the BCGs below redshift 1. Given that ICL and CG is often vaguely defined in those studies, our findings suggest that ICL growth happens at a much smaller radius (i.e., in the CG or at the CG outskirt) than we previously expected. On the other hand, our finding of little redshift evolution at large cluster radius is in excellent agreement with the hydrodynamic simulation study of IllustrisTNG \citep{2018MNRAS.475..648P}, which finds little redshift evolution in diffuse light stellar mass once the cluster's halo mass is fixed.

(4) We have measured additional properties of ICL: the color profile of diffuse light has a radial gradient, which becomes bluer at a larger radius, and also bluer in less rich/massive clusters. In addition, the diffuse light SB profiles appear to be ``self-similar'' after scaling by the cluster radius, and that ICL fraction in the total cluster stellar light appears to be dropping at a larger radius. %This result is also consistent with previous studies that investigate the color of diffuse light either through "stacking" clusters \citep[e,g.,][]{z05, 2022MNRAS.514.2692C}  or through observing individual cluster systems \citep[e,g.,][]{2018MNRAS.474.3009D, 2021MNRAS.508.2634Y}. It also corroborates spectroscopic results that ICL consists of a younger and metal-poorer stellar population than the cluster central galaxies \citep[e.g.,][]{2020MNRAS.491.2617E}. 
%We also find that the diffuse light color may be bluer in less rich/massive clusters, which is consistent with a disruption/stripping origin as those clusters tends to contain a bluer galaxy population \citep[e.g.,][]{2009ApJ...699.1333H, 2018A&A...613A..67S, 2020MNRAS.498.4303R}. 

Moving forward, there are multiple follow-up opportunities from our measurements. For example, in this paper, we have only studied the average properties of galaxy clusters in richness-redshift subsamples using a ``stacking'' method. As demonstrated in \cite{2022arXiv220905519G}, it is possible to acquire diffuse light measurements of individual galaxy clusters, especially within the 300 kpcs radial range. This would allow us to study how diffuse light properties may change with cluster ellipticity, dynamical state, or with CG properties. It may also be interesting to compare the diffuse light to other galaxy cluster measurements, such as their weak lensing signals as done in \cite{2021MNRAS.501.1300S}.

That said, there are also limitations in this study, especially related to the masking depth as discussed in Section 5. The redshift evolution results reported here are limited by the masking depth of cluster galaxies detected by DES. Faint or undetected cluster galaxies below the masking magnitude limit would have blended into our diffuse light measurements. In this analysis, we use the luminosity function and a test with a deeper magnitude limit to argue that the contribution from those faint galaxies does not affect our redshift evolution conclusions. However, this masking issue can be largely avoided by using a much deeper photometric catalog to more thoroughly mask the contribution of cluster galaxies. Future cosmic surveys like the Legacy Survey of Space and Time (LSST) from the Vera C. Rubin Observatory will be able to provide such a photometric catalog. 

On a different note, those future surveys will also provide many more photons, and a much larger cluster sample for this ``stacking'' (averaging) method, significantly improving the accuracy of  diffuse light measurements. Meanwhile, space-based cosmic survey programs like Euclid and the Nancy Grace Roman Telescope can provide imaging data that are less affected by sky background.  We look forward to using those data in the coming years.
%\green{I think it may be worth mentioning Euclid as well, in terms of space based science}

%\bibliography{reference}

\section*{Data Availability and Acknowledgements}

This paper is dedicated to the memory of the pioneering Egyptian/American astronomer, Sahar Allam, a woman whose wisdom, bravery, care, sensitivity, and sense of humor have guided and supported us in the past decade and during the most difficult times. We miss you dearly.

Our analyses are performed with a few software packages, including  \texttt{Astropy} \citep{astropy:2013, astropy:2018, astropy:2022}, \texttt{Numpy} \citep{harris2020array}, \texttt{Scipy} \citep{2020SciPy-NMeth},  and \texttt{Emcee} \citep{2013PASP..125..306F}. 
The data underlying this article were accessed from the DES and Illustris-TNG database. The derived data generated in this research will be shared on reasonable request to the corresponding author. 

The IllustrisTNG simulations were undertaken with compute time awarded by the Gauss Centre for Supercomputing (GCS) under GCS Large-Scale Projects GCS-ILLU and GCS-DWAR on the GCS share of the supercomputer Hazel Hen at the High Performance Computing Center Stuttgart (HLRS), as well as on the machines of the Max Planck Computing and Data Facility (MPCDF) in Garching, Germany.

Funding for the DES Projects has been provided by the U.S. Department of Energy, the U.S. National Science Foundation, the Ministry of Science and Education of Spain, 
the Science and Technology Facilities Council of the United Kingdom, the Higher Education Funding Council for England, the National Center for Supercomputing 
Applications at the University of Illinois at Urbana-Champaign, the Kavli Institute of Cosmological Physics at the University of Chicago, 
the Center for Cosmology and Astro-Particle Physics at the Ohio State University,
the Mitchell Institute for Fundamental Physics and Astronomy at Texas A\&M University, Financiadora de Estudos e Projetos, 
Funda{\c c}{\~a}o Carlos Chagas Filho de Amparo {\`a} Pesquisa do Estado do Rio de Janeiro, Conselho Nacional de Desenvolvimento Cient{\'i}fico e Tecnol{\'o}gico and 
the Minist{\'e}rio da Ci{\^e}ncia, Tecnologia e Inova{\c c}{\~a}o, the Deutsche Forschungsgemeinschaft and the Collaborating Institutions in the Dark Energy Survey. 

The Collaborating Institutions are Argonne National Laboratory, the University of California at Santa Cruz, the University of Cambridge, Centro de Investigaciones Energ{\'e}ticas, 
Medioambientales y Tecnol{\'o}gicas-Madrid, the University of Chicago, University College London, the DES-Brazil Consortium, the University of Edinburgh, 
the Eidgen{\"o}ssische Technische Hochschule (ETH) Z{\"u}rich, 
Fermi National Accelerator Laboratory, the University of Illinois at Urbana-Champaign, the Institut de Ci{\`e}ncies de l'Espai (IEEC/CSIC), 
the Institut de F{\'i}sica d'Altes Energies, Lawrence Berkeley National Laboratory, the Ludwig-Maximilians Universit{\"a}t M{\"u}nchen and the associated Excellence Cluster Universe, 
the University of Michigan, NSF's NOIRLab, the University of Nottingham, The Ohio State University, the University of Pennsylvania, the University of Portsmouth, 
SLAC National Accelerator Laboratory, Stanford University, the University of Sussex, Texas A\&M University, and the OzDES Membership Consortium.

Based in part on observations at Cerro Tololo Inter-American Observatory at NSF's NOIRLab (NOIRLab Prop. ID 2012B-0001; PI: J. Frieman), which is managed by the Association of Universities for Research in Astronomy (AURA) under a cooperative agreement with the National Science Foundation.

The DES data management system is supported by the National Science Foundation under Grant Numbers AST-1138766 and AST-1536171.
The DES participants from Spanish institutions are partially supported by MICINN under grants ESP2017-89838, PGC2018-094773, PGC2018-102021, SEV-2016-0588, SEV-2016-0597, and MDM-2015-0509, some of which include ERDF funds from the European Union. IFAE is partially funded by the CERCA program of the Generalitat de Catalunya.
Research leading to these results has received funding from the European Research
Council under the European Union's Seventh Framework Program (FP7/2007-2013) including ERC grant agreements 240672, 291329, and 306478.
We  acknowledge support from the Brazilian Instituto Nacional de Ci\^encia
e Tecnologia (INCT) do e-Universo (CNPq grant 465376/2014-2).

This manuscript has been authored by Fermi Research Alliance, LLC under Contract No. DE-AC02-07CH11359 with the U.S. Department of Energy, Office of Science, Office of High Energy Physics.

%%%%%%%%%%%%%%%%%%%% REFERENCES %%%%%%%%%%%%%%%%%%

% The best way to enter references is to use BibTeX:

\bibliographystyle{mnras}
\bibliography{reference} % if your bibtex file is called example.bib

\begin{thebibliography}{}
\makeatletter
\relax
\def\mn@urlcharsother{\let\do\@makeother \do\$\do\&\do\#\do\^\do\_\do\%\do\~}
\def\mn@doi{\begingroup\mn@urlcharsother \@ifnextchar [ {\mn@doi@}
  {\mn@doi@[]}}
\def\mn@doi@[#1]#2{\def\@tempa{#1}\ifx\@tempa\@empty \href
  {http://dx.doi.org/#2} {doi:#2}\else \href {http://dx.doi.org/#2} {#1}\fi
  \endgroup}
\def\mn@eprint#1#2{\mn@eprint@#1:#2::\@nil}
\def\mn@eprint@arXiv#1{\href {http://arxiv.org/abs/#1} {{\tt arXiv:#1}}}
\def\mn@eprint@dblp#1{\href {http://dblp.uni-trier.de/rec/bibtex/#1.xml}
  {dblp:#1}}
\def\mn@eprint@#1:#2:#3:#4\@nil{\def\@tempa {#1}\def\@tempb {#2}\def\@tempc
  {#3}\ifx \@tempc \@empty \let \@tempc \@tempb \let \@tempb \@tempa \fi \ifx
  \@tempb \@empty \def\@tempb {arXiv}\fi \@ifundefined
  {mn@eprint@\@tempb}{\@tempb:\@tempc}{\expandafter \expandafter \csname
  mn@eprint@\@tempb\endcsname \expandafter{\@tempc}}}

\bibitem[\protect\citeauthoryear{{Abbott} et~al.,}{{Abbott}
  et~al.}{2018}]{2018ApJS..239...18A}
{Abbott} T.~M.~C.,  et~al., 2018, \mn@doi [\apjs] {10.3847/1538-4365/aae9f0},
  \href {https://ui.adsabs.harvard.edu/abs/2018ApJS..239...18A} {239, 18}

\bibitem[\protect\citeauthoryear{{Abbott} et~al.,}{{Abbott}
  et~al.}{2020}]{2020PhRvD.102b3509A}
{Abbott} T.~M.~C.,  et~al., 2020, \mn@doi [\prd] {10.1103/PhysRevD.102.023509},
  \href {https://ui.adsabs.harvard.edu/abs/2020PhRvD.102b3509A} {102, 023509}

\bibitem[\protect\citeauthoryear{{Abbott} et~al.,}{{Abbott}
  et~al.}{2021}]{2021ApJS..255...20A}
{Abbott} T.~M.~C.,  et~al., 2021, \mn@doi [\apjs] {10.3847/1538-4365/ac00b3},
  \href {https://ui.adsabs.harvard.edu/abs/2021ApJS..255...20A} {255, 20}

\bibitem[\protect\citeauthoryear{{Abbott} et~al.,}{{Abbott}
  et~al.}{2022a}]{2022PhRvD.105b3520A}
{Abbott} T.~M.~C.,  et~al., 2022a, \mn@doi [\prd]
  {10.1103/PhysRevD.105.023520}, \href
  {https://ui.adsabs.harvard.edu/abs/2022PhRvD.105b3520A} {105, 023520}

\bibitem[\protect\citeauthoryear{{Abbott} et~al.,}{{Abbott}
  et~al.}{2022b}]{2022PhRvD.105d3512A}
{Abbott} T.~M.~C.,  et~al., 2022b, \mn@doi [\prd]
  {10.1103/PhysRevD.105.043512}, \href
  {https://ui.adsabs.harvard.edu/abs/2022PhRvD.105d3512A} {105, 043512}

\bibitem[\protect\citeauthoryear{{Abraham}, {van Dokkum}, {Conroy}, {Merritt},
  {Zhang}, {Lokhorst}, {Danieli}  \& {Mowla}}{{Abraham}
  et~al.}{2017}]{2017ASSL..434..333A}
{Abraham} R.,  {van Dokkum} P.,  {Conroy} C.,  {Merritt} A.,  {Zhang} J.,
  {Lokhorst} D.,  {Danieli} S.,   {Mowla} L.,  2017, in {Knapen} J.~H.,  {Lee}
  J.~C.,   {Gil de Paz} A.,  eds,  Astrophysics and Space Science Library Vol.
  434, Outskirts of Galaxies. p.~333 (\mn@eprint {arXiv} {1612.06415}),
  \mn@doi{10.1007/978-3-319-56570-5_10}

\bibitem[\protect\citeauthoryear{{Ahad}, {Bah{\'e}}  \& {Hoekstra}}{{Ahad}
  et~al.}{2023}]{2023MNRAS.518.3685A}
{Ahad} S.~L.,  {Bah{\'e}} Y.~M.,   {Hoekstra} H.,  2023, \mn@doi [\mnras]
  {10.1093/mnras/stac3357}, \href
  {https://ui.adsabs.harvard.edu/abs/2023MNRAS.518.3685A} {518, 3685}

\bibitem[\protect\citeauthoryear{{Alonso Asensio}, {Dalla Vecchia}, {Bah{\'e}},
  {Barnes}  \& {Kay}}{{Alonso Asensio} et~al.}{2020}]{2020MNRAS.494.1859A}
{Alonso Asensio} I.,  {Dalla Vecchia} C.,  {Bah{\'e}} Y.~M.,  {Barnes} D.~J.,
  {Kay} S.~T.,  2020, \mn@doi [\mnras] {10.1093/mnras/staa861}, \href
  {https://ui.adsabs.harvard.edu/abs/2020MNRAS.494.1859A} {494, 1859}

\bibitem[\protect\citeauthoryear{{Anbajagane}, {Evrard}, {Farahi}, {Barnes},
  {Dolag}, {McCarthy}, {Nelson}  \& {Pillepich}}{{Anbajagane}
  et~al.}{2020}]{2020MNRAS.495..686A}
{Anbajagane} D.,  {Evrard} A.~E.,  {Farahi} A.,  {Barnes} D.~J.,  {Dolag} K.,
  {McCarthy} I.~G.,  {Nelson} D.,   {Pillepich} A.,  2020, \mn@doi [\mnras]
  {10.1093/mnras/staa1147}, \href
  {https://ui.adsabs.harvard.edu/abs/2020MNRAS.495..686A} {495, 686}

\bibitem[\protect\citeauthoryear{{Arnaboldi} \& {Gerhard}}{{Arnaboldi} \&
  {Gerhard}}{2022}]{2022arXiv221209569A}
{Arnaboldi} M.,  {Gerhard} O.~E.,  2022, arXiv e-prints, \href
  {https://ui.adsabs.harvard.edu/abs/2022arXiv221209569A} {p. arXiv:2212.09569}

\bibitem[\protect\citeauthoryear{{Arnaboldi}, {Ventimiglia}, {Iodice},
  {Gerhard}  \& {Coccato}}{{Arnaboldi} et~al.}{2012}]{2012A&A...545A..37A}
{Arnaboldi} M.,  {Ventimiglia} G.,  {Iodice} E.,  {Gerhard} O.,   {Coccato} L.,
   2012, \mn@doi [\aap] {10.1051/0004-6361/201116752}, \href
  {https://ui.adsabs.harvard.edu/abs/2012A&A...545A..37A} {545, A37}

\bibitem[\protect\citeauthoryear{{Astropy Collaboration} et~al.,}{{Astropy
  Collaboration} et~al.}{2013}]{astropy:2013}
{Astropy Collaboration} et~al., 2013, \mn@doi [\aap]
  {10.1051/0004-6361/201322068}, \href
  {http://adsabs.harvard.edu/abs/2013A%26A...558A..33A} {558, A33}

\bibitem[\protect\citeauthoryear{{Astropy Collaboration} et~al.,}{{Astropy
  Collaboration} et~al.}{2018}]{astropy:2018}
{Astropy Collaboration} et~al., 2018, \mn@doi [\aj] {10.3847/1538-3881/aabc4f},
  \href {https://ui.adsabs.harvard.edu/abs/2018AJ....156..123A} {156, 123}

\bibitem[\protect\citeauthoryear{{Astropy Collaboration} et~al.,}{{Astropy
  Collaboration} et~al.}{2022}]{astropy:2022}
{Astropy Collaboration} et~al., 2022, \mn@doi [apj] {10.3847/1538-4357/ac7c74},
  \href {https://ui.adsabs.harvard.edu/abs/2022ApJ...935..167A} {935, 167}

\bibitem[\protect\citeauthoryear{{Barai}, {Brito}  \& {Martel}}{{Barai}
  et~al.}{2009}]{2009JApA...30....1B}
{Barai} P.,  {Brito} W.,   {Martel} H.,  2009, Journal of Astrophysics and
  Astronomy, \href {https://ui.adsabs.harvard.edu/abs/2009JApA...30....1B} {30,
  1}

\bibitem[\protect\citeauthoryear{{Barbosa}, {Arnaboldi}, {Coccato}, {Hilker},
  {Mendes de Oliveira}  \& {Richtler}}{{Barbosa}
  et~al.}{2016}]{2016A&A...589A.139B}
{Barbosa} C.~E.,  {Arnaboldi} M.,  {Coccato} L.,  {Hilker} M.,  {Mendes de
  Oliveira} C.,   {Richtler} T.,  2016, \mn@doi [\aap]
  {10.1051/0004-6361/201628137}, \href
  {https://ui.adsabs.harvard.edu/abs/2016A&A...589A.139B} {589, A139}

\bibitem[\protect\citeauthoryear{{Barfety} et~al.,}{{Barfety}
  et~al.}{2022}]{2022ApJ...930...25B}
{Barfety} C.,  et~al., 2022, \mn@doi [\apj] {10.3847/1538-4357/ac61dd}, \href
  {https://ui.adsabs.harvard.edu/abs/2022ApJ...930...25B} {930, 25}

\bibitem[\protect\citeauthoryear{{Behroozi}, {Wechsler}  \&
  {Conroy}}{{Behroozi} et~al.}{2013}]{2013ApJ...770...57B}
{Behroozi} P.~S.,  {Wechsler} R.~H.,   {Conroy} C.,  2013, \mn@doi [\apj]
  {10.1088/0004-637X/770/1/57}, \href
  {http://adsabs.harvard.edu/abs/2013ApJ...770...57B} {770, 57}

\bibitem[\protect\citeauthoryear{{Bernstein}}{{Bernstein}}{2007}]{2007ApJ...666..663B}
{Bernstein} R.~A.,  2007, \mn@doi [\apj] {10.1086/519824}, \href
  {http://adsabs.harvard.edu/abs/2007ApJ...666..663B} {666, 663}

\bibitem[\protect\citeauthoryear{{Bernstein} et~al.,}{{Bernstein}
  et~al.}{2017a}]{2017PASP..129g4503B}
{Bernstein} G.~M.,  et~al., 2017a, \mn@doi [\pasp] {10.1088/1538-3873/aa6c55},
  \href {https://ui.adsabs.harvard.edu/abs/2017PASP..129g4503B} {129, 074503}

\bibitem[\protect\citeauthoryear{{Bernstein} et~al.,}{{Bernstein}
  et~al.}{2017b}]{2017PASP..129k4502B}
{Bernstein} G.~M.,  et~al., 2017b, \mn@doi [\pasp] {10.1088/1538-3873/aa858e},
  \href {https://ui.adsabs.harvard.edu/abs/2017PASP..129k4502B} {129, 114502}

\bibitem[\protect\citeauthoryear{{Bernstein} et~al.,}{{Bernstein}
  et~al.}{2018}]{2018PASP..130e4501B}
{Bernstein} G.~M.,  et~al., 2018, \mn@doi [\pasp] {10.1088/1538-3873/aaa753},
  \href {https://ui.adsabs.harvard.edu/abs/2018PASP..130e4501B} {130, 054501}

\bibitem[\protect\citeauthoryear{{Bertin} \& {Arnouts}}{{Bertin} \&
  {Arnouts}}{1996}]{1996A&AS..117..393B}
{Bertin} E.,  {Arnouts} S.,  1996, \mn@doi [\aaps] {10.1051/aas:1996164}, \href
  {https://ui.adsabs.harvard.edu/abs/1996A&AS..117..393B} {117, 393}

\bibitem[\protect\citeauthoryear{{Bertin}, {Mellier}, {Radovich}, {Missonnier},
  {Didelon}  \& {Morin}}{{Bertin} et~al.}{2002}]{2002ASPC..281..228B}
{Bertin} E.,  {Mellier} Y.,  {Radovich} M.,  {Missonnier} G.,  {Didelon} P.,
  {Morin} B.,  2002, in {Bohlender} D.~A.,  {Durand} D.,   {Handley} T.~H.,
  eds,  Astronomical Society of the Pacific Conference Series Vol. 281,
  Astronomical Data Analysis Software and Systems XI. p.~228

\bibitem[\protect\citeauthoryear{{Bleem} et~al.,}{{Bleem}
  et~al.}{2020}]{2020ApJS..247...25B}
{Bleem} L.~E.,  et~al., 2020, \mn@doi [\apjs] {10.3847/1538-4365/ab6993}, \href
  {https://ui.adsabs.harvard.edu/abs/2020ApJS..247...25B} {247, 25}

\bibitem[\protect\citeauthoryear{{Burke}, {Collins}, {Stott}  \&
  {Hilton}}{{Burke} et~al.}{2012}]{2012MNRAS.425.2058B}
{Burke} C.,  {Collins} C.~A.,  {Stott} J.~P.,   {Hilton} M.,  2012, \mn@doi
  [\mnras] {10.1111/j.1365-2966.2012.21555.x}, \href
  {https://ui.adsabs.harvard.edu/abs/2012MNRAS.425.2058B} {425, 2058}

\bibitem[\protect\citeauthoryear{{Burke} et~al.,}{{Burke}
  et~al.}{2018}]{2018AJ....155...41B}
{Burke} D.~L.,  et~al., 2018, \mn@doi [\aj] {10.3847/1538-3881/aa9f22}, \href
  {https://ui.adsabs.harvard.edu/abs/2018AJ....155...41B} {155, 41}

\bibitem[\protect\citeauthoryear{{Ca{\~n}as}, {Lagos}, {Elahi}, {Power},
  {Welker}, {Dubois}  \& {Pichon}}{{Ca{\~n}as}
  et~al.}{2020}]{2020MNRAS.494.4314C}
{Ca{\~n}as} R.,  {Lagos} C. d.~P.,  {Elahi} P.~J.,  {Power} C.,  {Welker} C.,
  {Dubois} Y.,   {Pichon} C.,  2020, \mn@doi [\mnras] {10.1093/mnras/staa1027},
  \href {https://ui.adsabs.harvard.edu/abs/2020MNRAS.494.4314C} {494, 4314}

\bibitem[\protect\citeauthoryear{{Cannarozzo} et~al.,}{{Cannarozzo}
  et~al.}{2023}]{2023MNRAS.520.5651C}
{Cannarozzo} C.,  et~al., 2023, \mn@doi [\mnras] {10.1093/mnras/stac3023},
  \href {https://ui.adsabs.harvard.edu/abs/2023MNRAS.520.5651C} {520, 5651}

\bibitem[\protect\citeauthoryear{{Chen}, {Zu}, {Shao}  \& {Shan}}{{Chen}
  et~al.}{2022}]{2022MNRAS.514.2692C}
{Chen} X.,  {Zu} Y.,  {Shao} Z.,   {Shan} H.,  2022, \mn@doi [\mnras]
  {10.1093/mnras/stac1456}, \href
  {https://ui.adsabs.harvard.edu/abs/2022MNRAS.514.2692C} {514, 2692}

\bibitem[\protect\citeauthoryear{{Chun}, {Shin}, {Smith}, {Ko}  \&
  {Yoo}}{{Chun} et~al.}{2022}]{2022arXiv221202510C}
{Chun} K.,  {Shin} J.,  {Smith} R.,  {Ko} J.,   {Yoo} J.,  2022, arXiv
  e-prints, \href {https://ui.adsabs.harvard.edu/abs/2022arXiv221202510C} {p.
  arXiv:2212.02510}

\bibitem[\protect\citeauthoryear{{Coccato}, {Gerhard}  \&
  {Arnaboldi}}{{Coccato} et~al.}{2010a}]{2010MNRAS.407L..26C}
{Coccato} L.,  {Gerhard} O.,   {Arnaboldi} M.,  2010a, \mn@doi [\mnras]
  {10.1111/j.1745-3933.2010.00897.x}, \href
  {https://ui.adsabs.harvard.edu/abs/2010MNRAS.407L..26C} {407, L26}

\bibitem[\protect\citeauthoryear{{Coccato}, {Arnaboldi}, {Gerhard}, {Freeman},
  {Ventimiglia}  \& {Yasuda}}{{Coccato} et~al.}{2010b}]{2010A&A...519A..95C}
{Coccato} L.,  {Arnaboldi} M.,  {Gerhard} O.,  {Freeman} K.~C.,  {Ventimiglia}
  G.,   {Yasuda} N.,  2010b, \mn@doi [\aap] {10.1051/0004-6361/201014476},
  \href {https://ui.adsabs.harvard.edu/abs/2010A&A...519A..95C} {519, A95}

\bibitem[\protect\citeauthoryear{{Coccato}, {Gerhard}, {Arnaboldi}  \&
  {Ventimiglia}}{{Coccato} et~al.}{2011}]{2011A&A...533A.138C}
{Coccato} L.,  {Gerhard} O.,  {Arnaboldi} M.,   {Ventimiglia} G.,  2011,
  \mn@doi [\aap] {10.1051/0004-6361/201117546}, \href
  {https://ui.adsabs.harvard.edu/abs/2011A&A...533A.138C} {533, A138}

\bibitem[\protect\citeauthoryear{{Conroy}, {Wechsler}  \& {Kravtsov}}{{Conroy}
  et~al.}{2007}]{2007ApJ...668..826C}
{Conroy} C.,  {Wechsler} R.~H.,   {Kravtsov} A.~V.,  2007, \mn@doi [\apj]
  {10.1086/521425}, \href
  {https://ui.adsabs.harvard.edu/abs/2007ApJ...668..826C} {668, 826}

\bibitem[\protect\citeauthoryear{{Contini}}{{Contini}}{2021}]{2021Galax...9...60C}
{Contini} E.,  2021, \mn@doi [Galaxies] {10.3390/galaxies9030060}, \href
  {https://ui.adsabs.harvard.edu/abs/2021Galax...9...60C} {9, 60}

\bibitem[\protect\citeauthoryear{{Contini} \& {Gu}}{{Contini} \&
  {Gu}}{2021}]{2021ApJ...915..106C}
{Contini} E.,  {Gu} Q.,  2021, \mn@doi [\apj] {10.3847/1538-4357/ac01e6}, \href
  {https://ui.adsabs.harvard.edu/abs/2021ApJ...915..106C} {915, 106}

\bibitem[\protect\citeauthoryear{{Contini}, {De Lucia}, {Villalobos}  \&
  {Borgani}}{{Contini} et~al.}{2014}]{2014MNRAS.437.3787C}
{Contini} E.,  {De Lucia} G.,  {Villalobos} {\'A}.,   {Borgani} S.,  2014,
  \mn@doi [\mnras] {10.1093/mnras/stt2174}, \href
  {https://ui.adsabs.harvard.edu/abs/2014MNRAS.437.3787C} {437, 3787}

\bibitem[\protect\citeauthoryear{{Contini}, {Yi}  \& {Kang}}{{Contini}
  et~al.}{2018}]{2018MNRAS.479..932C}
{Contini} E.,  {Yi} S.~K.,   {Kang} X.,  2018, \mn@doi [\mnras]
  {10.1093/mnras/sty1518}, \href
  {https://ui.adsabs.harvard.edu/abs/2018MNRAS.479..932C} {479, 932}

\bibitem[\protect\citeauthoryear{{Contini}, {Yi}  \& {Kang}}{{Contini}
  et~al.}{2019}]{2019ApJ...871...24C}
{Contini} E.,  {Yi} S.~K.,   {Kang} X.,  2019, \mn@doi [\apj]
  {10.3847/1538-4357/aaf41f}, \href
  {https://ui.adsabs.harvard.edu/abs/2019ApJ...871...24C} {871, 24}

\bibitem[\protect\citeauthoryear{{Contini}, {Chen}  \& {Gu}}{{Contini}
  et~al.}{2022}]{2022ApJ...928...99C}
{Contini} E.,  {Chen} H.~Z.,   {Gu} Q.,  2022, \mn@doi [\apj]
  {10.3847/1538-4357/ac57c4}, \href
  {https://ui.adsabs.harvard.edu/abs/2022ApJ...928...99C} {928, 99}

\bibitem[\protect\citeauthoryear{{Cooper}, {Gao}, {Guo}, {Frenk}, {Jenkins},
  {Springel}  \& {White}}{{Cooper} et~al.}{2015}]{2015MNRAS.451.2703C}
{Cooper} A.~P.,  {Gao} L.,  {Guo} Q.,  {Frenk} C.~S.,  {Jenkins} A.,
  {Springel} V.,   {White} S.~D.~M.,  2015, \mn@doi [\mnras]
  {10.1093/mnras/stv1042}, \href
  {https://ui.adsabs.harvard.edu/abs/2015MNRAS.451.2703C} {451, 2703}

\bibitem[\protect\citeauthoryear{{Costanzi} et~al.,}{{Costanzi}
  et~al.}{2019}]{2019MNRAS.482..490C}
{Costanzi} M.,  et~al., 2019, \mn@doi [\mnras] {10.1093/mnras/sty2665}, \href
  {https://ui.adsabs.harvard.edu/abs/2019MNRAS.482..490C} {482, 490}

\bibitem[\protect\citeauthoryear{{Cui} et~al.,}{{Cui}
  et~al.}{2014}]{2014MNRAS.437..816C}
{Cui} W.,  et~al., 2014, \mn@doi [\mnras] {10.1093/mnras/stt1940}, \href
  {https://ui.adsabs.harvard.edu/abs/2014MNRAS.437..816C} {437, 816}

\bibitem[\protect\citeauthoryear{{DES Collaboration}}{{DES
  Collaboration}}{2005}]{DES2005}
{DES Collaboration} 2005, preprint, \href
  {http://adsabs.harvard.edu/abs/2005astro.ph.10346T} {} (\mn@eprint {arXiv}
  {astro-ph/0510346})

\bibitem[\protect\citeauthoryear{{DES Collaboration} et~al.,}{{DES
  Collaboration} et~al.}{2022}]{2022arXiv220705766D}
{DES Collaboration} et~al., 2022, \mn@doi [arXiv e-prints]
  {10.48550/arXiv.2207.05766}, \href
  {https://ui.adsabs.harvard.edu/abs/2022arXiv220705766D} {p. arXiv:2207.05766}

\bibitem[\protect\citeauthoryear{{De Lucia} \& {Blaizot}}{{De Lucia} \&
  {Blaizot}}{2007}]{2007MNRAS.375....2D}
{De Lucia} G.,  {Blaizot} J.,  2007, \mn@doi [\mnras]
  {10.1111/j.1365-2966.2006.11287.x}, \href
  {https://ui.adsabs.harvard.edu/abs/2007MNRAS.375....2D} {375, 2}

\bibitem[\protect\citeauthoryear{{DeMaio}}{{DeMaio}}{2017}]{2017PhDT.......215D}
{DeMaio} T.~N.,  2017, PhD thesis, University of Florida

\bibitem[\protect\citeauthoryear{{DeMaio}, {Gonzalez}, {Zabludoff}, {Zaritsky}
  \& {Brada{\v{c}}}}{{DeMaio} et~al.}{2015}]{2015MNRAS.448.1162D}
{DeMaio} T.,  {Gonzalez} A.~H.,  {Zabludoff} A.,  {Zaritsky} D.,
  {Brada{\v{c}}} M.,  2015, \mn@doi [\mnras] {10.1093/mnras/stv033}, \href
  {https://ui.adsabs.harvard.edu/abs/2015MNRAS.448.1162D} {448, 1162}

\bibitem[\protect\citeauthoryear{{DeMaio}, {Gonzalez}, {Zabludoff}, {Zaritsky},
  {Connor}, {Donahue}  \& {Mulchaey}}{{DeMaio}
  et~al.}{2018}]{2018MNRAS.474.3009D}
{DeMaio} T.,  {Gonzalez} A.~H.,  {Zabludoff} A.,  {Zaritsky} D.,  {Connor} T.,
  {Donahue} M.,   {Mulchaey} J.~S.,  2018, \mn@doi [\mnras]
  {10.1093/mnras/stx2946}, \href
  {https://ui.adsabs.harvard.edu/abs/2018MNRAS.474.3009D} {474, 3009}

\bibitem[\protect\citeauthoryear{{DeMaio} et~al.,}{{DeMaio}
  et~al.}{2020}]{2020MNRAS.491.3751D}
{DeMaio} T.,  et~al., 2020, \mn@doi [\mnras] {10.1093/mnras/stz3236}, \href
  {https://ui.adsabs.harvard.edu/abs/2020MNRAS.491.3751D} {491, 3751}

\bibitem[\protect\citeauthoryear{{Diehl} et~al.,}{{Diehl}
  et~al.}{2018}]{2018SPIE10704E..0DD}
{Diehl} H.~T.,  et~al., 2018, in Observatory Operations: Strategies, Processes,
  and Systems VII. p. 107040D, \mn@doi{10.1117/12.2312113}

\bibitem[\protect\citeauthoryear{{Dolag}, {Murante}  \& {Borgani}}{{Dolag}
  et~al.}{2010}]{2010MNRAS.405.1544D}
{Dolag} K.,  {Murante} G.,   {Borgani} S.,  2010, \mn@doi [\mnras]
  {10.1111/j.1365-2966.2010.16583.x}, \href
  {https://ui.adsabs.harvard.edu/abs/2010MNRAS.405.1544D} {405, 1544}

\bibitem[\protect\citeauthoryear{{Drlica-Wagner} et~al.,}{{Drlica-Wagner}
  et~al.}{2018}]{2018ApJS..235...33D}
{Drlica-Wagner} A.,  et~al., 2018, \mn@doi [\apjs] {10.3847/1538-4365/aab4f5},
  \href {https://ui.adsabs.harvard.edu/abs/2018ApJS..235...33D} {235, 33}

\bibitem[\protect\citeauthoryear{{Edwards}, {Alpert}, {Trierweiler}, {Abraham}
  \& {Beizer}}{{Edwards} et~al.}{2016}]{2016MNRAS.461..230E}
{Edwards} L.~O.~V.,  {Alpert} H.~S.,  {Trierweiler} I.~L.,  {Abraham} T.,
  {Beizer} V.~G.,  2016, \mn@doi [\mnras] {10.1093/mnras/stw1314}, \href
  {https://ui.adsabs.harvard.edu/abs/2016MNRAS.461..230E} {461, 230}

\bibitem[\protect\citeauthoryear{{Edwards} et~al.,}{{Edwards}
  et~al.}{2020}]{2020MNRAS.491.2617E}
{Edwards} L. O.~V.,  et~al., 2020, \mn@doi [\mnras] {10.1093/mnras/stz2706},
  \href {https://ui.adsabs.harvard.edu/abs/2020MNRAS.491.2617E} {491, 2617}

\bibitem[\protect\citeauthoryear{Efron}{Efron}{1982}]{efron1982jackknife}
Efron B.,  1982, The jackknife, the bootstrap and other resampling plans.
SIAM

\bibitem[\protect\citeauthoryear{{Farahi} et~al.,}{{Farahi}
  et~al.}{2019}]{2019MNRAS.490.3341F}
{Farahi} A.,  et~al., 2019, \mn@doi [\mnras] {10.1093/mnras/stz2689}, \href
  {https://ui.adsabs.harvard.edu/abs/2019MNRAS.490.3341F} {490, 3341}

\bibitem[\protect\citeauthoryear{{Flaugher} et~al.,}{{Flaugher}
  et~al.}{2015}]{2015AJ....150..150F}
{Flaugher} B.,  et~al., 2015, \mn@doi [\aj] {10.1088/0004-6256/150/5/150},
  \href {https://ui.adsabs.harvard.edu/abs/2015AJ....150..150F} {150, 150}

\bibitem[\protect\citeauthoryear{{Foreman-Mackey}, {Hogg}, {Lang}  \&
  {Goodman}}{{Foreman-Mackey} et~al.}{2013}]{2013PASP..125..306F}
{Foreman-Mackey} D.,  {Hogg} D.~W.,  {Lang} D.,   {Goodman} J.,  2013, \mn@doi
  [\pasp] {10.1086/670067}, \href
  {https://ui.adsabs.harvard.edu/abs/2013PASP..125..306F} {125, 306}

\bibitem[\protect\citeauthoryear{{Furnell} et~al.,}{{Furnell}
  et~al.}{2021}]{2021MNRAS.502.2419F}
{Furnell} K.~E.,  et~al., 2021, \mn@doi [\mnras] {10.1093/mnras/stab065}, \href
  {https://ui.adsabs.harvard.edu/abs/2021MNRAS.502.2419F} {502, 2419}

\bibitem[\protect\citeauthoryear{{Golden-Marx} et~al.,}{{Golden-Marx}
  et~al.}{2022}]{2022ApJ...928...28G}
{Golden-Marx} J.~B.,  et~al., 2022, \mn@doi [\apj] {10.3847/1538-4357/ac4cb4},
  \href {https://ui.adsabs.harvard.edu/abs/2022ApJ...928...28G} {928, 28}

\bibitem[\protect\citeauthoryear{{Golden-Marx} et~al.,}{{Golden-Marx}
  et~al.}{2023a}]{2022arXiv220905519G}
{Golden-Marx} J.~B.,  et~al., 2023a, \mn@doi [\mnras] {10.1093/mnras/stad469},
  \href {https://ui.adsabs.harvard.edu/abs/2023MNRAS.521..478G} {521, 478}

\bibitem[\protect\citeauthoryear{{Golden-Marx}, {Zu}, {Wang}, {Li}, {Zhang}  \&
  {Yang}}{{Golden-Marx} et~al.}{2023b}]{2023MNRAS.524.4455G}
{Golden-Marx} J.~B.,  {Zu} Y.,  {Wang} J.,  {Li} H.,  {Zhang} J.,   {Yang} X.,
  2023b, \mn@doi [\mnras] {10.1093/mnras/stad2174}, \href
  {https://ui.adsabs.harvard.edu/abs/2023MNRAS.524.4455G} {524, 4455}

\bibitem[\protect\citeauthoryear{{Gonzalez}, {Zabludoff}  \&
  {Zaritsky}}{{Gonzalez} et~al.}{2005}]{2005ApJ...618..195G}
{Gonzalez} A.~H.,  {Zabludoff} A.~I.,   {Zaritsky} D.,  2005, \mn@doi [\apj]
  {10.1086/425896}, \href
  {https://ui.adsabs.harvard.edu/abs/2005ApJ...618..195G} {618, 195}

\bibitem[\protect\citeauthoryear{{Gonzalez}, {Sivanandam}, {Zabludoff}  \&
  {Zaritsky}}{{Gonzalez} et~al.}{2013}]{2013ApJ...778...14G}
{Gonzalez} A.~H.,  {Sivanandam} S.,  {Zabludoff} A.~I.,   {Zaritsky} D.,  2013,
  \mn@doi [\apj] {10.1088/0004-637X/778/1/14}, \href
  {https://ui.adsabs.harvard.edu/abs/2013ApJ...778...14G} {778, 14}

\bibitem[\protect\citeauthoryear{{Gruen}, {Bernstein}, {Jarvis}, {Rowe},
  {Vikram}, {Plazas}  \& {Seitz}}{{Gruen} et~al.}{2015}]{2015JInst..10C5032G}
{Gruen} D.,  {Bernstein} G.~M.,  {Jarvis} M.,  {Rowe} B.,  {Vikram} V.,
  {Plazas} A.~A.,   {Seitz} S.,  2015, \mn@doi [Journal of Instrumentation]
  {10.1088/1748-0221/10/05/C05032}, \href
  {https://ui.adsabs.harvard.edu/abs/2015JInst..10C5032G} {10, C05032}

\bibitem[\protect\citeauthoryear{{Gu} et~al.,}{{Gu}
  et~al.}{2020}]{2020ApJ...894...32G}
{Gu} M.,  et~al., 2020, \mn@doi [\apj] {10.3847/1538-4357/ab845c}, \href
  {https://ui.adsabs.harvard.edu/abs/2020ApJ...894...32G} {894, 32}

\bibitem[\protect\citeauthoryear{{Hansen}, {Sheldon}, {Wechsler}  \&
  {Koester}}{{Hansen} et~al.}{2009}]{2009ApJ...699.1333H}
{Hansen} S.~M.,  {Sheldon} E.~S.,  {Wechsler} R.~H.,   {Koester} B.~P.,  2009,
  \mn@doi [\apj] {10.1088/0004-637X/699/2/1333}, \href
  {https://ui.adsabs.harvard.edu/abs/2009ApJ...699.1333H} {699, 1333}

\bibitem[\protect\citeauthoryear{{Harris} et~al.,}{{Harris}
  et~al.}{2017}]{2017MNRAS.467.4501H}
{Harris} K.~A.,  et~al., 2017, \mn@doi [\mnras] {10.1093/mnras/stx401}, \href
  {https://ui.adsabs.harvard.edu/abs/2017MNRAS.467.4501H} {467, 4501}

\bibitem[\protect\citeauthoryear{Harris et~al.,}{Harris
  et~al.}{2020}]{harris2020array}
Harris C.~R.,  et~al., 2020, \mn@doi [Nature] {10.1038/s41586-020-2649-2}, 585,
  357

\bibitem[\protect\citeauthoryear{{Hilker}, {Richtler}, {Barbosa}, {Arnaboldi},
  {Coccato}  \& {Mendes de Oliveira}}{{Hilker}
  et~al.}{2018}]{2018A&A...619A..70H}
{Hilker} M.,  {Richtler} T.,  {Barbosa} C.~E.,  {Arnaboldi} M.,  {Coccato} L.,
   {Mendes de Oliveira} C.,  2018, \mn@doi [\aap]
  {10.1051/0004-6361/201731737}, \href
  {https://ui.adsabs.harvard.edu/abs/2018A&A...619A..70H} {619, A70}

\bibitem[\protect\citeauthoryear{{Huang} et~al.,}{{Huang}
  et~al.}{2020}]{2020MNRAS.492.3685H}
{Huang} S.,  et~al., 2020, \mn@doi [\mnras] {10.1093/mnras/stz3314}, \href
  {https://ui.adsabs.harvard.edu/abs/2020MNRAS.492.3685H} {492, 3685}

\bibitem[\protect\citeauthoryear{{Huang} et~al.,}{{Huang}
  et~al.}{2022}]{2022MNRAS.515.4722H}
{Huang} S.,  et~al., 2022, \mn@doi [\mnras] {10.1093/mnras/stac1680}, \href
  {https://ui.adsabs.harvard.edu/abs/2022MNRAS.515.4722H} {515, 4722}

\bibitem[\protect\citeauthoryear{{Iodice} et~al.,}{{Iodice}
  et~al.}{2017}]{2017ApJ...851...75I}
{Iodice} E.,  et~al., 2017, \mn@doi [\apj] {10.3847/1538-4357/aa9b30}, \href
  {https://ui.adsabs.harvard.edu/abs/2017ApJ...851...75I} {851, 75}

\bibitem[\protect\citeauthoryear{{Joo} \& {Jee}}{{Joo} \&
  {Jee}}{2023}]{2023Natur.613...37J}
{Joo} H.,  {Jee} M.~J.,  2023, \mn@doi [\nat] {10.1038/s41586-022-05396-4},
  \href {https://ui.adsabs.harvard.edu/abs/2023Natur.613...37J} {613, 37}

\bibitem[\protect\citeauthoryear{{King}}{{King}}{1971}]{1971PASP...83..199K}
{King} I.~R.,  1971, \mn@doi [\pasp] {10.1086/129100}, \href
  {http://adsabs.harvard.edu/abs/1971PASP...83..199K} {83, 199}

\bibitem[\protect\citeauthoryear{{Kluge}, {Bender}, {Riffeser}, {Goessl},
  {Hopp}, {Schmidt}  \& {Ries}}{{Kluge} et~al.}{2021}]{2021ApJS..252...27K}
{Kluge} M.,  {Bender} R.,  {Riffeser} A.,  {Goessl} C.,  {Hopp} U.,  {Schmidt}
  M.,   {Ries} C.,  2021, \mn@doi [\apjs] {10.3847/1538-4365/abcda6}, \href
  {https://ui.adsabs.harvard.edu/abs/2021ApJS..252...27K} {252, 27}

\bibitem[\protect\citeauthoryear{{Ko} \& {Jee}}{{Ko} \&
  {Jee}}{2018}]{2018ApJ...862...95K}
{Ko} J.,  {Jee} M.~J.,  2018, \mn@doi [\apj] {10.3847/1538-4357/aacbda}, \href
  {https://ui.adsabs.harvard.edu/abs/2018ApJ...862...95K} {862, 95}

\bibitem[\protect\citeauthoryear{{Krick} \& {Bernstein}}{{Krick} \&
  {Bernstein}}{2007}]{2007AJ....134..466K}
{Krick} J.~E.,  {Bernstein} R.~A.,  2007, \mn@doi [\aj] {10.1086/518787}, \href
  {https://ui.adsabs.harvard.edu/abs/2007AJ....134..466K} {134, 466}

\bibitem[\protect\citeauthoryear{{Krick}, {Bernstein}  \& {Pimbblet}}{{Krick}
  et~al.}{2006}]{2006AJ....131..168K}
{Krick} J.~E.,  {Bernstein} R.~A.,   {Pimbblet} K.~A.,  2006, \mn@doi [\aj]
  {10.1086/498269}, \href
  {https://ui.adsabs.harvard.edu/abs/2006AJ....131..168K} {131, 168}

\bibitem[\protect\citeauthoryear{{Krick}, {Bridge}, {Desai}, {Mihos}, {Murphy},
  {Rudick}, {Surace}  \& {Neill}}{{Krick} et~al.}{2011}]{2011ApJ...735...76K}
{Krick} J.~E.,  {Bridge} C.,  {Desai} V.,  {Mihos} J.~C.,  {Murphy} E.,
  {Rudick} C.,  {Surace} J.,   {Neill} J.,  2011, \mn@doi [\apj]
  {10.1088/0004-637X/735/2/76}, \href
  {https://ui.adsabs.harvard.edu/abs/2011ApJ...735...76K} {735, 76}

\bibitem[\protect\citeauthoryear{{Kron}}{{Kron}}{1980}]{1980ApJS...43..305K}
{Kron} R.~G.,  1980, \mn@doi [\apjs] {10.1086/190669}, \href
  {https://ui.adsabs.harvard.edu/abs/1980ApJS...43..305K} {43, 305}

\bibitem[\protect\citeauthoryear{{Lan}, {M{\'e}nard}  \& {Mo}}{{Lan}
  et~al.}{2016}]{2016MNRAS.459.3998L}
{Lan} T.-W.,  {M{\'e}nard} B.,   {Mo} H.,  2016, \mn@doi [\mnras]
  {10.1093/mnras/stw898}, \href
  {http://adsabs.harvard.edu/abs/2016MNRAS.459.3998L} {459, 3998}

\bibitem[\protect\citeauthoryear{{Leung} et~al.,}{{Leung}
  et~al.}{2020}]{2020RNAAS...4..174L}
{Leung} Y.,  et~al., 2020, \mn@doi [Research Notes of the American Astronomical
  Society] {10.3847/2515-5172/abbd8d}, \href
  {https://ui.adsabs.harvard.edu/abs/2020RNAAS...4..174L} {4, 174}

\bibitem[\protect\citeauthoryear{{Li} et~al.,}{{Li}
  et~al.}{2019}]{2019MNRAS.490.2124L}
{Li} R.,  et~al., 2019, \mn@doi [\mnras] {10.1093/mnras/stz2565}, \href
  {https://ui.adsabs.harvard.edu/abs/2019MNRAS.490.2124L} {490, 2124}

\bibitem[\protect\citeauthoryear{{Li} et~al.,}{{Li}
  et~al.}{2022}]{2022MNRAS.515.5335L}
{Li} J.,  et~al., 2022, \mn@doi [\mnras] {10.1093/mnras/stac2121}, \href
  {https://ui.adsabs.harvard.edu/abs/2022MNRAS.515.5335L} {515, 5335}

\bibitem[\protect\citeauthoryear{{Lidman} et~al.,}{{Lidman}
  et~al.}{2012}]{2012MNRAS.427..550L}
{Lidman} C.,  et~al., 2012, \mn@doi [\mnras]
  {10.1111/j.1365-2966.2012.21984.x}, \href
  {https://ui.adsabs.harvard.edu/abs/2012MNRAS.427..550L} {427, 550}

\bibitem[\protect\citeauthoryear{{Lin}, {Brodwin}, {Gonzalez}, {Bode},
  {Eisenhardt}, {Stanford}  \& {Vikhlinin}}{{Lin}
  et~al.}{2013}]{2013ApJ...771...61L}
{Lin} Y.-T.,  {Brodwin} M.,  {Gonzalez} A.~H.,  {Bode} P.,  {Eisenhardt} P.
  R.~M.,  {Stanford} S.~A.,   {Vikhlinin} A.,  2013, \mn@doi [\apj]
  {10.1088/0004-637X/771/1/61}, \href
  {https://ui.adsabs.harvard.edu/abs/2013ApJ...771...61L} {771, 61}

\bibitem[\protect\citeauthoryear{{Longobardi}, {Arnaboldi}  \&
  {Gerhard}}{{Longobardi} et~al.}{2015a}]{2015Galax...3..212L}
{Longobardi} A.,  {Arnaboldi} M.,   {Gerhard} O.,  2015a, \mn@doi [Galaxies]
  {10.3390/galaxies3040212}, \href
  {https://ui.adsabs.harvard.edu/abs/2015Galax...3..212L} {3, 212}

\bibitem[\protect\citeauthoryear{{Longobardi}, {Arnaboldi}, {Gerhard}  \&
  {Hanuschik}}{{Longobardi} et~al.}{2015b}]{2015A&A...579A.135L}
{Longobardi} A.,  {Arnaboldi} M.,  {Gerhard} O.,   {Hanuschik} R.,  2015b,
  \mn@doi [\aap] {10.1051/0004-6361/201525773}, \href
  {https://ui.adsabs.harvard.edu/abs/2015A&A...579A.135L} {579, A135}

\bibitem[\protect\citeauthoryear{{Longobardi}, {Arnaboldi}, {Gerhard},
  {Pulsoni}  \& {S{\"o}ldner-Rembold}}{{Longobardi}
  et~al.}{2018a}]{2018A&A...620A.111L}
{Longobardi} A.,  {Arnaboldi} M.,  {Gerhard} O.,  {Pulsoni} C.,
  {S{\"o}ldner-Rembold} I.,  2018a, \mn@doi [\aap]
  {10.1051/0004-6361/201832729}, \href
  {https://ui.adsabs.harvard.edu/abs/2018A&A...620A.111L} {620, A111}

\bibitem[\protect\citeauthoryear{{Longobardi} et~al.,}{{Longobardi}
  et~al.}{2018b}]{2018ApJ...864...36L}
{Longobardi} A.,  et~al., 2018b, \mn@doi [\apj] {10.3847/1538-4357/aad3d2},
  \href {https://ui.adsabs.harvard.edu/abs/2018ApJ...864...36L} {864, 36}

\bibitem[\protect\citeauthoryear{{Mackie}}{{Mackie}}{1992}]{1992ApJ...400...65M}
{Mackie} G.,  1992, \mn@doi [\apj] {10.1086/171973}, \href
  {https://ui.adsabs.harvard.edu/abs/1992ApJ...400...65M} {400, 65}

\bibitem[\protect\citeauthoryear{{Marinacci} et~al.,}{{Marinacci}
  et~al.}{2018}]{2018MNRAS.480.5113M}
{Marinacci} F.,  et~al., 2018, \mn@doi [\mnras] {10.1093/mnras/sty2206}, \href
  {https://ui.adsabs.harvard.edu/abs/2018MNRAS.480.5113M} {480, 5113}

\bibitem[\protect\citeauthoryear{{Marini}, {Borgani}, {Saro}, {Murante},
  {Granato}, {Ragone-Figueroa}  \& {Taffoni}}{{Marini}
  et~al.}{2022}]{2022MNRAS.514.3082M}
{Marini} I.,  {Borgani} S.,  {Saro} A.,  {Murante} G.,  {Granato} G.~L.,
  {Ragone-Figueroa} C.,   {Taffoni} G.,  2022, \mn@doi [\mnras]
  {10.1093/mnras/stac1558}, \href
  {https://ui.adsabs.harvard.edu/abs/2022MNRAS.514.3082M} {514, 3082}

\bibitem[\protect\citeauthoryear{{Martel}, {Barai}  \& {Brito}}{{Martel}
  et~al.}{2012}]{2012ApJ...757...48M}
{Martel} H.,  {Barai} P.,   {Brito} W.,  2012, \mn@doi [\apj]
  {10.1088/0004-637X/757/1/48}, \href
  {https://ui.adsabs.harvard.edu/abs/2012ApJ...757...48M} {757, 48}

\bibitem[\protect\citeauthoryear{{Mart{\'\i}nez-Lombilla}
  et~al.,}{{Mart{\'\i}nez-Lombilla} et~al.}{2023}]{2023MNRAS.518.1195M}
{Mart{\'\i}nez-Lombilla} C.,  et~al., 2023, \mn@doi [\mnras]
  {10.1093/mnras/stac3119}, \href
  {https://ui.adsabs.harvard.edu/abs/2023MNRAS.518.1195M} {518, 1195}

\bibitem[\protect\citeauthoryear{{McClintock} et~al.,}{{McClintock}
  et~al.}{2019}]{2018arXiv180500039M}
{McClintock} T.,  et~al., 2019, \mn@doi [\mnras] {10.1093/mnras/sty2711}, \href
  {http://adsabs.harvard.edu/abs/2019MNRAS.482.1352M} {482, 1352}

\bibitem[\protect\citeauthoryear{{Melchior} et~al.,}{{Melchior}
  et~al.}{2017}]{2017MNRAS.469.4899M}
{Melchior} P.,  et~al., 2017, \mn@doi [\mnras] {10.1093/mnras/stx1053}, \href
  {https://ui.adsabs.harvard.edu/abs/2017MNRAS.469.4899M} {469, 4899}

\bibitem[\protect\citeauthoryear{{Mihos}}{{Mihos}}{2019}]{2019arXiv190909456M}
{Mihos} J.~C.,  2019, arXiv e-prints, \href
  {https://ui.adsabs.harvard.edu/abs/2019arXiv190909456M} {p. arXiv:1909.09456}

\bibitem[\protect\citeauthoryear{{Moffat}}{{Moffat}}{1969}]{1969A&A.....3..455M}
{Moffat} A.~F.~J.,  1969, \aap, \href
  {http://adsabs.harvard.edu/abs/1969A%26A.....3..455M} {3, 455}

\bibitem[\protect\citeauthoryear{{Monaco}, {Murante}, {Borgani}  \&
  {Fontanot}}{{Monaco} et~al.}{2006}]{2006ApJ...652L..89M}
{Monaco} P.,  {Murante} G.,  {Borgani} S.,   {Fontanot} F.,  2006, \mn@doi
  [\apjl] {10.1086/510236}, \href
  {https://ui.adsabs.harvard.edu/abs/2006ApJ...652L..89M} {652, L89}

\bibitem[\protect\citeauthoryear{{Montes}}{{Montes}}{2022}]{2022NatAs...6..308M}
{Montes} M.,  2022, \mn@doi [Nature Astronomy] {10.1038/s41550-022-01616-z},
  \href {https://ui.adsabs.harvard.edu/abs/2022NatAs...6..308M} {6, 308}

\bibitem[\protect\citeauthoryear{{Montes} \& {Trujillo}}{{Montes} \&
  {Trujillo}}{2014}]{2014ApJ...794..137M}
{Montes} M.,  {Trujillo} I.,  2014, \mn@doi [\apj]
  {10.1088/0004-637X/794/2/137}, \href
  {https://ui.adsabs.harvard.edu/abs/2014ApJ...794..137M} {794, 137}

\bibitem[\protect\citeauthoryear{{Montes} \& {Trujillo}}{{Montes} \&
  {Trujillo}}{2018}]{2018MNRAS.474..917M}
{Montes} M.,  {Trujillo} I.,  2018, \mn@doi [\mnras] {10.1093/mnras/stx2847},
  \href {https://ui.adsabs.harvard.edu/abs/2018MNRAS.474..917M} {474, 917}

\bibitem[\protect\citeauthoryear{{Montes} \& {Trujillo}}{{Montes} \&
  {Trujillo}}{2019}]{2019MNRAS.482.2838M}
{Montes} M.,  {Trujillo} I.,  2019, \mn@doi [\mnras] {10.1093/mnras/sty2858},
  \href {https://ui.adsabs.harvard.edu/abs/2019MNRAS.482.2838M} {482, 2838}

\bibitem[\protect\citeauthoryear{{Montes} \& {Trujillo}}{{Montes} \&
  {Trujillo}}{2022}]{2022ApJ...940L..51M}
{Montes} M.,  {Trujillo} I.,  2022, \mn@doi [\apjl] {10.3847/2041-8213/ac98c5},
  \href {https://ui.adsabs.harvard.edu/abs/2022ApJ...940L..51M} {940, L51}

\bibitem[\protect\citeauthoryear{{Montes}, {Brough}, {Owers}  \&
  {Santucci}}{{Montes} et~al.}{2021}]{2021ApJ...910...45M}
{Montes} M.,  {Brough} S.,  {Owers} M.~S.,   {Santucci} G.,  2021, \mn@doi
  [\apj] {10.3847/1538-4357/abddb6}, \href
  {https://ui.adsabs.harvard.edu/abs/2021ApJ...910...45M} {910, 45}

\bibitem[\protect\citeauthoryear{{Morganson} et~al.,}{{Morganson}
  et~al.}{2018}]{2018PASP..130g4501M}
{Morganson} E.,  et~al., 2018, \mn@doi [\pasp] {10.1088/1538-3873/aab4ef},
  \href {https://ui.adsabs.harvard.edu/abs/2018PASP..130g4501M} {130, 074501}

\bibitem[\protect\citeauthoryear{{Morishita}, {Abramson}, {Treu}, {Schmidt},
  {Vulcani}  \& {Wang}}{{Morishita} et~al.}{2017}]{2017ApJ...846..139M}
{Morishita} T.,  {Abramson} L.~E.,  {Treu} T.,  {Schmidt} K.~B.,  {Vulcani} B.,
    {Wang} X.,  2017, \mn@doi [\apj] {10.3847/1538-4357/aa8403}, \href
  {https://ui.adsabs.harvard.edu/abs/2017ApJ...846..139M} {846, 139}

\bibitem[\protect\citeauthoryear{{Murante} et~al.,}{{Murante}
  et~al.}{2004}]{2004ApJ...607L..83M}
{Murante} G.,  et~al., 2004, \mn@doi [\apjl] {10.1086/421348}, \href
  {https://ui.adsabs.harvard.edu/abs/2004ApJ...607L..83M} {607, L83}

\bibitem[\protect\citeauthoryear{{Murante}, {Giovalli}, {Gerhard}, {Arnaboldi},
  {Borgani}  \& {Dolag}}{{Murante} et~al.}{2007}]{2007MNRAS.377....2M}
{Murante} G.,  {Giovalli} M.,  {Gerhard} O.,  {Arnaboldi} M.,  {Borgani} S.,
  {Dolag} K.,  2007, \mn@doi [\mnras] {10.1111/j.1365-2966.2007.11568.x}, \href
  {https://ui.adsabs.harvard.edu/abs/2007MNRAS.377....2M} {377, 2}

\bibitem[\protect\citeauthoryear{{Naiman} et~al.,}{{Naiman}
  et~al.}{2018}]{2018MNRAS.477.1206N}
{Naiman} J.~P.,  et~al., 2018, \mn@doi [\mnras] {10.1093/mnras/sty618}, \href
  {https://ui.adsabs.harvard.edu/abs/2018MNRAS.477.1206N} {477, 1206}

\bibitem[\protect\citeauthoryear{{Nelson} et~al.,}{{Nelson}
  et~al.}{2018}]{2018MNRAS.475..624N}
{Nelson} D.,  et~al., 2018, \mn@doi [\mnras] {10.1093/mnras/stx3040}, \href
  {https://ui.adsabs.harvard.edu/abs/2018MNRAS.475..624N} {475, 624}

\bibitem[\protect\citeauthoryear{{Nelson} et~al.,}{{Nelson}
  et~al.}{2019}]{2019ComAC...6....2N}
{Nelson} D.,  et~al., 2019, \mn@doi [Computational Astrophysics and Cosmology]
  {10.1186/s40668-019-0028-x}, \href
  {https://ui.adsabs.harvard.edu/abs/2019ComAC...6....2N} {6, 2}

\bibitem[\protect\citeauthoryear{{Norberg}, {Baugh}, {Gazta{\~n}aga}  \&
  {Croton}}{{Norberg} et~al.}{2009}]{2009MNRAS.396...19N}
{Norberg} P.,  {Baugh} C.~M.,  {Gazta{\~n}aga} E.,   {Croton} D.~J.,  2009,
  \mn@doi [\mnras] {10.1111/j.1365-2966.2009.14389.x}, \href
  {https://ui.adsabs.harvard.edu/abs/2009MNRAS.396...19N} {396, 19}

\bibitem[\protect\citeauthoryear{{O'Donnell} et~al.,}{{O'Donnell}
  et~al.}{2021}]{2021arXiv211002418O}
{O'Donnell} J.~H.,  et~al., 2021, arXiv e-prints, \href
  {https://ui.adsabs.harvard.edu/abs/2021arXiv211002418O} {p. arXiv:2110.02418}

\bibitem[\protect\citeauthoryear{{Oser}, {Ostriker}, {Naab}, {Johansson}  \&
  {Burkert}}{{Oser} et~al.}{2010}]{2010ApJ...725.2312O}
{Oser} L.,  {Ostriker} J.~P.,  {Naab} T.,  {Johansson} P.~H.,   {Burkert} A.,
  2010, \mn@doi [\apj] {10.1088/0004-637X/725/2/2312}, \href
  {https://ui.adsabs.harvard.edu/abs/2010ApJ...725.2312O} {725, 2312}

\bibitem[\protect\citeauthoryear{{P{\'e}rez-Hern{\'a}ndez}, {Kemp},
  {Ramirez-Siordia}  \& {Nigoche-Netro}}{{P{\'e}rez-Hern{\'a}ndez}
  et~al.}{2022}]{2022MNRAS.511..201P}
{P{\'e}rez-Hern{\'a}ndez} E.,  {Kemp} S.~N.,  {Ramirez-Siordia} V.~H.,
  {Nigoche-Netro} A.,  2022, \mn@doi [\mnras] {10.1093/mnras/stab3785}, \href
  {https://ui.adsabs.harvard.edu/abs/2022MNRAS.511..201P} {511, 201}

\bibitem[\protect\citeauthoryear{{Pillepich} et~al.,}{{Pillepich}
  et~al.}{2018}]{2018MNRAS.475..648P}
{Pillepich} A.,  et~al., 2018, \mn@doi [\mnras] {10.1093/mnras/stx3112}, \href
  {https://ui.adsabs.harvard.edu/abs/2018MNRAS.475..648P} {475, 648}

\bibitem[\protect\citeauthoryear{{Plazas}, {Bernstein}  \& {Sheldon}}{{Plazas}
  et~al.}{2014}]{2014PASP..126..750P}
{Plazas} A.~A.,  {Bernstein} G.~M.,   {Sheldon} E.~S.,  2014, \mn@doi [\pasp]
  {10.1086/677682}, \href
  {https://ui.adsabs.harvard.edu/abs/2014PASP..126..750P} {126, 750}

\bibitem[\protect\citeauthoryear{{Presotto} et~al.,}{{Presotto}
  et~al.}{2014}]{2014A&A...565A.126P}
{Presotto} V.,  et~al., 2014, \mn@doi [\aap] {10.1051/0004-6361/201323251},
  \href {https://ui.adsabs.harvard.edu/abs/2014A&A...565A.126P} {565, A126}

\bibitem[\protect\citeauthoryear{{Puchwein}, {Springel}, {Sijacki}  \&
  {Dolag}}{{Puchwein} et~al.}{2010}]{2010MNRAS.406..936P}
{Puchwein} E.,  {Springel} V.,  {Sijacki} D.,   {Dolag} K.,  2010, \mn@doi
  [\mnras] {10.1111/j.1365-2966.2010.16786.x}, \href
  {https://ui.adsabs.harvard.edu/abs/2010MNRAS.406..936P} {406, 936}

\bibitem[\protect\citeauthoryear{{Puddu} et~al.,}{{Puddu}
  et~al.}{2021}]{2021A&A...645A...9P}
{Puddu} E.,  et~al., 2021, \mn@doi [\aap] {10.1051/0004-6361/202039259}, \href
  {https://ui.adsabs.harvard.edu/abs/2021A&A...645A...9P} {645, A9}

\bibitem[\protect\citeauthoryear{{Purcell}, {Bullock}  \& {Zentner}}{{Purcell}
  et~al.}{2007}]{2007ApJ...666...20P}
{Purcell} C.~W.,  {Bullock} J.~S.,   {Zentner} A.~R.,  2007, \mn@doi [\apj]
  {10.1086/519787}, \href
  {https://ui.adsabs.harvard.edu/abs/2007ApJ...666...20P} {666, 20}

\bibitem[\protect\citeauthoryear{{Racine}}{{Racine}}{1996}]{1996PASP..108..699R}
{Racine} R.,  1996, \mn@doi [\pasp] {10.1086/133788}, \href
  {http://adsabs.harvard.edu/abs/1996PASP..108..699R} {108, 699}

\bibitem[\protect\citeauthoryear{{Radovich} et~al.,}{{Radovich}
  et~al.}{2020}]{2020MNRAS.498.4303R}
{Radovich} M.,  et~al., 2020, \mn@doi [\mnras] {10.1093/mnras/staa2705}, \href
  {https://ui.adsabs.harvard.edu/abs/2020MNRAS.498.4303R} {498, 4303}

\bibitem[\protect\citeauthoryear{{Ragusa} et~al.,}{{Ragusa}
  et~al.}{2021}]{2021A&A...651A..39R}
{Ragusa} R.,  et~al., 2021, \mn@doi [\aap] {10.1051/0004-6361/202039921}, \href
  {https://ui.adsabs.harvard.edu/abs/2021A&A...651A..39R} {651, A39}

\bibitem[\protect\citeauthoryear{{Ragusa} et~al.,}{{Ragusa}
  et~al.}{2022}]{2022arXiv221206164R}
{Ragusa} R.,  et~al., 2022, \mn@doi [arXiv e-prints]
  {10.48550/arXiv.2212.06164}, \href
  {https://ui.adsabs.harvard.edu/abs/2022arXiv221206164R} {p. arXiv:2212.06164}

\bibitem[\protect\citeauthoryear{{Rozo} et~al.,}{{Rozo}
  et~al.}{2010}]{2010ApJ...708..645R}
{Rozo} E.,  et~al., 2010, \mn@doi [\apj] {10.1088/0004-637X/708/1/645}, \href
  {https://ui.adsabs.harvard.edu/abs/2010ApJ...708..645R} {708, 645}

\bibitem[\protect\citeauthoryear{Rudick, Mihos  \& McBride}{Rudick
  et~al.}{2006}]{Rudick_2006}
Rudick C.~S.,  Mihos J.~C.,   McBride C.,  2006, \mn@doi [The Astrophysical
  Journal] {10.1086/506176}, 648, 936

\bibitem[\protect\citeauthoryear{{Rudick}, {Mihos}, {Frey}  \&
  {McBride}}{{Rudick} et~al.}{2009}]{2009ApJ...699.1518R}
{Rudick} C.~S.,  {Mihos} J.~C.,  {Frey} L.~H.,   {McBride} C.~K.,  2009,
  \mn@doi [\apj] {10.1088/0004-637X/699/2/1518}, \href
  {https://ui.adsabs.harvard.edu/abs/2009ApJ...699.1518R} {699, 1518}

\bibitem[\protect\citeauthoryear{{Rudick}, {Mihos}  \& {McBride}}{{Rudick}
  et~al.}{2011}]{2011ApJ...732...48R}
{Rudick} C.~S.,  {Mihos} J.~C.,   {McBride} C.~K.,  2011, \mn@doi [\apj]
  {10.1088/0004-637X/732/1/48}, \href
  {https://ui.adsabs.harvard.edu/abs/2011ApJ...732...48R} {732, 48}

\bibitem[\protect\citeauthoryear{{Rykoff} et~al.,}{{Rykoff}
  et~al.}{2014}]{2014ApJ...785..104R}
{Rykoff} E.~S.,  et~al., 2014, \mn@doi [\apj] {10.1088/0004-637X/785/2/104},
  \href {http://adsabs.harvard.edu/abs/2014ApJ...785..104R} {785, 104}

\bibitem[\protect\citeauthoryear{{Rykoff} et~al.,}{{Rykoff}
  et~al.}{2016}]{2016ApJS..224....1R}
{Rykoff} E.~S.,  et~al., 2016, \mn@doi [\apjs] {10.3847/0067-0049/224/1/1},
  \href {http://adsabs.harvard.edu/abs/2016ApJS..224....1R} {224, 1}

\bibitem[\protect\citeauthoryear{{Sampaio-Santos} et~al.,}{{Sampaio-Santos}
  et~al.}{2021}]{2021MNRAS.501.1300S}
{Sampaio-Santos} H.,  et~al., 2021, \mn@doi [\mnras] {10.1093/mnras/staa3680},
  \href {https://ui.adsabs.harvard.edu/abs/2021MNRAS.501.1300S} {501, 1300}

\bibitem[\protect\citeauthoryear{{Saro} et~al.,}{{Saro}
  et~al.}{2015}]{2015MNRAS.454.2305S}
{Saro} A.,  et~al., 2015, \mn@doi [\mnras] {10.1093/mnras/stv2141}, \href
  {https://ui.adsabs.harvard.edu/abs/2015MNRAS.454.2305S} {454, 2305}

\bibitem[\protect\citeauthoryear{{Sarron}, {Martinet}, {Durret}  \&
  {Adami}}{{Sarron} et~al.}{2018}]{2018A&A...613A..67S}
{Sarron} F.,  {Martinet} N.,  {Durret} F.,   {Adami} C.,  2018, \mn@doi [\aap]
  {10.1051/0004-6361/201731981}, \href
  {https://ui.adsabs.harvard.edu/abs/2018A&A...613A..67S} {613, A67}

\bibitem[\protect\citeauthoryear{{Sevilla-Noarbe} et~al.,}{{Sevilla-Noarbe}
  et~al.}{2021}]{2021ApJS..254...24S}
{Sevilla-Noarbe} I.,  et~al., 2021, \mn@doi [\apjs] {10.3847/1538-4365/abeb66},
  \href {https://ui.adsabs.harvard.edu/abs/2021ApJS..254...24S} {254, 24}

\bibitem[\protect\citeauthoryear{{Sevilla} et~al.,}{{Sevilla}
  et~al.}{2011}]{2011arXiv1109.6741S}
{Sevilla} I.,  et~al., 2011, arXiv e-prints, \href
  {https://ui.adsabs.harvard.edu/abs/2011arXiv1109.6741S} {p. arXiv:1109.6741}

\bibitem[\protect\citeauthoryear{{Shin}, {Lee}, {Hwang}, {Song}, {Ko}, {Smith},
  {Kim}  \& {Yoo}}{{Shin} et~al.}{2022}]{2022ApJ...934...43S}
{Shin} J.,  {Lee} J.~C.,  {Hwang} H.~S.,  {Song} H.,  {Ko} J.,  {Smith} R.,
  {Kim} J.-W.,   {Yoo} J.,  2022, \mn@doi [\apj] {10.3847/1538-4357/ac7961},
  \href {https://ui.adsabs.harvard.edu/abs/2022ApJ...934...43S} {934, 43}

\bibitem[\protect\citeauthoryear{{Sommer-Larsen}}{{Sommer-Larsen}}{2006}]{2006MNRAS.369..958S}
{Sommer-Larsen} J.,  2006, \mn@doi [\mnras] {10.1111/j.1365-2966.2006.10352.x},
  \href {https://ui.adsabs.harvard.edu/abs/2006MNRAS.369..958S} {369, 958}

\bibitem[\protect\citeauthoryear{{Spavone} et~al.,}{{Spavone}
  et~al.}{2020}]{2020A&A...639A..14S}
{Spavone} M.,  et~al., 2020, \mn@doi [\aap] {10.1051/0004-6361/202038015},
  \href {https://ui.adsabs.harvard.edu/abs/2020A&A...639A..14S} {639, A14}

\bibitem[\protect\citeauthoryear{{Springel} et~al.,}{{Springel}
  et~al.}{2018}]{2018MNRAS.475..676S}
{Springel} V.,  et~al., 2018, \mn@doi [\mnras] {10.1093/mnras/stx3304}, \href
  {https://ui.adsabs.harvard.edu/abs/2018MNRAS.475..676S} {475, 676}

\bibitem[\protect\citeauthoryear{{Stott} et~al.,}{{Stott}
  et~al.}{2010}]{2010ApJ...718...23S}
{Stott} J.~P.,  et~al., 2010, \mn@doi [\apj] {10.1088/0004-637X/718/1/23},
  \href {https://ui.adsabs.harvard.edu/abs/2010ApJ...718...23S} {718, 23}

\bibitem[\protect\citeauthoryear{{Tal} \& {van Dokkum}}{{Tal} \& {van
  Dokkum}}{2011}]{2011ApJ...731...89T}
{Tal} T.,  {van Dokkum} P.~G.,  2011, \mn@doi [\apj]
  {10.1088/0004-637X/731/2/89}, \href
  {https://ui.adsabs.harvard.edu/abs/2011ApJ...731...89T} {731, 89}

\bibitem[\protect\citeauthoryear{{Tang}, {Lin}, {Cui}, {Kang}, {Wang},
  {Contini}  \& {Yu}}{{Tang} et~al.}{2018}]{2018ApJ...859...85T}
{Tang} L.,  {Lin} W.,  {Cui} W.,  {Kang} X.,  {Wang} Y.,  {Contini} E.,   {Yu}
  Y.,  2018, \mn@doi [\apj] {10.3847/1538-4357/aabd78}, \href
  {https://ui.adsabs.harvard.edu/abs/2018ApJ...859...85T} {859, 85}

\bibitem[\protect\citeauthoryear{{Ventimiglia}, {Arnaboldi}  \&
  {Gerhard}}{{Ventimiglia} et~al.}{2011}]{2011A&A...528A..24V}
{Ventimiglia} G.,  {Arnaboldi} M.,   {Gerhard} O.,  2011, \mn@doi [\aap]
  {10.1051/0004-6361/201015982}, \href
  {https://ui.adsabs.harvard.edu/abs/2011A&A...528A..24V} {528, A24}

\bibitem[\protect\citeauthoryear{Virtanen et~al.,}{Virtanen
  et~al.}{2020}]{2020SciPy-NMeth}
Virtanen P.,  et~al., 2020, \mn@doi [Nature Methods]
  {10.1038/s41592-019-0686-2}, \href {https://rdcu.be/b08Wh} {17, 261}

\bibitem[\protect\citeauthoryear{{Wetzell} et~al.,}{{Wetzell}
  et~al.}{2022}]{2022MNRAS.514.4696W}
{Wetzell} V.,  et~al., 2022, \mn@doi [\mnras] {10.1093/mnras/stac1623}, \href
  {https://ui.adsabs.harvard.edu/abs/2022MNRAS.514.4696W} {514, 4696}

\bibitem[\protect\citeauthoryear{{Wu} et~al.,}{{Wu}
  et~al.}{2022}]{2022MNRAS.515.4471W}
{Wu} H.-Y.,  et~al., 2022, \mn@doi [\mnras] {10.1093/mnras/stac2048}, \href
  {https://ui.adsabs.harvard.edu/abs/2022MNRAS.515.4471W} {515, 4471}

\bibitem[\protect\citeauthoryear{{Yang}, {Mo}  \& {van den Bosch}}{{Yang}
  et~al.}{2008}]{2008ApJ...676..248Y}
{Yang} X.,  {Mo} H.~J.,   {van den Bosch} F.~C.,  2008, \mn@doi [\apj]
  {10.1086/528954}, \href
  {https://ui.adsabs.harvard.edu/abs/2008ApJ...676..248Y} {676, 248}

\bibitem[\protect\citeauthoryear{{Yoo}, {Ko}, {Kim}  \& {Kim}}{{Yoo}
  et~al.}{2021}]{2021MNRAS.508.2634Y}
{Yoo} J.,  {Ko} J.,  {Kim} J.-W.,   {Kim} H.,  2021, \mn@doi [\mnras]
  {10.1093/mnras/stab2707}, \href
  {https://ui.adsabs.harvard.edu/abs/2021MNRAS.508.2634Y} {508, 2634}

\bibitem[\protect\citeauthoryear{{Zhang} et~al.,}{{Zhang}
  et~al.}{2016}]{2016ApJ...816...98Z}
{Zhang} Y.,  et~al., 2016, \mn@doi [\apj] {10.3847/0004-637X/816/2/98}, \href
  {https://ui.adsabs.harvard.edu/abs/2016ApJ...816...98Z} {816, 98}

\bibitem[\protect\citeauthoryear{{Zhang} et~al.,}{{Zhang}
  et~al.}{2019a}]{2019MNRAS.487.2578Z}
{Zhang} Y.,  et~al., 2019a, \mn@doi [\mnras] {10.1093/mnras/stz1361}, \href
  {https://ui.adsabs.harvard.edu/abs/2019MNRAS.487.2578Z} {487, 2578}

\bibitem[\protect\citeauthoryear{{Zhang} et~al.,}{{Zhang}
  et~al.}{2019b}]{2019MNRAS.488....1Z}
{Zhang} Y.,  et~al., 2019b, \mn@doi [\mnras] {10.1093/mnras/stz1612}, \href
  {https://ui.adsabs.harvard.edu/abs/2019MNRAS.488....1Z} {488, 1}

\bibitem[\protect\citeauthoryear{{Zhang} et~al.,}{{Zhang}
  et~al.}{2019c}]{2019ApJ...874..165Z}
{Zhang} Y.,  et~al., 2019c, \mn@doi [\apj] {10.3847/1538-4357/ab0dfd}, \href
  {https://ui.adsabs.harvard.edu/abs/2019ApJ...874..165Z} {874, 165}

\bibitem[\protect\citeauthoryear{{Zibetti}, {White}  \& {Brinkmann}}{{Zibetti}
  et~al.}{2004}]{2004MNRAS.347..556Z}
{Zibetti} S.,  {White} S. D.~M.,   {Brinkmann} J.,  2004, \mn@doi [\mnras]
  {10.1111/j.1365-2966.2004.07235.x}, \href
  {https://ui.adsabs.harvard.edu/abs/2004MNRAS.347..556Z} {347, 556}

\bibitem[\protect\citeauthoryear{{Zibetti}, {White}, {Schneider}  \&
  {Brinkmann}}{{Zibetti} et~al.}{2005}]{z05}
{Zibetti} S.,  {White} S. D.~M.,  {Schneider} D.~P.,   {Brinkmann} J.,  2005,
  \mn@doi [\mnras] {10.1111/j.1365-2966.2005.08817.x}, \href
  {https://ui.adsabs.harvard.edu/abs/2005MNRAS.358..949Z} {358, 949}

\bibitem[\protect\citeauthoryear{{Zwicky}}{{Zwicky}}{1951}]{1951PASP...63...61Z}
{Zwicky} F.,  1951, \mn@doi [\pasp] {10.1086/126318}, \href
  {https://ui.adsabs.harvard.edu/abs/1951PASP...63...61Z} {63, 61}

\bibitem[\protect\citeauthoryear{{Zwicky}}{{Zwicky}}{1952}]{1952PASP...64..242Z}
{Zwicky} F.,  1952, \mn@doi [\pasp] {10.1086/126484}, \href
  {https://ui.adsabs.harvard.edu/abs/1952PASP...64..242Z} {64, 242}

\bibitem[\protect\citeauthoryear{{van Dokkum} et~al.,}{{van Dokkum}
  et~al.}{2010}]{2010ApJ...709.1018V}
{van Dokkum} P.~G.,  et~al., 2010, \mn@doi [\apj]
  {10.1088/0004-637X/709/2/1018}, \href
  {https://ui.adsabs.harvard.edu/abs/2010ApJ...709.1018V} {709, 1018}

\makeatother
\end{thebibliography}

% Alternatively you could enter them by hand, like this:
% This method is tedious and prone to error if you have lots of references
%\begin{thebibliography}{99}
%\bibitem[\protect\citeauthoryear{Author}{2012}]{Author2012}
%Author A.~N., 2013, Journal of Improbable Astronomy, 1, 1
%\bibitem[\protect\citeauthoryear{Others}{2013}]{Others2013}
%Others S., 2012, Journal of Interesting Stuff, 17, 198
%\end{thebibliography}

%%%%%%%%%%%%%%%%%%%%%%%%%%%%%%%%%%%%%%%%%%%%%%%%%%

%%%%%%%%%%%%%%%%% APPENDICES %%%%%%%%%%%%%%%%%%%%%

\appendix
\section*{Affiliations}
{\small
$^{1}$ NSF's National Optical-Infrared Astronomy Research Laboratory, 950 N Cherry Ave, Tucson, AZ 85719, USA\\
$^{2}$ Mitchell Institute for Fundamental Physics and Astronomy, 4242 TAMU 576 University Dr, College Station, TX 77845, USA\\
$^{3}$ Department of Astronomy, Shanghai Jiao Tong University, Shanghai 200240, China\\
$^{4}$ Observat\'orio Nacional, Rua Gal. Jos\'e Cristino 77, Rio de Janeiro, RJ - 20921-400, Brazil\\
$^{5}$ Fermi National Accelerator Laboratory, P. O. Box 500, Batavia, IL 60510, USA\\
$^{6}$ Kavli Institute for Particle Astrophysics \& Cosmology, P. O. Box 2450, Stanford University, Stanford, CA 94305, USA\\
$^{7}$ SLAC National Accelerator Laboratory, Menlo Park, CA 94025, USA\\
$^{8}$ Laborat\'orio Interinstitucional de e-Astronomia - LIneA, Rua Gal. Jos\'e Cristino 77, Rio de Janeiro, RJ - 20921-400, Brazil\\
$^{9}$ Institute of Cosmology and Gravitation, University of Portsmouth, Portsmouth, PO1 3FX, UK\\
$^{10}$ University Observatory, Faculty of Physics, Ludwig-Maximilians-Universit\"at, Scheinerstr. 1, 81679 Munich, Germany\\
$^{11}$ Department of Physics \& Astronomy, University College London, Gower Street, London, WC1E 6BT, UK\\
$^{12}$ Instituto de Astrofisica de Canarias, E-38205 La Laguna, Tenerife, Spain\\
$^{13}$ Universidad de La Laguna, Dpto. Astrof\'isica, E-38206 La Laguna, Tenerife, Spain\\
$^{14}$ Institut de F\'{\i}sica d'Altes Energies (IFAE), The Barcelona Institute of Science and Technology, Campus UAB, 08193 Bellaterra (Barcelona) Spain\\
$^{15}$ Centre for Extragalactic Astronomy, Durham University, South Road, Durham DH1 3LE, UK\\
$^{16}$ Jodrell Bank Center for Astrophysics, School of Physics and Astronomy, University of Manchester, Oxford Road, Manchester, M13 9PL, UK\\
$^{17}$ University of Nottingham, School of Physics and Astronomy, Nottingham NG7 2RD, UK\\
$^{18}$ Astronomy Unit, Department of Physics, University of Trieste, via Tiepolo 11, I-34131 Trieste, Italy\\
$^{19}$ INAF-Osservatorio Astronomico di Trieste, via G. B. Tiepolo 11, I-34143 Trieste, Italy\\
$^{20}$ Institute for Fundamental Physics of the Universe, Via Beirut 2, 34014 Trieste, Italy\\
$^{21}$ Hamburger Sternwarte, Universit\"{a}t Hamburg, Gojenbergsweg 112, 21029 Hamburg, Germany\\
$^{22}$ School of Mathematics and Physics, University of Queensland,  Brisbane, QLD 4072, Australia\\
$^{23}$ Department of Physics, IIT Hyderabad, Kandi, Telangana 502285, India\\
$^{24}$ Institute of Theoretical Astrophysics, University of Oslo. P.O. Box 1029 Blindern, NO-0315 Oslo, Norway\\
$^{25}$ Kavli Institute for Cosmological Physics, University of Chicago, Chicago, IL 60637, USA\\
$^{26}$ Center for Astrophysical Surveys, National Center for Supercomputing Applications, 1205 West Clark St., Urbana, IL 61801, USA\\
$^{27}$ Department of Astronomy, University of Illinois at Urbana-Champaign, 1002 W. Green Street, Urbana, IL 61801, USA\\
$^{28}$ Santa Cruz Institute for Particle Physics, Santa Cruz, CA 95064, USA\\
$^{29}$ Center for Cosmology and Astro-Particle Physics, The Ohio State University, Columbus, OH 43210, USA\\
$^{30}$ Department of Physics, The Ohio State University, Columbus, OH 43210, USA\\
$^{31}$ Center for Astrophysics $\vert$ Harvard \& Smithsonian, 60 Garden Street, Cambridge, MA 02138, USA\\
$^{32}$ Australian Astronomical Optics, Macquarie University, North Ryde, NSW 2113, Australia\\
$^{33}$ Lowell Observatory, 1400 Mars Hill Rd, Flagstaff, AZ 86001, USA\\
$^{34}$ Jet Propulsion Laboratory, California Institute of Technology, 4800 Oak Grove Dr., Pasadena, CA 91109, USA\\
$^{35}$ Departamento de F\'isica Matem\'atica, Instituto de F\'isica, Universidade de S\~ao Paulo, CP 66318, S\~ao Paulo, SP, 05314-970, Brazil\\
$^{36}$ Centro de Investigaciones Energ\'eticas, Medioambientales y Tecnol\'ogicas (CIEMAT), Madrid, Spain\\
$^{37}$ Instituci\'o Catalana de Recerca i Estudis Avan\c{c}ats, E-08010 Barcelona, Spain\\
$^{38}$ Department of Physics, Carnegie Mellon University, Pittsburgh, Pennsylvania 15312, USA\\
$^{39}$ Department of Physics and Astronomy, Pevensey Building, University of Sussex, Brighton, BN1 9QH, UK\\
$^{40}$ School of Physics and Astronomy, University of Southampton,  Southampton, SO17 1BJ, UK\\
$^{41}$ Computer Science and Mathematics Division, Oak Ridge National Laboratory, Oak Ridge, TN 37831\\
$^{42}$ Department of Physics, University of Michigan, Ann Arbor, MI 48109, USA\\
$^{43}$ Lawrence Berkeley National Laboratory, 1 Cyclotron Road, Berkeley, CA 94720, USA\\
}

% Don't change these lines
\bsp	% typesetting comment
\label{lastpage}
\end{document}